\PassOptionsToPackage{comma,numbers,sort&compress}{natbib}
\documentclass[preprint,10pt]{elsarticle}
\usepackage[a4paper,margin=1.5cm]{geometry}
\usepackage{amssymb}
\usepackage{amsmath}
\usepackage{hyperref}
\usepackage{booktabs}
\usepackage{subcaption}
\usepackage{multirow}
\usepackage{float}
\usepackage{enumitem}
\usepackage{multicol}
\usepackage{mdframed}
\usepackage{array}
\usepackage{etoolbox}
\usepackage{longtable}
\usepackage{caption}
\usepackage[section]{placeins}
\usepackage{tabularx}
\usepackage{rotating}
\usepackage{pdflscape}
\usepackage{makecell}
\usepackage{afterpage}
\usepackage{textcomp}
\usepackage{graphicx}

\hypersetup{
    pdftitle={Block-modular grey-box identification of a short-stroke seawater-pump power take-off for wave energy: loss decomposition and multi-regime validation},
    pdfauthor={Mahdi Tayyebati},
    pdfsubject={Wave energy; power take-off; grey-box identification},
    pdfkeywords={wave energy converter, power take-off, seawater pump, grey-box identification, loss decomposition},
    pageanchor=false,
    colorlinks=true,
    linkcolor=blue,
    citecolor=blue,
    urlcolor=blue
}

\journal{Renewable Energy}

\makeatletter
\AtBeginDocument{\check@mathfonts}
\makeatother

\begin{document}
\begin{frontmatter}
\title{Block-modular grey-box identification of a short-stroke seawater-pump power take-off for wave energy: loss decomposition and multi-regime validation}

\author[inst1]{Mahdi Tayyebati\corref{cor1}}
\ead[cor1]{matay@dtu.dk}
\cortext[cor1]{Corresponding author.}
\author[inst1]{Amir Hamza Siddiqui}

\author[inst2]{Mian Masoud}

\author[inst2]{Kristian Glejb{\o}l}

\author[inst1]{Christian Berggreen}

\affiliation[inst1]{organization={Department of Civil and Mechanical Engineering, Technical University of Denmark},
            addressline={DK-2800},
            city={Kongens Lyngby},
            country={Denmark}}

\affiliation[inst2]{organization={Wavepiston A/S},
            addressline={DK-3000},
            city={Helsing{\o}r},
            country={Denmark}}

\begin{abstract}
Physically interpretable power take-off models support wave-energy development at every stage. For the seawater-pump class, few have been identified on a real machine and validated at the system level. This paper presents a block-modular grey-box model of a short-stroke seawater pump, identified from bench data alone. The pump decomposes into chamber, relief valve, piston seal and tip check valve, each structured by component physics, and the few parameters are identified block by block from terminal pressure, displacement, force and discharged-mass measurements, with a closed-cycle mass balance that decouples the degenerate seal and valve losses and closes to within 1\% of the inflow. Over the sinusoidal campaign, the pressure-energised piston seal carries 84\% of the lost mass and 97\% of the lost energy, and the pump delivers about four-fifths of its input as high-pressure flow. Generalisation is demonstrated rather than assumed. With parameters fixed on ramps and sinusoids, the model reproduces 360\,s sea-state records in free run to about 10\% of range, captures 80\% of several thousand peaks within 5\,bar, transfers to a re-sprung valve on cracking strokes once its crack pressure is identified, follows thousand-cycle endurance records that drift below 0.1\,bar, and reproduces the force and cycle-mean absorbed power, 0.1 to 5.7\,kW, to a 4\% median error. A practical-identifiability and uncertainty analysis confirms the terminal data resolve the retained parameters; thirty-seven repeated records set the experimental floor. The result is a compact, physically labelled model for design, energy assessment and model-based control of seawater-pump power take-offs.
\end{abstract}


\end{frontmatter}

\section*{Nomenclature}

\begingroup
\renewcommand{\theHtable}{nomenclature.\arabic{table}}
\footnotesize
\setlength{\tabcolsep}{4pt}
\begin{longtable}{>{\centering\arraybackslash}p{2.9cm}>{\centering\arraybackslash}p{1.6cm}p{11.1cm}}
\hline
\textbf{Variable} & \textbf{Unit} & \textbf{Description} \\
\hline
\endfirsthead
\hline
\textbf{Variable} & \textbf{Unit} & \textbf{Description} \\
\hline
\endhead
\hline
\endfoot
\hline
\endlastfoot
$A_{\mathrm{eff}}(p)$ & m$^2$ & Relief-valve effective flow area \\
$A_P$ & m$^2$ & Effective piston area \\
$a$ & m$^2$ & Valve area scale (effective area at 10\,bar over-crack) \\
$b$ & -- & Valve area exponent \\
$C$ & m$^2$ & Energised-seal coefficient \\
$C_b$ & m$^3$\,s$^{-1}$ & Seal blow-by coefficient (flow at 10\,bar over-onset) \\
$C_t$ & m$^2$ & Tip discharge coefficient \\
$F$ & N & Actuator (rig) force \\
$F_{\mathrm{fric}}$ & N & Coulomb friction magnitude \\
$\langle F\dot{x}\rangle$ & W & Cycle-mean mechanical (rod) power \\
$g_{\mathrm{tip}}$ & -- & Travel-gated tip back-flow window indicator \\
$H_s$ & mm & Displacement significant height ($4\sigma_x$) \\
$k$ & m$^3$\,mm\,s$^{-1}$ & Tip dead-band constant \\
$m_{20}$ & kg & Measured 20-cycle valve discharge mass \\
$m_b$ & -- & Seal blow-by exponent \\
$m_{\mathrm{in}}$ & kg & Inflow mass over the window \\
$m_{\mathrm{nonvalve}}$ & kg & Non-valve loss mass, $m_{\mathrm{in}}-m_{20}$ \\
$m_{\mathrm{seal}}$ & kg & Clean seal mass target \\
$m_{\mathrm{tipback}},\ m_{\mathrm{tipleak}}$ & kg & Tip re-seat and tip back-leak masses \\
$P_{\mathrm{atm}}$ & Pa & Atmospheric pressure \\
$P_{\mathrm{back}},\ P_{\mathrm{tank}}$ & Pa & Valve back and tank pressures (equal to $P_{\mathrm{atm}}$ on the rig) \\
$P_{\mathrm{ref}}$ & Pa & Seal-exponent normaliser \\
$P_{\mathrm{ref},v}$ & Pa & Valve-area normaliser (10\,bar) \\
$p$ & Pa & Absolute chamber pressure (model state) \\
$p_{\mathrm{crack}}$ & bar (g) & Valve crack (set) pressure \\
$p_g$ & bar & Chamber gauge pressure, $(p-P_{\mathrm{atm}})/10^5$ \\
$p_{\mathrm{on}}$ & bar (g) & Seal blow-by onset pressure \\
$Q_{\mathrm{blow}}$ & m$^3$\,s$^{-1}$ & Seal blow-by flow \\
$Q_{\mathrm{film}}$ & m$^3$\,s$^{-1}$ & Pressure-energised seal film leak \\
$Q_{\mathrm{tip,back}}$ & m$^3$\,s$^{-1}$ & Motion-driven tip re-seating diversion \\
$Q_{\mathrm{tip,leak}}$ & m$^3$\,s$^{-1}$ & Steady pressure-driven tip back-leak (valve unseated) \\
$Q_{\mathrm{valve}}$ & m$^3$\,s$^{-1}$ & Relief-valve discharge flow \\
$r$ & -- & Entrained-gas compression ratio, $(P_{\mathrm{atm}}/p)^{1/\kappa}$ \\
$s$ & -- & Tip check-valve seating state ($0$ open, $1$ seated) \\
$T_m$ & s & Displacement-spectrum mean period ($m_0/m_1$) \\
$T_p$ & s & Displacement-spectrum peak period \\
$V(x)$ & m$^3$ & Instantaneous chamber volume, $V_0-A_P x$ \\
$V_0$ & m$^3$ & Chamber dead volume (includes connecting volumes) \\
$V_{\varepsilon}$ & m\,s$^{-1}$ & Tip velocity threshold (noise reject at reversal) \\
$v_{\mathrm{peak}}$ & mm\,s$^{-1}$ & Peak piston velocity \\
$v_{\mathrm{rev}}$ & mm\,s$^{-1}$ & Piston reversal speed \\
$v_{\mathrm{rms}}$ & mm\,s$^{-1}$ & Root-mean-square piston velocity \\
$x,\ \dot{x}$ & m,\ m\,s$^{-1}$ & Piston displacement and velocity (measured inputs) \\
$\alpha$ & -- & Entrained-air fraction parameter \\
$\beta_{\mathrm{eff}}(p)$ & Pa & Effective bulk modulus of the chamber fluid \\
$\beta_L$ & Pa & Bulk modulus of pure water \\
$\gamma$ & -- & Seal softening exponent \\
$\delta V,\ \delta x$ & m$^3$,\ m & Per-cycle dead-band volume and travel \\
$\Delta p_s$, $\Delta p_t$, $\Delta p_v$ & Pa & Seal, tip and valve over-pressures \\
$\kappa$ & -- & Adiabatic gas exponent \\
$\rho$ & kg\,m$^{-3}$ & Fluid (water) density \\

\end{longtable}
\endgroup
\setcounter{table}{0}

\section*{Abbreviations}

\begingroup
\renewcommand{\theHtable}{abbreviations.\arabic{table}}
\footnotesize
\setlength{\tabcolsep}{4pt}
\begin{longtable}{>{\centering\arraybackslash}p{2.2cm}p{13.7cm}}
\hline
\textbf{Abbreviation} & \textbf{Description} \\
\hline
\endfirsthead
\hline
\textbf{Abbreviation} & \textbf{Description} \\
\hline
\endhead
\hline
\endfoot
\hline
\endlastfoot
WEC   & Wave energy converter \\
PTO   & Power take-off \\
EC    & Energy collector (Wavepiston A/S oscillating plate) \\
EU    & European Union \\
SHY   & Seawater HYdraulic PTO using dynamic passive controller for WECs (EU project) \\
DN    & Nominal diameter (pipework and valve connections) \\
PT    & Pressure-transmitter balloon in Fig.~\ref{Figure_1} \\
WT    & Force-transducer balloon in Fig.~\ref{Figure_1} \\
ZT    & Position-transducer balloon in Fig.~\ref{Figure_1} \\
POM   & Polyoxymethylene (piston seal material) \\
CFD   & Computational fluid dynamics \\
ODE   & Ordinary differential equation \\
LSODA & Livermore solver for ODEs with automatic method switching \\
OLS   & Ordinary least squares \\
RMS   & Root mean square \\
RMSE  & Root-mean-square error \\
NRMSE & Normalised root-mean-square error (range-normalised) \\
SD    & Standard deviation \\
UQ    & Uncertainty quantification \\
OOD   & Out-of-distribution \\
IEC   & International Electrotechnical Commission \\
ISO   & International Organization for Standardization \\

\end{longtable}
\endgroup

\setcounter{table}{0}

\section{Introduction}

The power take-off (PTO) of a wave energy converter (WEC) turns slow, reciprocating and highly variable wave-induced motion into useful output, and it carries a large share of the conversion losses, the failure modes and the cost of the device. Reliable PTO models are therefore needed at every stage of development: for design and sizing, for energy assessment, for model-based control and, finally, for condition monitoring. These needs are particularly pressing for the seawater-pump PTO class, in which the absorber drives a reciprocating pump that pressurises seawater directly, an architecture of growing interest for wave-powered pumping and reverse-osmosis desalination \cite{simmons2025,zhang2019}. The pressurised flow is the generating medium of this architecture. Onshore it drives a turbine-generator set, or it feeds reverse-osmosis membranes directly, so hydraulic power that the pump fails to deliver is power the device never generates. Devices of this class have operated at sea, most prominently the Oyster flap, which pumped high-pressure seawater ashore to a hydroelectric power station \cite{whittaker2012}, but their pumps have generally been represented by design-level models rather than identified from measurements of the machine.

A first, well-established modelling tradition addresses the PTO within whole-system wave-to-wire models \cite{penalba2016}. It was found early on, from component-level hydraulic PTO models driven by realistic sea data, that over-simplified PTO representations distort power estimates and can mislead design decisions \cite{cargo2016}. High-fidelity wave-to-wire platforms have since coupled numerical wave tanks based on computational fluid dynamics (CFD) to detailed PTO and generator models \cite{penalba2018,asiikkis2024}; nonlinearity measures have been proposed to decide which dynamics a given application requires \cite{penalba2019}; and systematic complexity-reduction methodologies now derive fast, application-tailored variants from such reference models \cite{penalba2020}. A balanced wave-to-wire model carrying every conversion stage from the wave to the grid has been assembled for several device and hydraulic-PTO topologies \cite{penalba2019w2w}, and the demands that such a PTO places on its electrical drive have been measured in hardware-in-the-loop tests \cite{gaspar2017}. Statistical simulation studies have also connected hydraulic operating parameters to sea-state characteristics across full scatter diagrams \cite{zeinali2024}. Two traits are shared across this tradition. Loss terms are constructed forward, from catalogue data, manufacturer coefficients or generic loss models, rather than identified from measurements of the machine they describe. Validation is also largely simulation-based; experimental confirmation, when present, is typically drawn from component-level studies rather than conducted on the assembled system.

A second tradition models a single component in isolation at high resolution. Validated flow and dynamic models of direct spring-loaded relief valves \cite{zong2020}, CFD studies of self-acting pump valves under transient flow \cite{menendezblanco2019}, and elastohydrodynamic lubrication models of reciprocating seals \cite{schmidt2010} resolve the internal physics of each element in detail. Such models are accurate within their scope, but they are computationally expensive; they treat one component decoupled from the circuit that excites it, and their internal parameters cannot be recovered from the terminal measurements available on an assembled machine. They are therefore a source of model structure for system-level work rather than a substitute for it.

A third and smaller body of work identifies WEC dynamics directly from measured data. Linear and polynomial nonlinear model structures have been identified from real wave-tank experiments \cite{giorgi2019}, and deep operator networks have recently been trained on tank data as continuous-time surrogates for a hinged-raft device \cite{zhang2023}. Work on estimation and control has also moved towards realistic conditions, and this shift has been deliberate. Nonlinear Kalman filtering has been applied to experimental data \cite{davis2020}, and the extent to which model mismatch and prediction errors erode the theoretical control gains has been quantified \cite{hillis2020}. The identification strand, however, has remained largely black-box. The resulting models contain few physically meaningful parameters, provide no decomposition of where energy is lost, and are usually validated within the distribution of the training campaign; deliberate out-of-distribution (OOD) tests, in which the hardware itself is altered between training and validation, are rare.

A gap therefore remains between these traditions. The literature reviewed above does not contain a parsimonious, physically interpretable grey-box model of a real seawater WEC pump identified and validated at the system level against measured data. Bench and wave-tank validation of assembled hydraulic power take-offs has been reported \cite{dang2020,giorgi2024}, but for architectures other than a seawater pump and without a physically interpretable, data-identified decomposition of the losses. Nor has such a model been shown to transfer across excitation classes, to remain predictive after a deliberate hardware modification, and to hold throughout extended operation. The closest application studies, which are design analyses of hydraulic PTO architectures for wave-powered reverse-osmosis desalination \cite{simmons2025,simmons2023}, rely on idealised component models and are not identified against measurements from a physical machine. This paper addresses that gap for a physical seawater pump on a test rig, where a laboratory hydraulic actuator imposes motion profiles that include displacement records derived from irregular sea states. The device is not deployed at sea, and no claims are made regarding field performance. The pump comprises a direct spring-loaded relief valve with a crack pressure of approximately 60 bar, a pressure-energised polymer piston seal and a passive sliding check valve, drawing from and discharging to a single low-pressure tank. Its pumping cycle has a distinctive shape: suction, a dead band, a sharp compression to the crack pressure, a discharge plateau regulated by the relief-valve pressure-override characteristic, and decompression of the trapped fluid at motion reversal. The model is block-modular and grey-box: the structure of each block is fixed by the component physics, in the spirit of the detailed component literature, and only a small number of parameters per block are identified from data. The identified model decomposes the losses by mechanism. Seal leakage is represented as two regimes of a single path, a pressure-energised film leak and a pressure-gated blow-by term, and this combined channel is found to be the dominant loss. The remaining channels are relief-valve discharge with a pressure-dependent effective area, and check-valve back-flow driven by motion and by pressure. The validation programme is built to expose failures of generalisation. With every parameter frozen on ramps and sinusoids, the model reproduces the response under irregular sea-state profiles, remains predictive on an OOD test in which the relief valve is deliberately re-sprung, and holds throughout extended operation of approximately 1000 pumping cycles.

The contribution of the paper is this combination. A block-modular grey-box identification methodology is applied to a complete reciprocating seawater-pump PTO, in which the structure of each block is fixed from physics and the few parameters per block are identified from terminal measurements. The loss budget is resolved into physical mechanisms, with shares recovered from the bench data themselves, not assumed in advance from catalogue coefficients, and it separates the seal channel, with its film leak and blow-by, from the relief-valve discharge and from the check-valve back-flow. The multi-regime generalisation protocol spans ramp and sinusoidal excitation, real irregular sea-state profiles, an OOD re-sprung relief valve and extended operation of approximately 1000 cycles. A practical-identifiability analysis with parameter uncertainty quantification (UQ) completes the assessment, together with a validation approach for irregular excitation that rests on peak values, spectra and distributions rather than on point-by-point error. The body of the paper establishes these claims in the order given, on measurements from one physical machine, and every identified parameter, loss share and error bar can be traced back to the bench records that produced it.

\section{Experimental system and test programme}
\subsection{Test rig and pump}
\label{sec:rig}

\begin{figure}[!ht]\centering
\includegraphics[width=\linewidth]{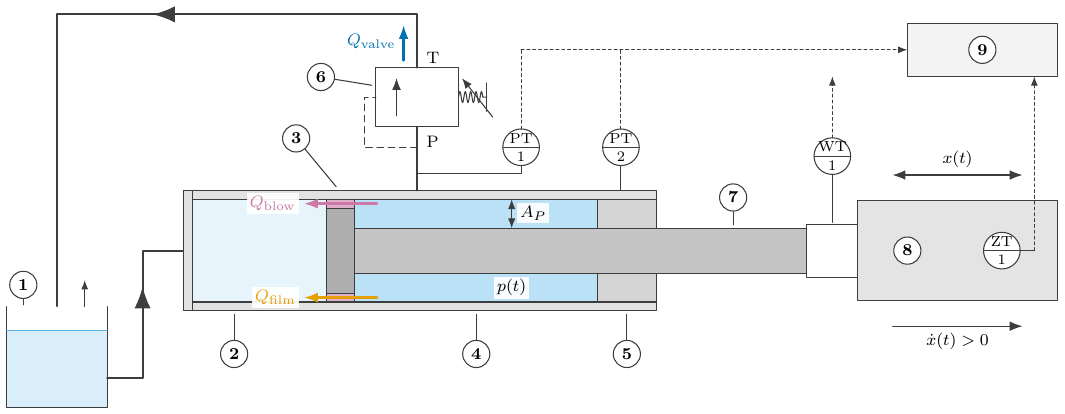}
\caption{Bench test rig, not to scale: 1 low-pressure reservoir, 2 head-side volume, 3 annular piston carrying the seal pack and the tip check valve, 4 annular pumping chamber, 5 discharge block and gland, 6 relief valve, 7 piston rod, 8 servo-hydraulic actuator, 9 data acquisition. Continuous lines carry water; dashed lines mark the valve pilot connection and instrument signals; balloons PT, WT and ZT are the pressure, force and position instruments.}
\label{Figure_1}\end{figure}

The device under test is a concept-demonstration prototype of a short-stroke reciprocating seawater pump developed as the power take-off (PTO) of the Wavepiston A/S wave energy converter, in which an oscillating energy collector (EC) plate drives the pump directly. All testing reported here was performed on a laboratory bench rig. The pump is mounted in a servo-hydraulic test frame, and an MTS 204.70S servo-hydraulic actuator (force capacity $\pm$97.9\,kN, stroke 914\,mm) imposes a prescribed displacement on the piston under closed-loop displacement control. The rig therefore reproduces the kinematic boundary condition that the EC would impose at sea, while all other quantities are measured under controlled laboratory conditions. The imposed motion spans deterministic profiles (rate-controlled ramps and constant-amplitude sinusoids) and realistic irregular records: piston displacement time series generated from a numerical simulation of a single EC for the planned offshore deployment of the EU SHY project, Froude-scaled to the reduced stroke of the short-stroke pump at a length scale factor of 12 (project motion specification, SHY project). The seven records derive from seven reference sea states at a nominal site, spanning full-scale significant heights of 1.0 to 5.0\,m and energy periods of 5 to 12\,s. Scaled back by the square root of the length factor, the imposed peak periods of Table~\ref{Table_2} correspond to about 6 to 14\,s at full scale, which sits above the quoted energy periods in the way a peak-period measure should. The EC simulation embeds its own PTO reaction law, and the records are used here purely as imposed kinematics; the achieved strokes stay within the $\pm$200\,mm test range. That law represents the full-scale Wavepiston A/S pump, whose working area and stroke both differ from the bench unit, and it assumes the same 60\,bar working pressure used here. The similitude is mixed by design. The motion is Froude-scaled while the pump operates at its full pressure settings rather than the twelvefold-reduced pressure that strict Froude scaling would impose, and any transfer of the identified losses to another scale must respect this. The working fluid on the bench is water. The rig, its components and the sensor locations are shown schematically in Fig.~\ref{Figure_1}.

\subsection{Pump construction and components}
\label{sec:pump}

The pump is a single-acting positive-displacement unit built around a lumped chamber. On the inlet side, the chamber draws water from a low-pressure tank through a passive sliding tip check valve mounted in the piston tip. The valve is an in-house design by Wavepiston A/S rather than a catalogue component. Water enters through an annular seat recess of 78\,mm inner and 82\,mm outer diameter, and the slide travels a nominal 6.5\,mm between its unseated and seated positions. The valve opens on retraction and must travel to re-seat when the motion reverses. The piston carries a pressure-energised polymer seal, a polyoxymethylene (POM) set of nominal 90\,mm diameter with a dynamic seal ring, a seal support ring and a guide and slide ring, running water-lubricated against the liner bore, and the sealing contact tightens as the chamber pressure rises. On the discharge side, the chamber is closed by a LESER Type 5264 direct spring-loaded relief valve, DN 25, with a nominal crack pressure of 60\,bar gauge, which discharges back into the same tank and presents a regulated, near-constant back-pressure load to the pump during discharge. The rig runs on a single tank, which feeds the suction side and receives the valve discharge. It is open to atmosphere, so the tank and valve back pressures equal atmospheric pressure, as assumed in Section~\ref{sec:model}. With the relief valve seated, the chamber is sealed apart from its leakage paths, which is the configuration exploited by the ramp tests of Section~\ref{sec:programme}.

Two quantities anchor the model, and neither is a bare drawing dimension. The piston area, $A_P=3.559\times10^{-3}$\,m$^2$, is an effective area: it sits close to the annulus between the liner bore and the rod, is identified from the measured rod force (Section~\ref{sec:ident}), and is then held fixed throughout the mass balance and the force readout. The chamber dead volume, $V_0=7.090\times10^{-3}$\,m$^3$, includes the connecting volumes and is identified from the bench pressure data as described in Section~\ref{sec:model}. The actuator imposes displacement amplitudes of up to $\pm$200\,mm about mid-stroke, so the largest stroke travels 400\,mm and sweeps approximately 1.4\,L. The dead volume is therefore five to ten times the swept volume across the test amplitudes, and this ratio shapes the characteristic pumping cycle described in the Introduction. Part of each compression stroke is spent compressing the chamber contents through the dead band and the compression ramp before the relief valve cracks and the regulated discharge plateau is reached, with the stored compression returning as spring-back at motion reversal.

\subsection{Instrumentation and data acquisition}
\label{sec:daq}

All channels are sampled at 1024\,Hz by the test-frame data-acquisition system. The recorded channel set varies slightly by test class. The ramp records carry the time base, the commanded and achieved piston displacement, the actuator force and the two chamber pressure channels. The cyclic records (sinusoidal and long-duration) carry the command frequency, the time base, a cycle-count channel used to delimit individual pumping cycles, the achieved displacement, the force and the two pressures. The sea-state records carry the time base, the commanded and achieved displacement, the force, the two pressure channels and the two water-temperature channels. The logged water temperature ranges from 21 to 27\,\textdegree C and drifts by less than 0.8\,K within a record. The whole programme was run on the same bench under the same laboratory conditions. Each run is therefore close to isothermal, and neither the temperature dependence of the seal nor that of the water viscosity is exercised within it. The water density is taken as 1000\,kg\,m$^{-3}$ throughout; across the logged temperatures the true value differs from this by less than 0.3\%, well inside the closure residual of Section~\ref{sec:res-loss}. Force is logged in kN. The pressure channels, Press4A and Press4B, are logged in kN\,mm$^{-2}$ and converted to Pa in processing. The chamber pressure used throughout this work is the average of the two transducers, which suppresses uncorrelated transducer noise. The two channels use ifm pressure transmitters of different span, a PT5402 with a 0--100\,bar range and a PT5401 with a 0--250\,bar range. Both are specified to a characteristics deviation below $\pm$0.5\% of span, a repeatability below $\pm$0.05\% of span, a linearity deviation below $\pm$0.1\% and a long-term stability below $\pm$0.1\% per six months. Over a full sea-state record the two track each other to a mean difference of 0.06\,bar and a largest difference of 0.41\,bar. That is well inside the $\pm$0.52\,bar peak repeatability of the bench reported in Section~\ref{sec:res-repeat}, which is what justifies averaging them. The step response is 1\,ms against a sample interval of 0.98\,ms. Transducer and sampling rate therefore limit the record at comparable frequencies, both an order of magnitude shorter than the compression-front timing offsets of Section~\ref{sec:res-seastate}, whose median absolute value is 23\,ms. The front timing quoted there reflects the model, not the instrumentation. Piston displacement is measured by the actuator-internal MTS 204.70S transducer, of $\pm$450\,mm range, and force by an MTS 661.21A-03 load cell of 100\,kN capacity. Both channels run through MTS 494.26 signal conditioners. The instrument balloons PT, WT and ZT in Fig.~\ref{Figure_1} mark the pressure, force and displacement channels, respectively. Piston velocity is not measured directly but obtained by Savitzky--Golay differentiation of the measured displacement (window 11 samples, polynomial order 3, Section~\ref{sec:numerics}).

The identification target for the relief-valve block is the mass discharged through the valve over the 20 cycles of each sinusoidal test, $m_{20}$. For these tests, the valve outlet was diverted from its tank return into a separate collection vessel standing on a balance, so the 20-cycle discharge was weighed without disturbing the single-tank circuit. The resulting masses are recorded per test in a mass log.

\subsection{Test programme and test cases}
\label{sec:programme}

The campaign comprises several classes of test, used for different purposes and referred to by clean case names throughout the paper; Table~\ref{Table_1} summarises them. The identification uses only the ramp and sinusoidal classes, and all other classes are reserved for validation.

\begin{table}[!ht]
\centering
\caption{Bench test programme, excluding the repeat runs. The ramp and sinusoidal classes identify the model; the remaining classes validate it. Three of the sea states are also repeated to quantify experimental uncertainty.}
\label{Table_1}
\footnotesize
\setlength{\tabcolsep}{4pt}
\begin{tabularx}{\linewidth}{@{}l c >{\raggedright\arraybackslash}X >{\raggedright\arraybackslash}X@{}}
\toprule
Test class & Runs & Range & Purpose \\
\midrule
Rate-controlled ramps          & 57 & 50--500\,mm\,s$^{-1}$, to 55\,bar          & identify chamber, seal, tip, force \\
Constant-amplitude sinusoids   & 26 & 100--200\,mm, 0.05--0.9\,Hz                & identify valve, blow-by, dead band; validate \\
Irregular sea states, 60\,bar  & 7  & seven states, 360\,s each                  & validate generalisation \\
Irregular sea states, 32\,bar  & 7  & re-sprung valve, 360\,s each               & validate transfer \\
Re-sprung relief valve         & 4  & 200\,mm, 0.05--0.50\,Hz                    & out-of-distribution transfer \\
Endurance                      & 2  & 200\,mm, 0.25 and 0.50\,Hz, $\sim$1000 cyc & validate stationarity \\
\bottomrule
\end{tabularx}
\end{table}

\subsubsection{Rate-controlled ramps}
With the relief valve kept seated (chamber pressure below the crack pressure), the actuator drove the piston in displacement ramps at six rates, 50 to 500\,mm\,s$^{-1}$, in stepped segments towards a 50\,bar target, with peak recorded pressures approaching 55\,bar, giving 57 ramp segments in total. Because the valve passes no flow in this class, the pressure response isolates the chamber compliance and the leakage paths. These data identify the chamber, energised-seal and tip-valve parameters, as well as the force-readout parameters (Section~\ref{sec:ident}).

\subsubsection{Sinusoidal tests}
Twenty-six datasets cover constant-amplitude displacement cycling with a sine-tapered envelope, named by amplitude and frequency, for example ``200\,mm, 0.05\,Hz''. Each test runs 20 cycles. These tests exercise the relief valve through its full discharge range and provide the 20-cycle valve mass $m_{20}$. They identify the valve area parameters, the dead-band constant and the blow-by parameters, and they serve as the pressure-trace validation set for the identified model.

\subsubsection{Irregular sea states}
For validation of the pump model, seven irregular displacement records, named Sea state 1 to Sea state 7, were generated from the EC motion simulation and Froude-scaled as described in Section~\ref{sec:rig}. Each record lasted 360\,s and was run at two relief-valve settings, a 60\,bar case and a 32\,bar case, giving fourteen primary runs. These data are used exclusively for validation. No parameter is identified from them. Three of the sea states, Sea states 1, 2 and 5, were repeated several times at each setting, and the whole 60\,bar set was rerun after the pressure controller had been reset, to characterise experimental repeatability and reproducibility (Section~\ref{sec:res-repeat}). Because the test specification defines the cases by the source sea states rather than by achieved wave statistics, Table~\ref{Table_2} characterises each case by imposed-displacement statistics derived from the achieved displacement records: the displacement significant height $H_{s}$ (four times the standard deviation), the spectral peak and mean periods $T_p$ and $T_m$, and the peak and root-mean-square piston velocity.

\begin{table}[!ht]
\centering
\caption{Irregular sea-state cases, characterised by statistics of the achieved piston displacement. These describe the imposed motion, not a wave field; each case was run at both valve settings.}
\label{Table_2}
\footnotesize
\setlength{\tabcolsep}{6pt}
\begin{tabular}{lccccc}
\toprule
Sea state & $H_s$ (mm) & $T_p$ (s) & $T_m$ (s) & $v_{\mathrm{peak}}$ (mm\,s$^{-1}$) & $v_{\mathrm{rms}}$ (mm\,s$^{-1}$) \\
\midrule
Sea state 1 & 93  & 1.78 & 1.65 & 540  & 103 \\
Sea state 2 & 60  & 1.78 & 1.72 & 495  & 66  \\
Sea state 3 & 171 & 2.00 & 2.00 & 733  & 153 \\
Sea state 4 & 297 & 2.13 & 1.93 & 1105 & 270 \\
Sea state 5 & 115 & 2.67 & 2.18 & 643  & 100 \\
Sea state 6 & 286 & 2.91 & 2.37 & 991  & 219 \\
Sea state 7 & 270 & 4.00 & 2.90 & 812  & 172 \\
\bottomrule
\end{tabular}
\end{table}

\subsubsection{Repeat runs for measurement uncertainty}
Sea states 1, 2 and 5 were each acquired five times per session at the 60\,bar setting, in the two sessions separated by the controller reset, and three times at the 32\,bar setting. Together with the single second-session re-runs of the other four 60\,bar states, this gives thirty-seven records beyond the fourteen primaries, in nine repeated groups (Section~\ref{sec:res-repeat}). These physical repeats bound the run-to-run repeatability of the bench and put the free-run validation errors of Section~\ref{sec:results} in context. The parameter-uncertainty analysis itself resamples the identification records directly (Section~\ref{sec:res-uq}).

\subsubsection{Long-duration runs}
Two endurance runs cycle the pump at 200\,mm amplitude for 1000 analysed cycles each, at 0.25\,Hz (66.7\,min) and at 0.50\,Hz (33.3\,min). They are used exclusively to assess the stationarity of the bench response throughout extended operation.

\subsubsection{Re-sprung relief valve}
In a final configuration, the relief-valve spring was adjusted by hand towards a nominal 40\,bar setting and the 200\,mm sinusoids were repeated at four frequencies (0.05, 0.15, 0.25 and 0.50\,Hz). The crack pressure actually realised by the hand adjustment is identified from the data in Section~\ref{sec:res-seastate}. This deliberately altered hardware provides the out-of-distribution diagnostic case analysed in the validation section.

\section{Block-modular model and identification methodology}
\label{sec:model}

\subsection{Modelling approach and assumptions}
\label{sec:approach}

The model follows the grey-box, block-modular philosophy outlined in the Introduction. The pump is decomposed into physical blocks (chamber, relief valve, piston seal, tip check valve, force readout); the structure of each block is fixed in advance from the component physics, and only a small number of parameters per block are identified from measured data. Throughout, the identified parameters are keyed to physical states, principally pressure and piston velocity, never to the labels of individual test runs. A parameter learned as a function of pressure applies to any excitation that visits that pressure, which lets the model generalise across excitation classes rather than interpolate between test conditions. The control-volume and compressibility formulation follows established hydraulic modelling practice \cite{merritt1967}, and the identification viewpoint follows the grey-box framework of system identification \cite{ljung1999}, applied to wave-energy devices from measured data \cite{giorgi2019}.

The model rests on the following assumptions. The pump chamber is treated as a single lumped control volume with uniform pressure. The working fluid is water carrying a small entrained-air fraction, which gives the chamber a pressure-dependent compliance. The entrained gas is assumed to compress adiabatically. The chamber walls are rigid, and all compliance is attributed to the fluid. The moving-mass inertia of the piston, the rod and the entrained water column is not carried in the force readout, which is quasi-static. On suction, the chamber is assumed to refill completely without cavitation, and the pressure state is accordingly held at or above atmospheric pressure. The final model is a single-state ordinary differential equation (ODE) in the absolute chamber pressure $p$, with the tip check valve treated as an ideal element switched by the piston velocity and the piston displacement $x(t)$ and velocity $\dot{x}(t)$ supplied as measured inputs.

Each loss term is driven by its own over-pressure. With $P_{\mathrm{atm}}$, $P_{\mathrm{tank}}$ and $P_{\mathrm{back}}$ the atmospheric, tank and valve back pressures (all equal on this rig),
\begin{equation}
\begin{aligned}
\Delta p_s&=\max(p-P_{\mathrm{atm}},0),\\
\Delta p_t&=\max(p-P_{\mathrm{tank}},0),\\
\Delta p_v&=\max(p-P_{\mathrm{back}},0),
\end{aligned}
\label{eq:overpressures}
\end{equation}
for the seal, tip valve and relief valve, respectively, Eq.~\eqref{eq:overpressures}. The relief-valve area gate uses the absolute crack pressure $p_{\mathrm{crack}}=60\,\mathrm{bar}+P_{\mathrm{atm}}$, whereas the seal blow-by gate uses the gauge onset pressure $p_{\mathrm{on}}=58$\,bar. The two gates therefore sit in different reference frames and are kept distinct throughout. The same discipline applies to the data: the pressure transducers read gauge pressure, the model state is absolute, and the conversion is made once in the data layer. The constitutive laws are therefore always evaluated in terms of absolute pressure, and every comparison with measurement is made in the sensor gauge frame.

\subsection{Chamber pressure dynamics}
\label{sec:chamber}

Mass conservation in the lumped chamber, with the compliance expressed through the effective bulk modulus $\beta_{\mathrm{eff}}(p)$ of Section~\ref{sec:bulk}, gives the pressure dynamics
\begin{equation}
\dot{p}=\frac{\beta_{\mathrm{eff}}(p)}{V(x)}\left(A_P\,\dot{x}\,s
- Q_{\mathrm{tip,back}} - Q_{\mathrm{tip,leak}} - Q_{\mathrm{film}} - Q_{\mathrm{blow}} - Q_{\mathrm{valve}}\right),
\label{eq:pressureODE}
\end{equation}
where $A_P$ is the piston area and the instantaneous chamber volume is
\begin{equation}
V(x)=V_0-A_P\,x,
\label{eq:volume}
\end{equation}
with $V_0$ the dead volume ($V$ is floored numerically, Section~\ref{sec:numerics}). The first term in Eq.~\eqref{eq:pressureODE} is the swept inflow $A_P\dot{x}s$. The remaining terms are the loss channels developed in Section~\ref{sec:losses}.

The gating variable $s$ is the seating state of the tip check valve, with $s=0$ open and $s=1$ seated. The sign of the swept term follows $\dot{x}$ directly, positive on compression and negative on retraction, and $s$ determines whether the swept flow pressurises the chamber: with the valve open the displaced fluid returns to the tank and the swept term is gated off, and with the valve seated the full $A_P\dot{x}$ acts on the chamber. The valve responds quickly relative to the stroke period and is represented as an ideal check element,
\begin{equation}
s=
\begin{cases}
1, & \dot{x}>V_{\varepsilon},\\[2pt]
0, & \text{otherwise},
\end{cases}
\label{eq:tipgate}
\end{equation}
with $V_{\varepsilon}$ a small velocity threshold that rejects sensor noise around reversal. The gate is a velocity-keyed seating convention: during the brief re-seating window at each reversal, the diverted swept flow is carried entirely by the motion-driven back-flow term $Q_{\mathrm{tip,back}}$ of Section~\ref{sec:losses}, and outside compression the tip path is the pressure-driven leak of the same section, so the idealisation does not lose the finite re-seating travel.

\subsection{Effective bulk modulus and chamber compliance}
\label{sec:bulk}

\begin{figure}[!ht]\centering
\includegraphics[width=\linewidth]{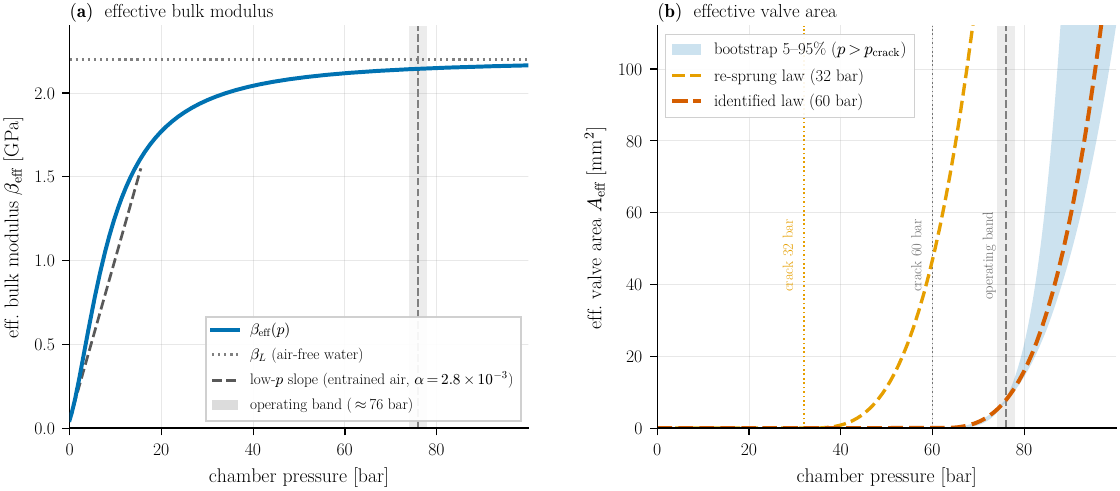}
\caption{The two identified constitutive laws. (a) Effective bulk modulus of Eq.~\eqref{eq:betaeff} with the identified entrained-air fraction. (b) Relief-valve effective area of Eq.~\eqref{eq:valve} above the 60\,bar crack, with its 5--95\% bootstrap band; the same law shifted to the identified 32\,bar crack of the re-sprung valve is overlaid.}
\label{Figure_2}\end{figure}

The chamber compliance is dominated by the entrained air rather than by the water itself. Following an established entrained-air effective bulk-modulus formulation \cite{merritt1967,gholizadeh2014}, with adiabatic compression of the gas fraction,
\begin{equation}
\beta_{\mathrm{eff}}(p)=\frac{1+\alpha\,r}{\dfrac{1}{\beta_L}+\dfrac{\alpha\,r}{\kappa\,p}},
\qquad
r=\left(\frac{P_{\mathrm{atm}}}{p}\right)^{1/\kappa},
\label{eq:betaeff}
\end{equation}
where $\beta_L$ is the bulk modulus of pure water, $\alpha$ is the entrained-air fraction parameter (identified), $\kappa$ is the adiabatic exponent, and $p$ is clamped to $p\ge P_{\mathrm{atm}}$. Near atmospheric pressure ($r\to1$) the gas term dominates the denominator and $\beta_{\mathrm{eff}}$ is far softer than $\beta_L$: the entrained air governs the compliance, which is what produces the sharp but finite compression ramp observed at the start of each pressurisation. At high pressure ($r\to0$) the gas is compressed away and $\beta_{\mathrm{eff}}\to\beta_L$, recovering the stiff response of pure water on the discharge plateau. The compressibility of the water itself is retained in full, and on its own accounts for a volume change near 0.3\% at 60\,bar. The entrained-air term sits in parallel with it and governs the compliance only at low pressure. The identified $\alpha$ is an effective compliance parameter that absorbs any unmodelled mechanical compliance of hoses and components, and with $\kappa$ fixed, the low-pressure compliance scales as $\alpha/\kappa$. Hence, its value is conditional on the adiabatic choice. The identified law is plotted in Fig.~\ref{Figure_2}a.

\subsection{Flow and loss channels}
\label{sec:losses}

\subsubsection{Relief-valve discharge}
The direct spring-loaded relief valve is modelled as an orifice with a pressure-dependent effective area above the crack pressure, consistent with the flow and lift modelling of direct spring-loaded valves \cite{wuc2021,zong2020} and with their experimentally observed pressure-dependent operation and blowdown \cite{zhangj2018,xu2018}:
\begin{equation}
\begin{aligned}
A_{\mathrm{eff}}(p)&=a\left(\frac{\max(p-p_{\mathrm{crack}},0)}{P_{\mathrm{ref},v}}\right)^{b},\\
Q_{\mathrm{valve}}(p)&=A_{\mathrm{eff}}(p)\sqrt{\frac{2\,\Delta p_v}{\rho}}\;\mathbf{1}\!\left[p>p_{\mathrm{crack}}\right],
\end{aligned}
\label{eq:valve}
\end{equation}
where $a$ is the area scale (the effective area at 10\,bar over-crack), $b$ is the area exponent, $P_{\mathrm{ref},v}$ is a fixed 10\,bar normaliser, $\rho$ is the fluid density and $\mathbf{1}[\cdot]$ is the indicator function. The valve is exactly closed for $p\le p_{\mathrm{crack}}$. The pressure dependence of $A_{\mathrm{eff}}$ carries the pressure-override characteristic of the valve. When the chamber pressure rises above the set point, the disc lifts further, and the flow area grows with it. The power-law form should be read as a semi-empirical closure, not as a lift law derived from first principles. A purely spring-set poppet would give near-linear lift, whereas the identified exponent $b\approx3.2$ lumps the lift, discharge-coefficient and Reynolds dependences that the terminal discharge data do not separate.

\subsubsection{Seal leakage: one path, two additive regimes}
The pressure-energised polymer piston seal is one physical leakage path represented as the sum of two regimes. The structure follows the mixed-lubrication account of reciprocating seals, developed as a transient finite-element film model for rod seals \cite{schmidt2010}, extended to the structural-thermal coupling that builds up over sustained cycling \cite{lichen2025}, and condensed into duty-parameter maps of leakage, friction and wear \cite{nikas2023}. The closest published analogue to the present seal is a water-lubricated Glyd-ring piston pair modelled under the sinusoidal rod velocity of a reciprocating pump \cite{li2024seal}. Two further strands bear on the high-pressure regime. Vibration is reported to change contact pressure and leakage \cite{wangw2023}, and seal degradation has been tracked from monitoring data rather than from a lubrication model \cite{zhaox2022}. The first regime is the pressure-energised film leak, present at all pressures,
\begin{equation}
Q_{\mathrm{film}}(p)=C\,\exp\!\left(-\gamma\,\frac{\Delta p_s}{P_{\mathrm{ref}}}\right)\sqrt{\frac{2\,\Delta p_s}{\rho}},
\label{eq:leak}
\end{equation}
an orifice flow modulated by a softening exponential. Rising pressure energises the seal lip against its counterface, and the leak decreases as $p$ rises ($\gamma>0$, with $C$ an effective leakage area and $P_{\mathrm{ref}}$ a fixed 60\,bar normaliser). The exponential is a semi-empirical closure that captures the observed monotonic tightening of the energised seal without resolving the underlying mixed-lubrication film. The second regime is a pressure-gated blow-by term (borrowing the engine term for flow past the piston sealing element), identically zero below the gauge onset $p_{\mathrm{on}}$,
\begin{equation}
Q_{\mathrm{blow}}(p)=C_b\left(\frac{\max(p_g-p_{\mathrm{on}},0)}{10\,\mathrm{bar}}\right)^{m_b},
\label{eq:blowby}
\end{equation}
where $p_g$ is the chamber gauge pressure expressed in bar, $C_b$ is the blow-by coefficient (the flow at 10\,bar over-onset) and $m_b$ is the blow-by exponent. The bracket is dimensionless, which leaves $C_b$ carrying the full flow units. Physically, the onset is consistent with the energised seal beginning to open a leak path at high pressure, whether by recoverable gap-breathing of the lip or by incipient extrusion cannot be resolved from terminal measurements, and the stationary endurance response of Section~\ref{sec:res-longdur} favours a recoverable mechanism. At the plateau flow, the identified law corresponds to an equivalent orifice gap of roughly 5\,$\mu$m around the bore circumference. The onset value is not resolved independently of the terminal data and is pinned just above the peak pressures reached in the ramp tests, so the film--blow-by split is conditional on it; the total seal loss is not. The ramps are therefore governed by the energised term alone, while the higher sinusoidal discharge plateaus activate the blow-by term, which then dominates the seal loss. The summed seal law is therefore non-monotonic in pressure, and the combined seal channel is the dominant source of loss in the pump.

\subsubsection{Check-valve flows: two flows through one passive valve}
The sliding tip check valve produces two flows, both passive. Valves of this kind are usually treated as fluid-structure interaction problems, whether for the stability of spring and shim valves under oscillatory forcing \cite{schickhofer2022} or for the self-acting suction and discharge valves of a diaphragm pump, where the computed delivery was checked against measured performance curves \cite{menendezblanco2019}. Closer to the present geometry, valve-plate motion has been coupled to chamber pressure pulsation in a reciprocating compressor \cite{zhaob2018}, and a reciprocating pump has been simulated with the piston motion imposed as an input and the valve dynamics resolved \cite{qiu2023}, which is the boundary condition used here. The first is the pressure-driven tip leak through the valve while it is open,
\begin{equation}
Q_{\mathrm{tip,leak}}(p,s)=(1-s)\,C_t\sqrt{\frac{2\,\Delta p_t}{\rho}},
\label{eq:tipleak}
\end{equation}
with $C_t$ the tip discharge coefficient, gated by the check state of Eq.~\eqref{eq:tipgate}. The velocity-keyed gate is a convention, and it has a cost that should be stated. At reversals and during dwells the physical valve is held seated by the pressure, yet the model keeps the tip open there, and the open-tip term becomes the path through which the chamber decompresses. It stands in for the expansion and refill dynamics that the single-state model does not resolve. The tip-leak entry of the loss budget (Section~\ref{sec:res-loss}) reports this path, not a measured seat leak. The second is the motion-driven re-seating back-flow, the volume a check valve passes while it re-seats with a finite lag \cite{bacak2023}. The sliding element requires travel to re-seat on flow reversal, and during this travel-limited dead band at the start of each compression, the swept inflow is diverted to the tank:
\begin{equation}
Q_{\mathrm{tip,back}}=A_P\,\max(\dot{x},0)\,g_{\mathrm{tip}}(t),
\label{eq:tipback}
\end{equation}
where $g_{\mathrm{tip}}(t)\in\{0,1\}$ is a travel-gated window indicator (an input built from the measured displacement, not a state), equal to one from the start of a compression until the piston has advanced the dead-band travel $\delta x=\delta V/A_P$, and zero thereafter. The dead-band volume follows an inverse law in the reversal speed,
\begin{equation}
\delta V=\frac{k}{v_{\mathrm{rev}}},
\label{eq:deadband}
\end{equation}
with $v_{\mathrm{rev}}$ the piston reversal speed in mm\,s$^{-1}$ and $k$ a constant with units m$^3\cdot$(mm\,s$^{-1}$): the faster the reversal, the sooner the valve seats. The inverse form was selected against constant and linear alternatives on the measured per-cycle dead-band volumes. The constant $k$ is then refined on the measured pressure trajectories. The value returned by the per-cycle dead-band volumes alone over-predicts the re-seating travel at low reversal speed. It delays the modelled compression, so $k$ is moved onto the flat minimum of the pooled trajectory error (Section~\ref{sec:res-uq}), which halves the dead band and removes the low-frequency phase lag. Equations~\eqref{eq:tipleak} and \eqref{eq:tipback} thus describe pressure-driven back-leak while unseated and swept flow lost during re-seating, respectively.

\subsubsection{Rejected hold-time creep term}
A dwell-dependent hold-time creep term, in which sustained dwell above the blow-by onset produces an additional seal loss proportional to the hold time, was evaluated as a candidate sixth channel. It was rejected: on the closed-cycle mass balance, it accounted for only about 1\% of the swept inflow (5\% of the seal loss) and returned unphysical negative values on 9 of the 26 sinusoidal datasets. The two-regime seal law of Eqs.~\eqref{eq:leak}--\eqref{eq:blowby} therefore represents the seal loss in full, and no creep term is retained. This keeps the model parsimonious and avoids overfitting.

\subsection{Output force}
\label{sec:force}

The rig measures the actuator force, which the model reproduces through an algebraic readout,
\begin{equation}
F=A_P\,\max(p-P_{\mathrm{atm}},0)+F_{\mathrm{fric}}\,\mathrm{sign}(\dot{x}),
\label{eq:force}
\end{equation}
in N: the piston pressure force on the gauge pressure, and a Coulomb friction term of magnitude $F_{\mathrm{fric}}$ that flips with the sign of the piston velocity at each reversal. A constant friction term is the coarsest of the available descriptions. Dynamic friction in hydraulic cylinders has been identified from experiments with a modified LuGre model \cite{tran2012}, and the traction that the lubricating film exerts back on the seal has been resolved numerically \cite{peng2020}. Two experimental studies bound what a lumped term can capture: friction measured from position and chamber pressure alone across velocity, load and seal profile \cite{pan2021}, and friction of polymer coaxial seals measured against pressure and sliding velocity \cite{sarbu2024}. The force cell carries a constant zero offset ($-0.604$\,kN on this rig), which is removed from the measured trace as a sensor tare rather than modelled. The readout is quasi-static, carrying the pressure and friction terms but not the moving-mass inertia, which appears only as a small under-prediction of the peak force at the compression front. It is validated against the measured force cell, and used to recover the mechanical power the pump absorbs, in Section~\ref{sec:res-force}.

\subsection{Identification methodology}
\label{sec:ident}

\subsubsection{Block order and fixed versus identified parameters}
The model is identified block by block, in an order chosen so that each fit conditions on quantities already pinned down. Freezing earlier blocks before estimating later ones mirrors the staged identification of wave-energy models from experimental data \cite{giorgi2019,jaramillo2020}. First, the chamber core, energised seal and tip valve ($V_0$, $\alpha$, $C$, $\gamma$, $C_t$) are identified by a global ODE fit of Eq.~\eqref{eq:pressureODE} to the ramp tests, and then frozen. The effective piston area and friction parameters ($A_P$, $F_{\mathrm{fric}}$) are identified from the measured force in a companion fit of the same campaign and held fixed thereafter. Second, the relief-valve area parameters ($a$, $b$) are identified from the measured valve discharge mass of the 26 sinusoidal datasets. Third, the blow-by parameters ($C_b$, $m_b$) are identified on the clean seal mass target defined below. Finally, the dead-band constant $k$ is refined on the measured pressure trajectories, as described with Eq.~\eqref{eq:deadband}. The physical constants ($\beta_L$, $\kappa$, $\rho$, $P_{\mathrm{atm}}$) and the normalisers ($P_{\mathrm{ref}}$, $P_{\mathrm{ref},v}$) are fixed, the crack pressure $p_{\mathrm{crack}}$ is fixed from the valve design specification at the nominal 60\,bar setting, and the blow-by onset $p_{\mathrm{on}}$ is pinned just above the ramp peak pressures, as described in Section~\ref{sec:losses}.

\subsubsection{Closed-cycle mass-balance target}
The central identification device is a closed-cycle mass balance. The training target is the measured valve discharge mass over 20 cycles, $m_{20}$ [kg]. Over a closed periodic cycle, the chamber storage term $\int (V/\beta_{\mathrm{eff}})\,\dot{p}\,\mathrm{d}t$ cancels, because the pressure (and the piston position) return to their starting values. The construction of the training target therefore requires neither $\beta_{\mathrm{eff}}$ nor $A_{\mathrm{eff}}$: the target is assembled purely from integrals of the measured pressure and displacement, and the candidate loss laws are subsequently evaluated on the measured pressure trace. The inflow mass over the window, Eq.~\eqref{eq:inflow}, is
\begin{equation}
m_{\mathrm{in}}=\rho\,A_P\!\int \dot{x}^{+}\,\mathrm{d}t,
\label{eq:inflow}
\end{equation}
with $\dot{x}^{+}=\max(\dot{x},0)$, and the total non-valve loss mass follows by difference,
\begin{equation}
m_{\mathrm{nonvalve}}=m_{\mathrm{in}}-m_{20}.
\label{eq:nonvalve}
\end{equation}
The clean seal target also subtracts both tip contributions: the motion-driven back-flow mass, $m_{\mathrm{tipback}}=\rho\,A_P\,\delta x\cdot 20$, with $\delta x$ the per-cycle dead-band travel of Section~\ref{sec:losses}, and the pressure-driven back-leak mass, $m_{\mathrm{tipleak}}=\rho\int Q_{\mathrm{tip,leak}}(p_{\mathrm{meas}},s)\,\mathrm{d}t$. The seal blow-by parameters are then fitted so that the modelled two-regime seal loss matches this clean seal target:
\begin{equation}
m_{\mathrm{seal}}=m_{\mathrm{in}}-m_{20}-m_{\mathrm{tipback}}-m_{\mathrm{tipleak}}
\;=\;\rho\!\int\!\left(Q_{\mathrm{film}}+Q_{\mathrm{blow}}\right)(p_{\mathrm{meas}})\,\mathrm{d}t.
\label{eq:targets}
\end{equation}
Both tip contributions are removed from the seal target so that the seal channel is isolated cleanly and the forward model does not double-count the tip back-leak. Like every target quantity, $m_{\mathrm{tipleak}}$ is evaluated on the measured pressure trace (with the check state taken from the measured velocity), so the closed-cycle cancellation argument is unaffected. All predicted masses entering these fits are likewise evaluated on the measured signals, for example $\hat{m}_{\mathrm{valve}}=\rho\int Q_{\mathrm{valve}}(p_{\mathrm{meas}})\,\mathrm{d}t$, so no ODE integration is required inside the optimisation loops. Because the seal target depends on the dead-band constant through $m_{\mathrm{tipback}}$, the blow-by pair is re-identified whenever $k$ changes. With the trajectory-refined $k$, the aggregate budget closes to within 0.8\% of the swept inflow (Section~\ref{sec:res-loss}).

\subsubsection{Objectives and optimiser}
The valve parameters are fitted in log-mass space over all 26 sinusoidal datasets simultaneously, with residuals $r_i=\log_{10}(\hat{m}_{\mathrm{valve},i}+\epsilon)-\log_{10}(m_{20,i}+\epsilon)$, where $\epsilon=10^{-9}$ guards the logarithm, using a trust-region-reflective least-squares solver with a soft-L1 loss and free variables $[\log_{10}a,\,b]$ bounded to $[-9,-3]\times[0.2,6]$. Parameter covariance is estimated by the Gauss--Newton approximation $\Sigma=\sigma^{2}(J^{\top}J)^{-1}$ at the solution, with the variance scaled by the residual cost, and the uncertainty on $a$ recovered by the delta method. The blow-by parameters are fitted with the same solver on the clean seal target of Eq.~\eqref{eq:targets}. The chamber-stage parameters are identified by the global ODE fit that minimises the root-mean-square (RMS) pressure error across the ramp tests.

\subsubsection{Identifiability findings}
Three findings fixed the final structure and are summarised here. First, a constant valve area fails: a single effective area fitted per dataset varies by more than an order of magnitude across conditions and collapses onto pressure rather than excitation frequency, and forward simulations with a constant area diverge to non-physical pressures. This forces the pressure-dependent area of Eq.~\eqref{eq:valve}. Second, the seal and the valve are degenerate above the crack pressure: a joint fit of ($C$, $\gamma$, $a$, $b$) is unidentifiable because seal and valve losses trade off freely once both channels are active. The degeneracy is broken not by regularising the joint fit but by the closed-cycle decoupling: the valve is fitted to the measured valve mass $m_{20}$ and the seal to the non-valve residual of Eq.~\eqref{eq:nonvalve}, which separates the channels by construction. Third, a leak that scales with speed cannot be seen by the cyclic mass target. Whatever the frequency, $\int C_v|\dot{x}|\,\mathrm{d}t$ integrates to the same stroke distance over a cycle, so every test returns the same mass and the coefficient has nothing to push against. In the fits it drifts towards zero. The seal loss is accordingly keyed to pressure through the two-regime law of Eqs.~\eqref{eq:leak}--\eqref{eq:blowby}, and no velocity-proportional leakage term is retained. Within these targets, a velocity-keyed component is unidentifiable. The pressure-only law is a modelling choice validated in free-run, not evidence that the seal loss is speed-independent.

\subsection{Numerical implementation}
\label{sec:numerics}

The forward simulations of the ramp and sinusoidal tests integrate the pressure ODE, Eq.~\eqref{eq:pressureODE}, with the LSODA solver (\texttt{scipy.integrate.solve\_ivp}), with relative tolerance $10^{-5}$, absolute tolerance $10^{3}$\,Pa on $p$, and a maximum step of $2\times10^{-3}$\,s. The check state $s$ is evaluated algebraically from the measured velocity through Eq.~\eqref{eq:tipgate} at each evaluation of the right-hand side. For speed, the time grid is sub-sampled by a factor of four and the simulated pressure is linearly interpolated back onto the full measurement grid. The initial condition $p_0$ is the first measured pressure sample converted to absolute (or $P_{\mathrm{atm}}$, whichever is larger). The inputs $x(t)$ and $\dot{x}(t)$ are interpolants of the measured displacement and its Savitzky--Golay derivative (window 11 samples, polynomial order 3), and the tip gate $g_{\mathrm{tip}}(t)$ is likewise an externally clocked input built from the measured signals. The chamber volume in Eq.~\eqref{eq:volume} carries an analytic floor of $10^{-9}$\,m$^{3}$ and a second floor of $10^{-6}$\,m$^{3}$ inside the right-hand side, which puts the effective floor in the simulations at $10^{-6}$\,m$^{3}$.

The irregular sea-state and endurance records are stiff for an explicit adaptive solver, because the right-hand side switches between compliant filling and near-incompressible discharge many times per record. These records are integrated with a linearly-implicit fixed-step Euler scheme, $p_{n+1}=p_n+h\,f/(1-h\,J)$ with $h=10^{-3}$\,s and $J$ the numerical pressure Jacobian, which is unconditionally stable for the dissipative dynamics here ($J<0$). Two implementations of this scheme, a reference version with callable inputs and an accelerated version with inputs pre-sampled on the sub-step grid, agree to better than $10^{-3}$\,bar on the validation records. The accelerated version is what makes whole-record validation of the 360\,s sea states and the 4000\,s endurance runs practical. The identification and validation scripts are archived with the data so that every reported number can be regenerated from the raw records.

\begin{table}[!ht]
\centering
\caption{Consolidated model parameters. Fixed parameters are set from physics, fluid properties or the component design specification. Identified parameters are estimated from the bench data in the block order of Section~\ref{sec:ident}; the blow-by onset is assumed (pinned), not identified. Gauss--Newton standard errors are given for the valve and blow-by pairs. The crack pressure is specified in gauge and used as absolute, $60\,\mathrm{bar}+P_{\mathrm{atm}}$, in Eq.~\eqref{eq:valve}.}
\label{Table_3}
\footnotesize
\setlength{\tabcolsep}{4pt}
\renewcommand{\arraystretch}{1.15}
\begin{tabular}{@{}llrll@{}}
\toprule
Symbol & Meaning & Value & Unit & Type \\
\midrule
$A_P$              & Effective piston area & $3.559\times10^{-3}$  & m$^2$               & Identified \\
$V_0$              & Dead volume                & $7.090\times10^{-3}$  & m$^3$               & Identified \\
$\alpha$           & Entrained-air fraction     & $2.827\times10^{-3}$  & --                  & Identified \\
$\beta_L$          & Liquid bulk modulus        & $2.2\times10^{9}$     & Pa                  & Fixed \\
$\kappa$           & Adiabatic gas exponent     & $1.4$                 & --                  & Fixed \\
$P_{\mathrm{atm}}$ & Atmospheric pressure       & $1.013\times10^{5}$   & Pa                  & Fixed \\
$\rho$             & Fluid density              & $1000$                & kg\,m$^{-3}$        & Fixed \\
$P_{\mathrm{ref}}$ & Seal-exponent normaliser   & $6.0\times10^{6}$     & Pa                  & Fixed \\
$C$                & Energised-seal coefficient & $3.254\times10^{-6}$  & m$^2$               & Identified \\
$\gamma$           & Seal softening exponent    & $6.435$               & --                  & Identified \\
$C_t$              & Tip discharge coefficient  & $2.932\times10^{-6}$  & m$^2$               & Identified \\
$V_{\varepsilon}$  & Tip velocity threshold     & $2.0\times10^{-3}$    & m\,s$^{-1}$         & Fixed \\
$a$                & Valve area scale           & $(1.746\pm0.145)\times10^{-6}$ & m$^2$      & Identified \\
$b$                & Valve area exponent        & $3.197\pm0.200$       & --                  & Identified \\
$p_{\mathrm{crack}}$ & Valve crack pressure (nominal) & $60$              & bar\,(g)            & Fixed \\
$p_{\mathrm{crack}}^{\mathrm{low}}$ & Re-sprung valve crack & $32$            & bar\,(g)            & Identified \\
$P_{\mathrm{ref},v}$ & Valve-area normaliser    & $1.0\times10^{6}$     & Pa                  & Fixed \\
$C_b$              & Seal blow-by coefficient   & $(1.239\pm0.065)\times10^{-4}$ & m$^3$\,s$^{-1}$ & Identified \\
$m_b$              & Seal blow-by exponent      & $0.642\pm0.116$       & --                  & Identified \\
$p_{\mathrm{on}}$  & Seal blow-by onset         & $58.0$                & bar\,(g)            & Pinned \\
$k$                & Tip dead-band constant     & $4.5\times10^{-3}$    & m$^3$\,mm\,s$^{-1}$ & Identified \\
$F_{\mathrm{fric}}$ & Coulomb friction          & $56.4$                & N                   & Identified \\
\bottomrule
\end{tabular}
\end{table}

\section{Results and Discussion}
\label{sec:results}

This section reports the block-by-block identification, the recovered loss decomposition, and the validation and generalisation of the identified model. The primary metric is the free-run pressure error: the model is integrated forward on the measured motion alone, with no pressure feedback, and compared with the measured chamber pressure. Errors are reported as the range-normalised root-mean-square error (NRMSE), $\mathrm{NRMSE}=100\,\mathrm{RMSE}/(\max p_{\mathrm{meas}}-\min p_{\mathrm{meas}})$, together with the root-mean-square error (RMSE) in bar, per record and by regime. All validation errors are computed over whole records: the full 20-cycle sinusoidal traces including run-in, the full 360\,s sea-state records and the full endurance records, not selected windows. The model state is absolute pressure, and the transducers read gauge pressure, so every comparison is made in the sensor gauge frame. Because a free-run error on an irregular signal is penalised by small phase and timing offsets even when the amplitudes are correct, the sea-state assessment is corroborated by statistics that do not depend on point-by-point alignment: per-wave peak capture, spectra and exceedance. This validation philosophy, which benchmarks against measured records and quantifies experimental uncertainty, follows established practice for wave-energy bench and tank testing \cite{bacelli2019,orphin2021}. It also mirrors the incremental validation of a numerical wave tank against the measured power production of a deployed point-absorber device \cite{windt2020}, and it aligns with the intent of the International Electrotechnical Commission (IEC) guidance on pre-prototype wave-energy testing \cite{iec62600103}. All parameter values quoted here are the identified set of Table~\ref{Table_3}; the numerical verification underpinning them is summarised in Section~\ref{sec:res-verify}.

\subsection{Block-by-block identification}
\label{sec:res-ident}

\begin{figure}[!ht]\centering
\includegraphics[width=\linewidth]{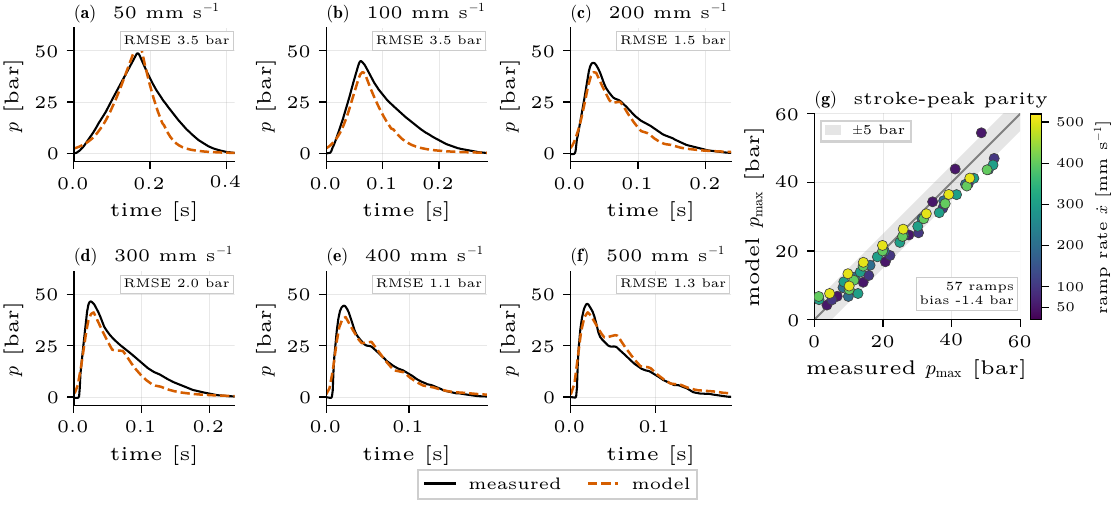}
\caption{Ramp identification. (a--f) Measured (solid) and free-run model (dashed) chamber pressure for one full stroke at each of the six rates (50--500\,mm\,s$^{-1}$), with the relief valve seated. (g) Stroke-peak parity over all 57 segments, coloured by ramp rate, with the $\pm$5\,bar band shaded.}
\label{Figure_3}\end{figure}

The chamber core, energised seal and tip valve were identified together by the global ODE fit of Eq.~\eqref{eq:pressureODE} to the 57 ramp segments. Run forward on the measured motion, the identified chamber block reproduces the ramp pressure histories with a pooled RMSE of 1.7\,bar over all segments (pooled NRMSE 6.6\%, Fig.~\ref{Figure_3}), across peak pressures from a few bar to 55\,bar and rise rates from 50 to 500\,mm\,s$^{-1}$, and it reproduces the stroke peaks to a mean $-1.4$\,bar bias over all 57 segments (Fig.~\ref{Figure_3}g); over the 47 strokes exceeding 10\,bar the bias is $-2.1$\,bar. The identified entrained-air fraction $\alpha=2.83\times10^{-3}$ sets the chamber compliance through Eq.~\eqref{eq:betaeff}: it softens $\beta_{\mathrm{eff}}$ by more than an order of magnitude near atmospheric pressure and produces the finite compression ramp seen at the start of each pressurisation, then stiffens towards $\beta_L$ on the plateau (Fig.~\ref{Figure_2}a). Read physically, the low-pressure offset of $\beta_{\mathrm{eff}}$ measures how much air the chamber carries into a stroke, and the high-pressure tail measures the stiffness the chamber recovers once that air has been compressed away. The seal-softening exponent $\gamma=6.44$ confirms a strongly pressure-energised leak. The visible misfit sits at the two slowest rates, where the modelled decompression runs ahead of the measured tail (Fig.~\ref{Figure_3}a,b), so the identified leak laws overstate the mid-pressure loss on those strokes. Retaining the pressure-driven tip back-leak $Q_{\mathrm{tip,leak}}$ is justified by a leave-one-term test at the identification stage: dropping the term degraded the pooled ramp error of the chamber fit by roughly a third, so it is kept in the final structure.

\begin{figure}[!ht]\centering
\includegraphics[width=\linewidth]{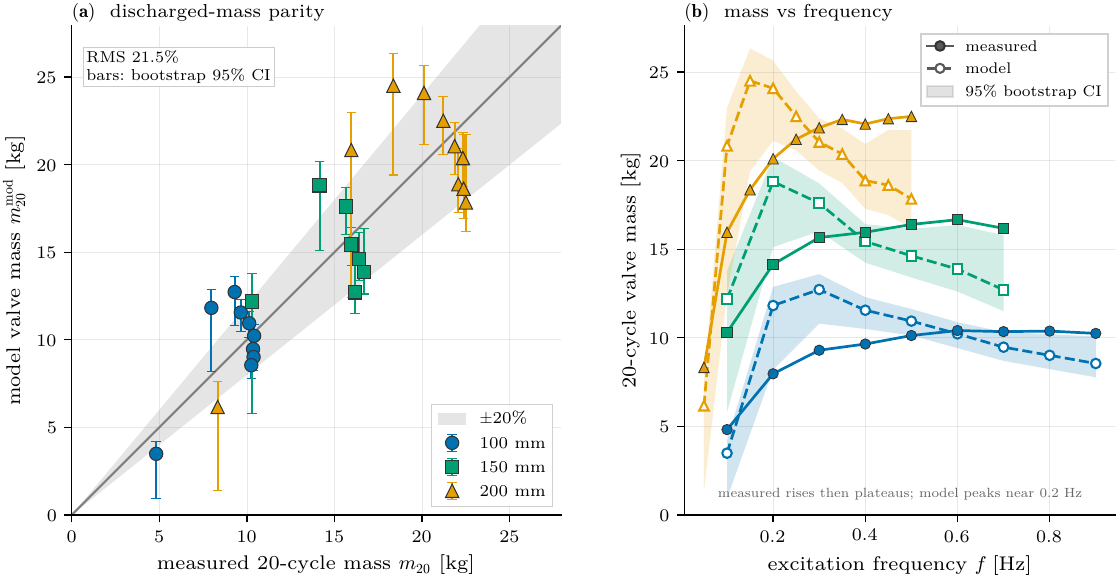}
\caption{Relief-valve discharged-mass validation. (a) Measured versus predicted 20-cycle mass $m_{20}$ for the 26 sinusoidal datasets with per-point 95\% bootstrap intervals (RMS 21.5\%, shaded band $\pm$20\%). (b) The same masses against frequency by amplitude, measured (solid) and model (dashed) with its 95\% bootstrap band.}
\label{Figure_4}\end{figure}

The relief-valve area parameters were then identified from the measured 20-cycle discharge mass $m_{20}$ over the 26 sinusoidal datasets, fitted in log-mass space. The identified area law is strongly pressure-dependent, with scale $a=1.75\times10^{-6}\,\mathrm{m}^2$ and exponent $b=3.20$, and it matches the measured valve mass with an RMS of 21.5\% across the full amplitude and frequency range (Fig.~\ref{Figure_4}a). The first identifiability finding of Section~\ref{sec:ident} shows up here in the data. A single constant effective area could not reconcile the discharged masses across the conditions, and the pressure-dependent area of Eq.~\eqref{eq:valve} can. The 21.5\% should not be read as random scatter. It is a structured residual, and the measurement itself sets part of it. The static area law over-predicts the discharged mass in the 0.1--0.3\,Hz band and under-predicts it above 0.4\,Hz and on the two near-crack partial strokes (Fig.~\ref{Figure_4}b), the signature of the valve dynamics and Reynolds dependence that a pressure-only effective area omits, and part of the spread comes from weighing the small near-crack discharge masses. The same shortfall recurs as the plateau over-prediction of the out-of-distribution test (Section~\ref{sec:res-ood}) and the high-frequency roll-off of the accuracy map, so it is one limitation of the valve block, not three separate defects. The identified law, with its bootstrap uncertainty band, is shown in Fig.~\ref{Figure_2}b. On the clean seal target, the blow-by law identifies a coefficient $C_b=1.24\times10^{-4}$\,m$^3$\,s$^{-1}$ and an exponent $m_b=0.64\pm0.12$; the summed seal channel matches the per-record seal mass target to an RMS of 9.7\%. The dead-band constant was first estimated from the per-cycle dead-band volumes against reversal speed ($\delta V=k/v_{\mathrm{rev}}$) and then refined on the measured pressure trajectories to $k=4.5\times10^{-3}$; Section~\ref{sec:res-uq} shows the trajectory objective is convex in $k$ and rejects the mass-balance estimate. The identified dead band is a volume surrogate rather than a statement about slide position, but it is worth placing against the geometry. Dividing $\delta V$ by the piston area gives a piston travel of a few millimetres over most of the campaign. It equals the nominal 6.5\,mm slide travel of the valve at a reversal speed near 195\,mm\,s$^{-1}$, midway through the tested range. Faster reversals give less, 4.0\,mm at 314\,mm\,s$^{-1}$ and 2.0\,mm at 628\,mm\,s$^{-1}$. The slowest strokes give more than the geometric value, 20\,mm at 63\,mm\,s$^{-1}$, consistent with the dead band also absorbing back-leak through the closing gap when the valve takes longer to seat. Because the closed-cycle targets are built the way they are, the valve and blow-by objectives do not interact, and a joint four-parameter re-fit lands on the same optimum as the sequential one. This was run as a consistency check on the implementation. It says nothing on its own about whether the blocks are separable.

\subsection{Loss decomposition}
\label{sec:res-loss}

\begin{figure}[!ht]\centering
\includegraphics[width=\linewidth]{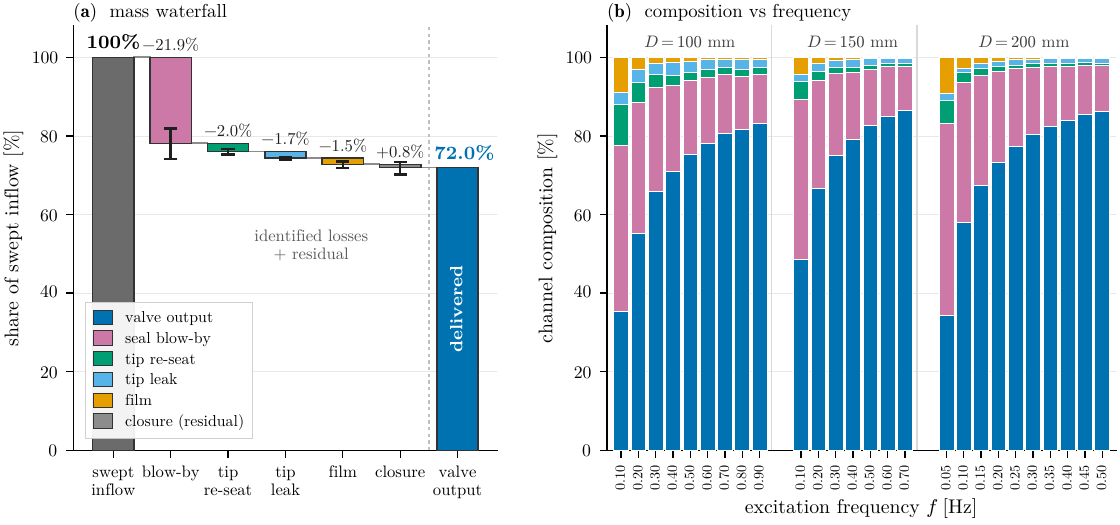}
\caption{Identified mass-loss decomposition over the sinusoidal campaign. (a) Budget from swept inflow to valve output, each loss channel with its 95\% bootstrap interval and the closure residual as its own bar. (b) Channel composition of each cycle against excitation frequency, by amplitude.}
\label{Figure_5}\end{figure}

Because each channel is keyed to a physical mechanism, the identified model resolves the pump loss budget by mechanism instead of as a lumped efficiency. Aggregated over the 26 sinusoidal datasets, the total inflow mass is 533\,kg, of which 384\,kg (72.0\%) leaves through the relief valve as useful discharge and 149\,kg (28\%) is lost. The four attributed channels sum to 27.1\% of the inflow, and the closure residual carries the remaining 0.8 percentage points. Per record, the residual spans $-6$\% to $+15$\% (median 0.3\%), so the aggregate reflects some cancellation across conditions. The seal dominates the loss. The closed-cycle balance defines the seal channel as the non-valve residual after the two tip contributions are removed, so it aggregates any unmetered leakage path; the attribution to the piston seal rests on the pressure signature of the identified laws, and Section~\ref{sec:res-discuss} lists the direct checks that would corroborate it. The pressure-gated blow-by term alone accounts for 21.9\% of the swept inflow, the single largest loss, while the pressure-energised film leak (1.5\%), the motion-driven tip re-seat diversion (2.0\%) and the pressure-driven tip leak (1.7\%) each contribute only a few percent (Fig.~\ref{Figure_5}a). Summed over its two regimes, the seal carries 84\% of everything that is lost. Its two regimes act in different parts of the cycle: the energised film leak governs the compression and low-plateau phases, while the blow-by term activates above its 58\,bar onset and overtakes the film leak on the high-pressure discharge plateau, producing the non-monotonic seal law of Section~\ref{sec:losses}. The frequency dependence in Fig.~\ref{Figure_5}b makes the mechanism visible: at 0.05--0.1\,Hz the pump spends most of each cycle on the plateau where blow-by is active, and the seal claims up to half of the swept inflow, falling to about 12\% at the highest frequencies. The composition of the test matrix therefore weights the aggregate shares; a deployment-weighted budget would follow the joint distribution of stroke and period at the site. This budget comes out of the closed-cycle mass balance itself, with no catalogue coefficient assumed in advance, which is a decomposition that whole-system models have so far constructed forward rather than identified \cite{cargo2016,penalba2018}.

\subsection{Energy budget and delivered efficiency}
\label{sec:res-energy}

\begin{figure}[!ht]\centering
\includegraphics[width=\linewidth]{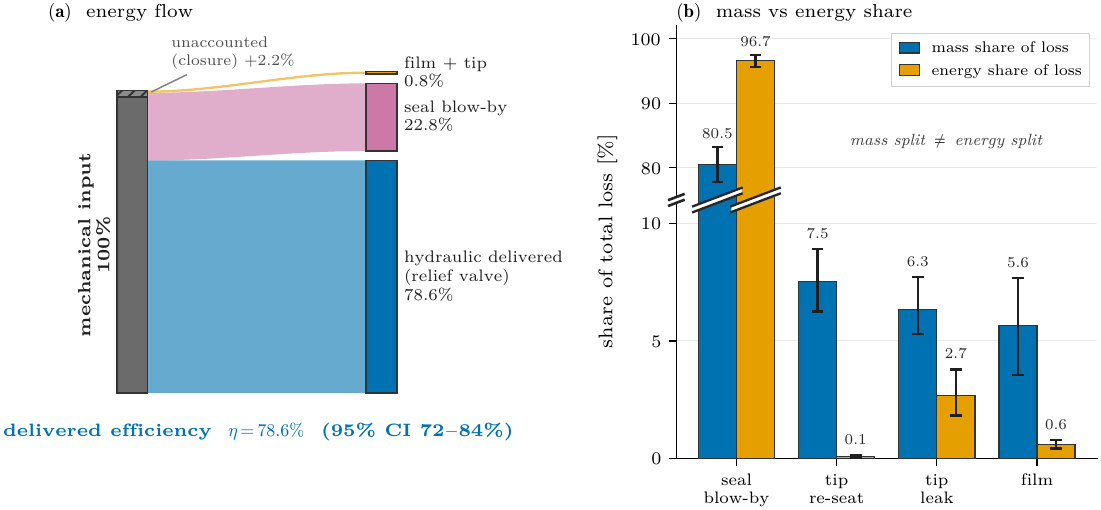}
\caption{Energy budget over the sinusoidal campaign. (a) Mechanical indicator work flowing to useful high-pressure discharge (78.6\%) and to the loss channels, with the closure residual. (b) Share of dissipated energy against share of lost mass by channel, on a broken y-axis.}
\label{Figure_6}\end{figure}

The mass budget records where the water goes, not where the work goes, and the two differ because a kilogram lost through the seal at 70\,bar carries far more energy than a kilogram diverted at the tip near atmospheric pressure. Integrating the pressure against the flow over the campaign yields the energetic counterpart (Fig.~\ref{Figure_6}). Of the mechanical indicator work delivered to the fluid, 78.6\% (bootstrap 95\% interval 72--84\%) leaves as useful high-pressure discharge through the relief valve. Hence, the pump converts mechanical input to hydraulic output with an efficiency near 79\% during the sinusoidal campaign, with the balance dissipated through the loss channels. The compression energy is returned as spring-back at each reversal. The energy split is far more lopsided than the mass split. The seal carries 97\% of the dissipated energy against 84\% of the lost mass, almost all of it through the blow-by regime acting on the high-pressure plateau. In contrast, the tip flows that account for 13\% of the lost mass cost almost nothing energetically at their near-atmospheric pressures. The energy budget closes to within 2.2\% of the input work. The delivered fraction is operating-point-dependent, mirroring Fig.~\ref{Figure_5}b: per record, it ranges from about 30\% on the slowest strokes to above 90\% at mid frequencies, with 78.6\% as the campaign aggregate. For a power take-off, whose product is high-pressure flow, this identifies the piston seal as the single loss worth engineering and establishes a concrete conversion efficiency at a stated operating mix for design and energy assessment. Because that flow is what a downstream turbine or membrane converts, the 21.9\% of the swept inflow lost past the seal is the largest single deduction from the energy the device can generate.

\subsection{In-regime validation: ramps and sinusoids}
\label{sec:res-inregime}

\begin{figure}[!ht]\centering
\includegraphics[width=\linewidth]{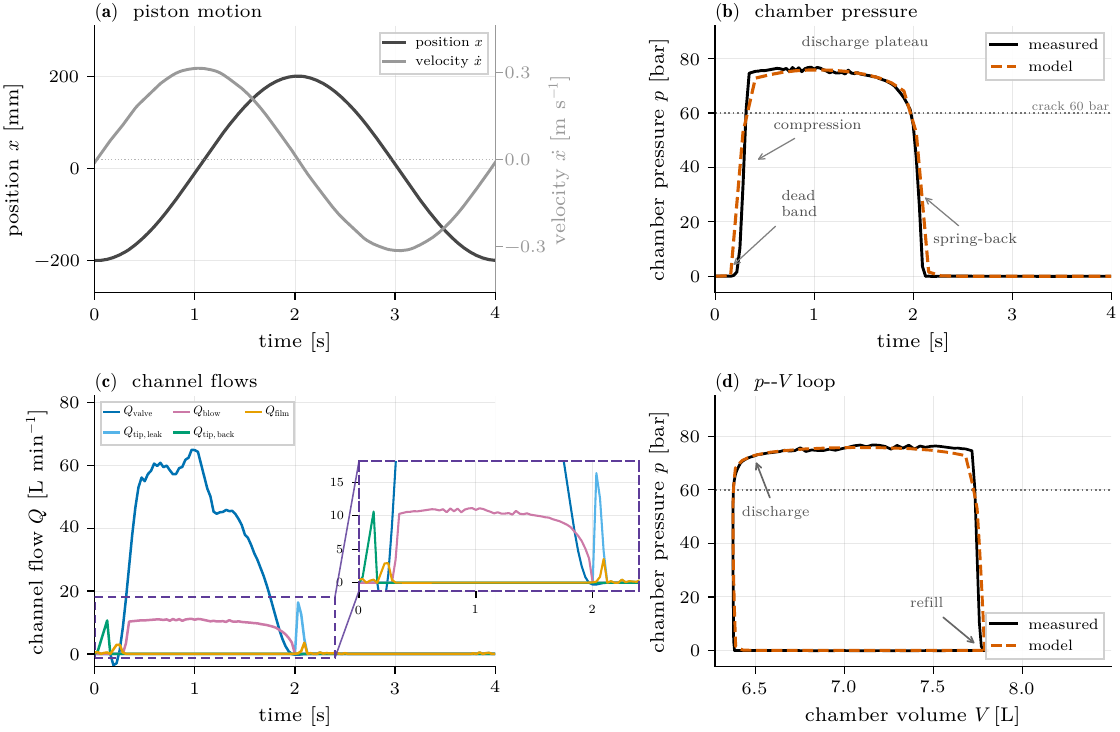}
\caption{Anatomy of one pumping cycle (200\,mm, 0.25\,Hz, 60\,bar), windowed so compression begins at $t=0$. (a) Imposed position and velocity. (b) Measured and free-run model pressure. (c) Model channel flows on the measured pressure, low-flow window magnified in the inset. (d) Measured and modelled pressure--volume loop.}
\label{Figure_7}\end{figure}

\begin{figure}[!ht]\centering
\includegraphics[width=\linewidth]{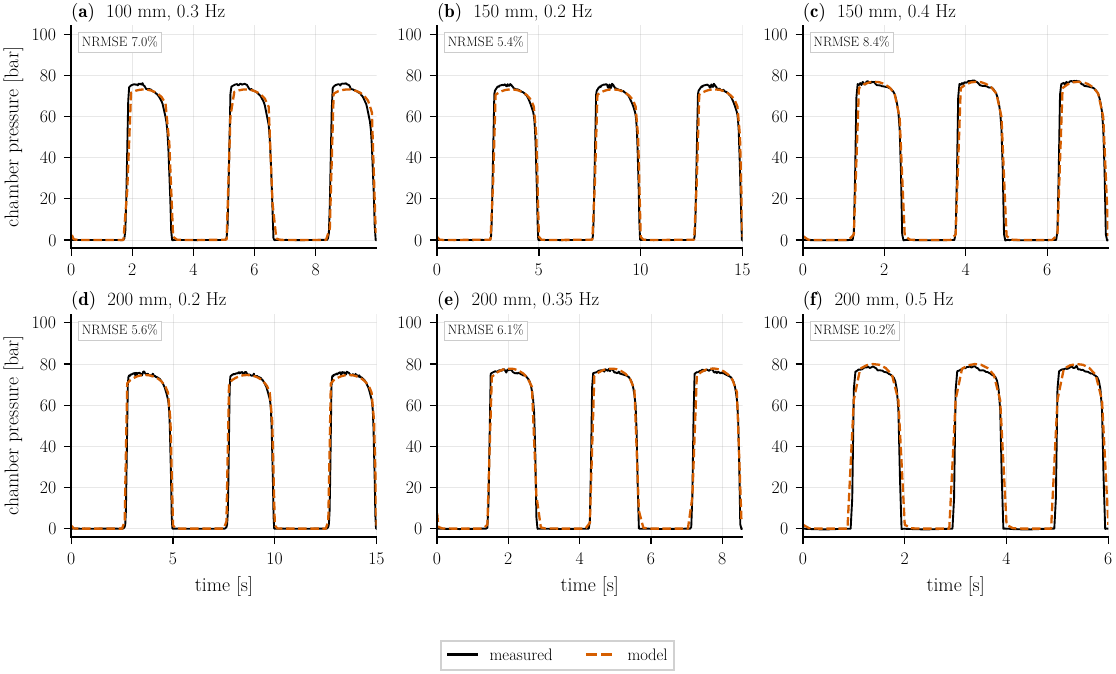}
\caption{Sinusoidal validation. Measured and free-run model chamber pressure over three steady cycles for six representative amplitude--frequency combinations; NRMSE values refer to the complete 20-cycle records.}
\label{Figure_8}\end{figure}

\begin{figure}[!ht]\centering
\includegraphics[width=\linewidth]{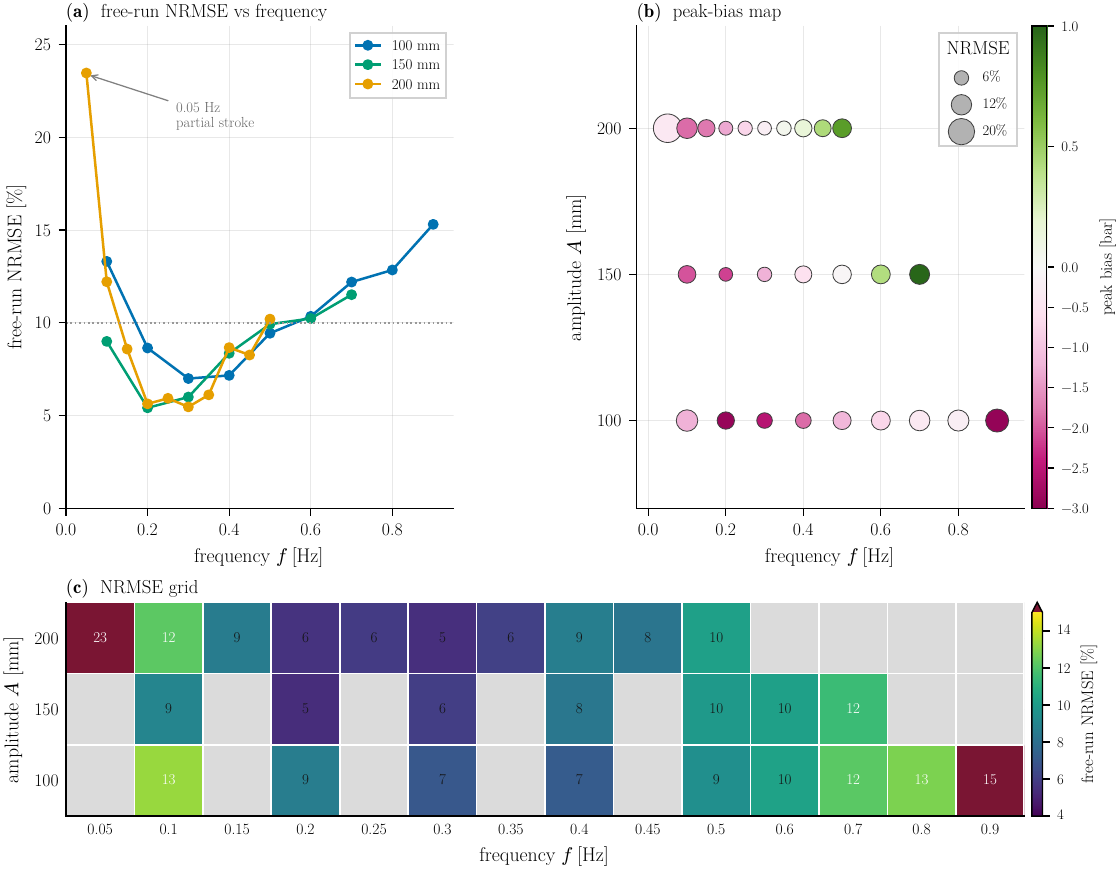}
\caption{Free-run accuracy across the sinusoidal campaign (whole-record NRMSE). (a) NRMSE against excitation frequency, by amplitude. (b) The same runs, sized by NRMSE and coloured by signed peak bias, positive where the model over-cracks. (c) Gridded map over frequency and amplitude; unsampled cells are greyed and the 0.05\,Hz partial stroke is over-ranged.}
\label{Figure_9}\end{figure}

Run forward on the measured motion, the model reproduces the full pumping cycle across the sinusoidal set, tracking the measured pressure over three steady cycles for six representative amplitude--frequency combinations (Fig.~\ref{Figure_8}). Figure~\ref{Figure_7} dissects one 200\,mm, 0.25\,Hz case. The suction phase, travel-limited dead band, sharp compression to the crack pressure, regulated discharge plateau and decompression spring-back all occur at the correct pressures and piston positions. The channel flows show the valve discharge and plateau-gated blow-by in their expected phases, and the pressure--volume loop closes on the measured indicator diagram. Over the whole set the free-run error on the complete 20-cycle records is 8.8\% median NRMSE (mean 9.7\%, Table~\ref{Table_4}), and 623 of the 624 individual stroke peaks, counted over the complete records including their run-in and run-out strokes, fall within 5\,bar of the measured peak, the single exception being one stroke of the fastest 100\,mm case. Restricting the error to a steady four-cycle window changes the median by only half a percentage point, which confirms that the run-in and run-out transients are not hiding a start-up artefact. The accuracy map in Fig.~\ref{Figure_9} locates the residual structure. Accuracy is best at mid frequency (5--7\% between 0.2 and 0.4\,Hz), rises to about 15\% by 0.9\,Hz, and is worst for the 0.05\,Hz partial stroke (23\%), whose slow compression spends the longest time in the low-pressure regime where the free-run phase penalty accumulates. The signed peak bias in Fig.~\ref{Figure_9}b remains within a few bar across the campaign and is mostly negative, with the largest values on the 100\,mm strokes.

\begin{table}[!ht]
\centering
\caption{Free-run pressure error and per-stroke peak capture by regime, computed over whole records (20-cycle sinusoidal traces, 360\,s sea states). NRMSE is range-normalised on the measured pressure. Ramps and sinusoids are identification classes (the sinusoidal trajectories also refined the dead-band constant); the sea states are validation, conditional at 32\,bar on the crack pressure identified from the plateaux of those records. The low-pressure warm-up segments inflate per-ramp NRMSE; the pooled ramp error is 1.7\,bar (6.6\%).}
\label{Table_4}
\footnotesize
\setlength{\tabcolsep}{5pt}
\begin{tabular}{lccccc}
\toprule
Regime & $n$ & NRMSE median (\%) & NRMSE mean (\%) & RMSE mean (bar) & peaks within 5\,bar \\
\midrule
Ramps                 & 57 & 7.0  & 9.8  & 1.8 & -- \\
Sinusoids             & 26 & 8.8  & 9.7  & 7.4 & 99.8\% \\
Sea states (60\,bar)  & 7  & 10.7 & 10.4 & 8.3 & 80\% \\
Sea states (32\,bar)  & 7  & 11.4 & 11.7 & 6.0 & 83\% \\
\bottomrule
\end{tabular}
\end{table}

\subsection{Generalisation to irregular sea states}
\label{sec:res-seastate}

\begin{figure}[!ht]\centering
\includegraphics[width=\linewidth]{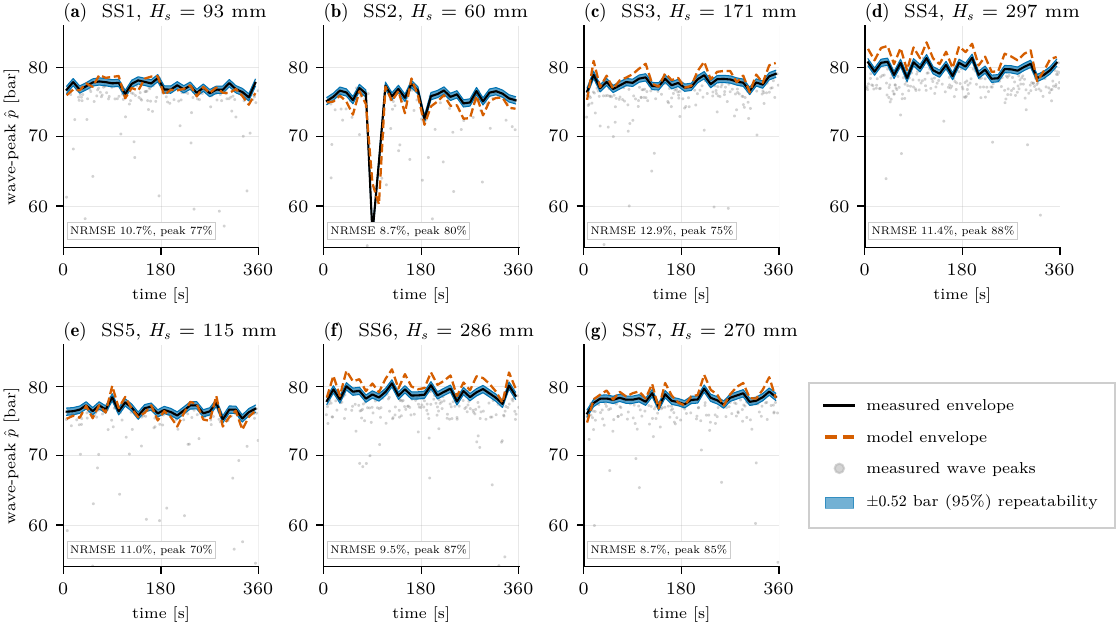}
\caption{Irregular sea-state validation, all seven 60\,bar states. (a--g) Per-wave peak-pressure envelope against time: measured envelope (black) with its 95\% repeatability band, measured peaks (grey) and free-run model (vermillion dashed). Each panel carries its NRMSE and peak capture.}
\label{Figure_10}\end{figure}

\begin{figure}[!ht]\centering
\includegraphics[width=\linewidth]{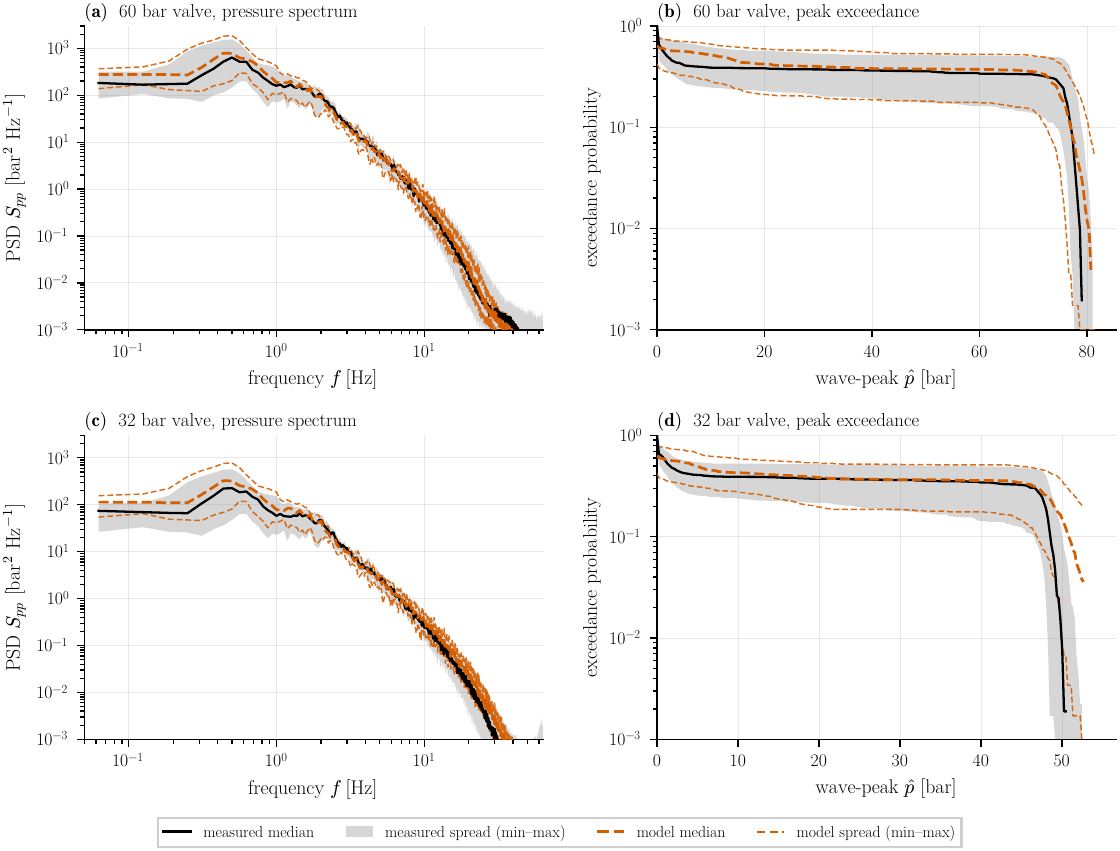}
\caption{Spectral and extreme-value validation across the seven sea states, (a,b) at 60\,bar and (c,d) at 32\,bar. (a,c) Chamber-pressure spectral density; (b,d) wave-peak exceedance. Medians (measured black, model vermillion dashed) with the min--max envelopes across states (grey band and dashed lines).}
\label{Figure_11}\end{figure}

\begin{figure}[!ht]\centering
\includegraphics[width=\linewidth]{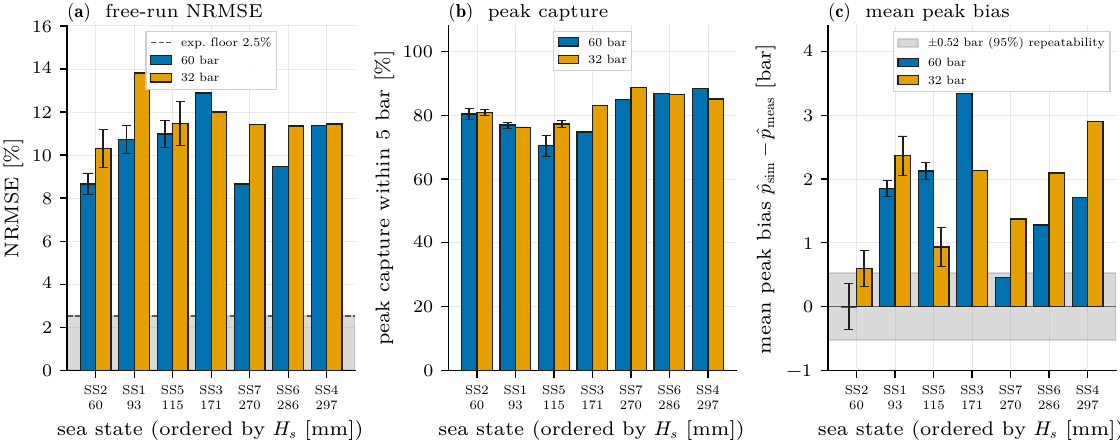}
\caption{Whole-record free-run error (a), peak capture (b) and signed peak bias (c) for all fourteen sea-state runs, ordered by imposed height (blue 60\,bar, orange 32\,bar). Error bars on the repeated states are run-to-run standard deviations. The shaded bands are the experimental floor and the 95\% peak-repeatability band.}
\label{Figure_12}\end{figure}

The central validation is on the fourteen irregular sea-state runs. On the 60\,bar set, across seven full 360\,s records, the identified model achieves a mean free-run NRMSE of 10.4\% (pressure RMSE of 8.3\,bar; Table~\ref{Table_4}). The model captures the irregular pressure history over the whole record and wave by wave (Fig.~\ref{Figure_10}). It reproduces the per-wave statistics: of the 3\,523 individual pumping events in the 60\,bar set, 80\% of peaks fall within 5\,bar and 71\% within 2\,bar of the measured value, with no run producing a non-physical over-pressure. Errors are near-flat across the fourteen conditions (Fig.~\ref{Figure_12}): the free-run NRMSE and the signed peak bias change little with sea state, so the residual behaves as a floor set by the free-run phase penalty, not as a regime-specific failure. The worst 60\,bar case is Sea state 3, at 12.9\% NRMSE and 75\% peak capture (Fig.~\ref{Figure_10}c); its residual concentrates at the compression fronts, where a fraction of a cycle of timing offset displaces a 75\,bar edge, while the discharge plateaux track within the repeatability band of Section~\ref{sec:res-repeat}. Matching the 30\,bar up-crossings of model and measurement across the 60\,bar set puts a number on this: the median absolute front-timing offset is 23\,ms (90th percentile 55\,ms over 1\,264 fronts), and at front slopes of 0.4--0.8\,bar\,ms$^{-1}$ the timing alone accounts for errors of several bar at the fronts. Pooled over the 7\,141 waves of the campaign, the peak error is small in the median, 0.7\,bar at the 60\,bar setting and 1.1\,bar at 32\,bar, but heavy-tailed and signed. The model over-predicts the peaks of the small partial waves (mean $+2.2$\,bar below 20\,bar measured peak) and is nearly unbiased on the full strokes (mean $-0.6$\,bar above 65\,bar); the positive run means of Fig.~\ref{Figure_12}c ($+1.5$\,bar at 60\,bar, $+1.7$\,bar at 32\,bar) are pulled up by the small-wave tail. This over-cracking of the near-crack strokes is the same partial-stroke mechanism that limits the slowest sinusoid and the out-of-distribution test below. The pressure spectrum is matched out to several times the wave band, and the peak-exceedance curves of model and measurement overlie one another across all seven states at both valve settings (Fig.~\ref{Figure_11}). Figure~\ref{Figure_12} summarises both metrics against the run-to-run repeatability.

The re-sprung valve was found to crack at about 32\,bar rather than its nominal 40\,bar target, identified from the regulated plateaux of these records themselves. With that one scalar, the 32\,bar set achieves a mean NRMSE of 11.7\% and 83\% peak capture (59\% within 2\,bar; the fixed thresholds are relatively looser against the lower plateau), on a par with the 60\,bar set. Because the crack pressure comes from the same records, the 32\,bar rows of Table~\ref{Table_4} are transfer results, conditional on that scalar; every other parameter is carried over unchanged. The essential result is that a model whose parameters were fixed on ramps and sinusoids follows unseen irregular motion derived from real sea states at both valve settings, with no retuning beyond the crack pressure of the altered spring. The Supplementary Material shows every run at both settings, each record in its entirety with a magnified window alongside.

\subsection{Experimental repeatability and the measurement floor}
\label{sec:res-repeat}

\begin{figure}[!ht]\centering
\includegraphics[width=\linewidth]{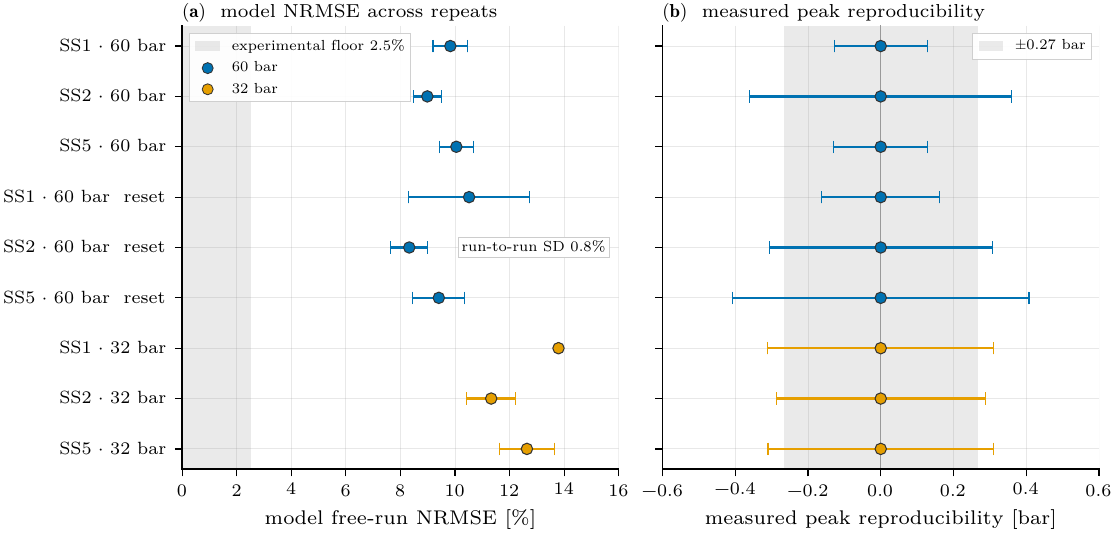}
\caption{Experimental repeatability over the thirty-seven repeated sea-state records, acquired in two sessions separated by a pressure-controller reset. (a) Model free-run NRMSE per repeated group against the point-wise experimental floor. (b) Measured peak-pressure reproducibility per group ($\pm$1\,SD). Blue 60\,bar, orange 32\,bar.}
\label{Figure_13}\end{figure}

A free-run error is only meaningful against the spread of the experiment itself. The test programme repeats three of the reference sea states, Sea states 1, 2 and 5, several times at each setting, and runs the whole 60\,bar set a second time after the pressure controller has been reset; thirty-seven such records were acquired and none was used in the identification. Because every repeat replays the same Froude-scaled target displacement, the run-to-run spread of the measured pressure is a demanding experimental reference for any free-run metric, though not a strict lower bound, since part of the spread is achieved-motion difference that the model observes through its input. In amplitude the rig repeats closely. The discharge-plateau pressure repeats to 0.18\,bar and the record peak pressure to 0.27\,bar across all nine repeated groups, including across the controller reset (Fig.~\ref{Figure_13}b), so the intermediate precision in the sense of ISO~5725-3, across sessions on one rig, is barely wider than the within-session repeatability. The point-wise spread is larger, at 2.5\% of range within a session. The reason is timing. Two nominally identical irregular records never line up exactly at the steep compression fronts, and the small run-to-run jitter there is the same phase penalty that inflates the free-run NRMSE. Since the amplitude figures repeat to well under a bar, the spread is in the timing, not the amplitude.

Run on each repeat, the model reproduces the measured pressure to a free-run NRMSE that is stable from run to run, 10.5\% on average with a run-to-run standard deviation of 0.8\% (Fig.~\ref{Figure_13}a). The residual free-run error is therefore set by the same run-to-run timing jitter of the compression fronts that governs the point-wise experimental floor, not by the amplitude reproducibility, which repeats to sub-bar, and it does not depend on which realisation of a sea state is used. Read with the near-flat error map and the sub-bar amplitude reproducibility, this locates the residual where the peak and spectral statistics already place it: in the free-run timing of the compression fronts, not in the amplitudes the model is asked to predict.

\subsection{Out-of-distribution test: re-sprung relief valve}
\label{sec:res-ood}

In the out-of-distribution test, the relief-valve spring was physically reset to a nominally lower crack pressure, and the 200\,mm sinusoids were repeated at four frequencies, 0.05 to 0.50\,Hz. The model is applied with only the crack pressure changed to the 32\,bar value identified in Section~\ref{sec:res-seastate}, and all other parameters held at their 60\,bar values. On the three frequencies whose strokes crack the valve, the model predicts the altered response at 12.7--19.8\% NRMSE and reproduces the peak pressures to between 0.6 and 4.4\,bar (Fig.~\ref{Figure_14}; the four runs are tabulated individually in the Supplementary Material). The 0.05\,Hz partial stroke does not transfer. Its measured peak of 26.9\,bar never reaches the identified crack, while the model over-cracks to 43.0\,bar and the whole-record error grows to 63.6\%. This is the near-crack regime already found weakest on the 60\,bar sinusoids, exercised here where the area law is pure extrapolation, and it bounds the transfer claim to strokes that reach the regulated plateau. For the cracking cases, the residual is a mild plateau over-prediction that grows with frequency, consistent with the valve area law $(a,b)$ having been identified at the original spring setting. The 5--95\% predictive band, propagated from the parameter bootstrap and a $\pm$2\,bar crack-pressure uncertainty, lies outside the measured plateau (Fig.~\ref{Figure_14}), which places the residual in the area characteristic of the re-sprung valve rather than in the identified parameters. Lowering a single physically meaningful parameter therefore moves the model onto the re-sprung hardware for every stroke that reaches the plateau.

\begin{figure}[!ht]\centering
\includegraphics[width=\linewidth]{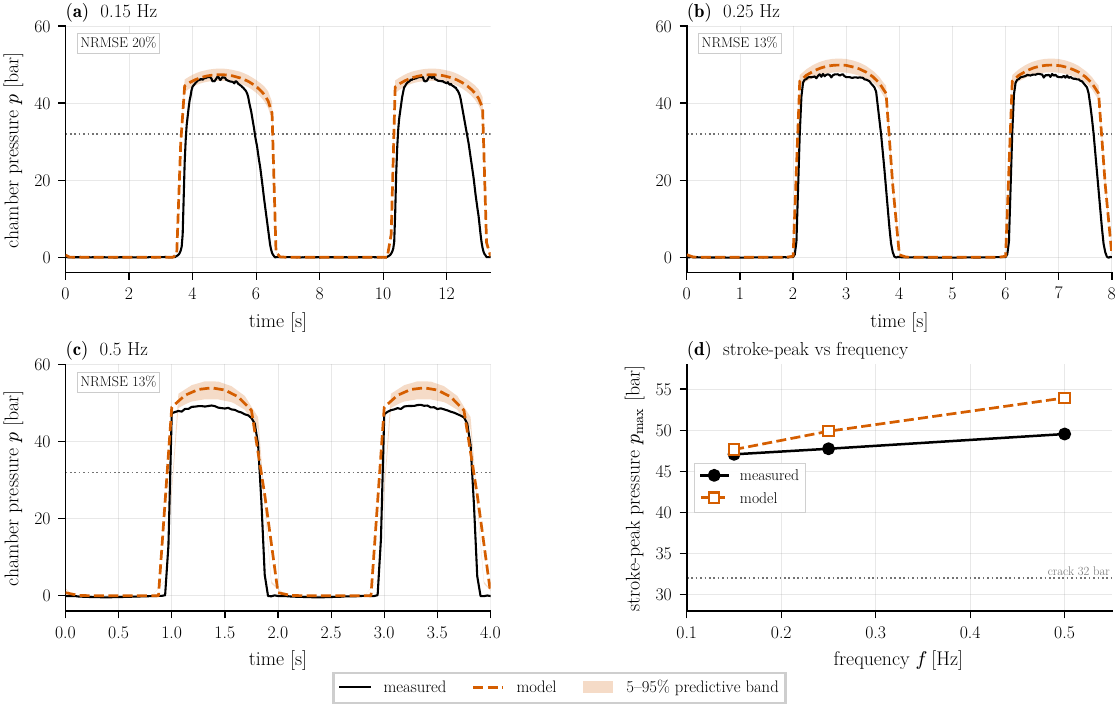}
\caption{Out-of-distribution transfer to the re-sprung relief valve, with only the crack pressure changed to the identified 32\,bar. (a--c) Measured and model chamber pressure at the three cracking frequencies, with the 5--95\% predictive band from the parameter bootstrap and the crack-pressure uncertainty. (d) Stroke-peak pressure against frequency, measured and model.}
\label{Figure_14}\end{figure}

\subsection{Long-duration stationarity}
\label{sec:res-longdur}

\begin{figure}[!ht]\centering
\includegraphics[width=\linewidth]{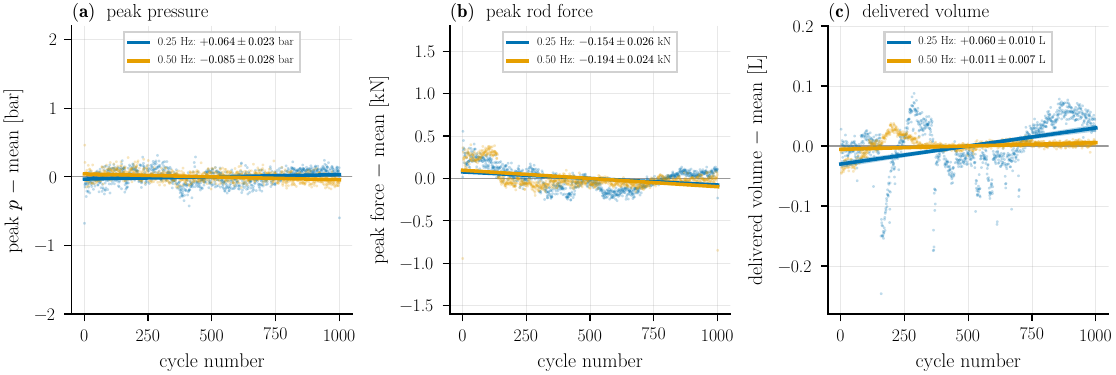}
\caption{Long-duration stationarity over the two $\sim$1000-cycle endurance runs (0.25 and 0.50\,Hz): (a) peak chamber pressure, (b) peak rod force and (c) valve-delivered volume per cycle, each referenced to its mean, with the ordinary-least-squares drift line and its 95\% confidence band.}
\label{Figure_15}\end{figure}

The two endurance runs test whether the pump response holds throughout extended operation. The model was run in free-run over the complete records: 1000 cycles at 0.25\,Hz (66.7\,min) and 1000 cycles at 0.50\,Hz (33.3\,min), with whole-record NRMSE values of 3.9\% and 5.1\%, respectively. The error is lower than in the 20-cycle tests because the long records are periodic and the phase penalty does not accumulate. A deterministic model on a periodic input cannot drift, so stationarity here is a property of the bench, not a further test of the model. Fitting each per-cycle quantity against cycle number (Fig.~\ref{Figure_15}; the delivered volume is inferred from the identified valve law on the measured pressure, the other two are direct measurements) resolves only negligible drift over the thousand cycles: the peak pressure by $+0.064\pm0.023$\,bar at 0.25\,Hz and $-0.085\pm0.028$\,bar at 0.50\,Hz (below 0.1\,bar in a 77\,bar peak), the peak rod force by $-0.15$ and $-0.19$\,kN (below 0.7\% of the $\sim$28\,kN peak), and the valve-delivered volume per cycle by $+0.06$ and $+0.01$\,L (a few percent of the $\sim$1\,L delivered). The ordinary-least-squares (OLS) intervals treat cycles as independent, so, with serial correlation, they are indicative; the magnitudes carry the conclusion. The trends are negligible, so the rig shows no sign of a slow change such as progressive seal wear on this hour-scale check, in pressure, mechanical load or delivered volume, and the identified parameters remain valid throughout; durability at sea is a separate question that the bench does not address. The constant peak offset of about 1\,bar is the same plateau bias seen in the sinusoidal set.

\subsection{Force readout and mechanical power}
\label{sec:res-force}

\begin{figure}[!ht]\centering
\includegraphics[width=\linewidth]{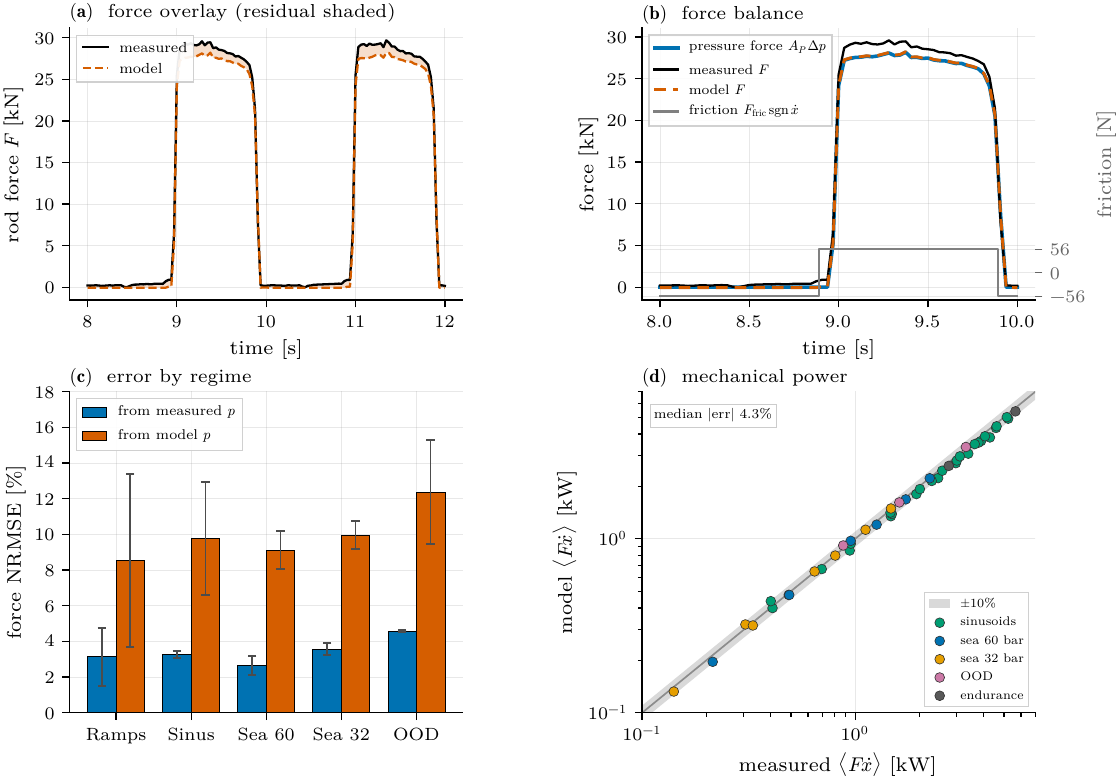}
\caption{Force readout, decomposition and mechanical power. (a) Measured and free-run model rod force over two cycles (200\,mm, 0.5\,Hz), residual shaded. (b) One cycle decomposed into pressure force (left axis) and Coulomb friction reversal (right axis), measured force de-tared. (c) Force NRMSE by regime, readout law against end to end, mean $\pm$1\,SD. (d) Cycle-mean power $\langle F\dot{x}\rangle$, model against measurement, with the $\pm$10\% band.}
\label{Figure_16}\end{figure}

The single model state is chamber pressure, and the rod force is the algebraic readout of Eq.~\eqref{eq:force}. On the validation records it is therefore a prediction to be checked, not a fitted output. Evaluated on the measured pressure, the readout matches the measured actuator force to a median NRMSE of 2.6--3.6\% across ramps, sinusoids and sea states, and 4.6\% on the re-sprung valve (Table~\ref{Table_5}), which validates the readout, with its fixed piston area and identified friction constant, independently of the chamber dynamics. Run end to end, with the force computed from the free-run pressure of the model itself, the error rises to 7--10\% (13\% on the re-sprung valve), because the readout then carries the free-run pressure error on top of its own, and tracks the underlying pressure fidelity almost exactly (Fig.~\ref{Figure_16}a,b). The pressure force $A_P(p-P_{\mathrm{atm}})$ dominates the rod load, and the friction reversal of $\pm56$\,N against a $\sim$29\,kN peak is a sub-percent correction. The systematic departure is a 1--2\,kN under-prediction of the rod force on the sinusoids (median peak bias $-1.8$\,kN, 6\% of the peak), and it persists at about 1\,kN along the discharge plateau, where the piston barely accelerates, so the moving-mass inertia that the quasi-static readout omits explains only the compression-front part. The plateau-level deficit behaves as a pressure-dependent seal-friction component that the constant Coulomb term does not carry. The seal is pressure-energised by design. A friction component that grows with the sealed pressure is expected rather than surprising. This component is collinear with the piston pressure force in these data, so the split is not identifiable from the rig force alone, and, unlike an inertial term, it dissipates net work, consistent in sign with the cycle-mean power bias below.

That mechanical power is the quantity of direct interest for a power take-off. The cycle-mean power delivered to the pump rod, $\langle F\dot{x}\rangle$, spans about 0.14 to 5.7\,kW over the cyclic records. The model reproduces it to a median relative error of 4.3\% (mean 6.7\%, Fig.~\ref{Figure_16}d): $-5.1\%$ on the sinusoids, $-2.7\%$ and $-0.4\%$ on the two sea-state sets and $-4.5\%$ on the endurance runs, and to within 2.1\% on the three cracking frequencies of the re-sprung valve. Resolving the load into work also closes the energy picture left implicit by the mass budget in Section~\ref{sec:res-loss}. On the 60\,bar sinusoids, the median record delivers 76\% of its rod power as high-pressure discharge at the relief valve, consistent with the 78.6\% campaign-aggregate indicator-work fraction of Section~\ref{sec:res-energy}, the balance being the compression spring-back returned at reversal together with the seal and valve losses already itemised. The identified model therefore reproduces the internal chamber pressure, the force the actuator must impose and the power the pump absorbs, across four excitation regimes and both valve settings.

\begin{table}[!ht]
\centering
\caption{Force readout and mechanical-power validation by regime, over whole records. The readout-law column evaluates Eq.~\eqref{eq:force} on the measured pressure, and the end-to-end column evaluates it on the free-run pressure of the model. The peak bias is the median model-minus-measured stroke peak force, and the last column is the mean relative error in the cycle-mean rod power $\langle F\dot{x}\rangle$.}
\label{Table_5}
\footnotesize
\setlength{\tabcolsep}{6pt}
\begin{tabular}{lccccc}
\toprule
Regime & $n$ & NRMSE law (\%) & NRMSE end-to-end (\%) & peak bias (kN) & $\langle F\dot{x}\rangle$ error (\%) \\
\midrule
Ramps                & 57 & 2.6 & 7.3  & $+0.6$ & --\phantom{$+0.0$} \\
Sinusoids            & 26 & 3.2 & 9.2  & $-1.8$ & $-5.1$ \\
Sea states (60\,bar) & 7  & 2.6 & 9.5  & $-0.7$ & $-2.7$ \\
Sea states (32\,bar) & 7  & 3.6 & 9.8  & $-0.3$ & $-0.4$ \\
Re-sprung valve      & 4  & 4.6 & 13.5 & $-0.5$ & $+2.1$\textsuperscript{a} \\
Endurance            & 2  & 2.2 & --   & $-1.1$ & $-4.5$ \\
\bottomrule
\end{tabular}

\vspace{2pt}
{\footnotesize\textsuperscript{a}For the three re-sprung-valve frequencies that reach the crack pressure.}
\end{table}

\subsection{Uncertainty quantification and practical identifiability}
\label{sec:res-uq}

\begin{figure}[!ht]\centering
\includegraphics[width=\linewidth]{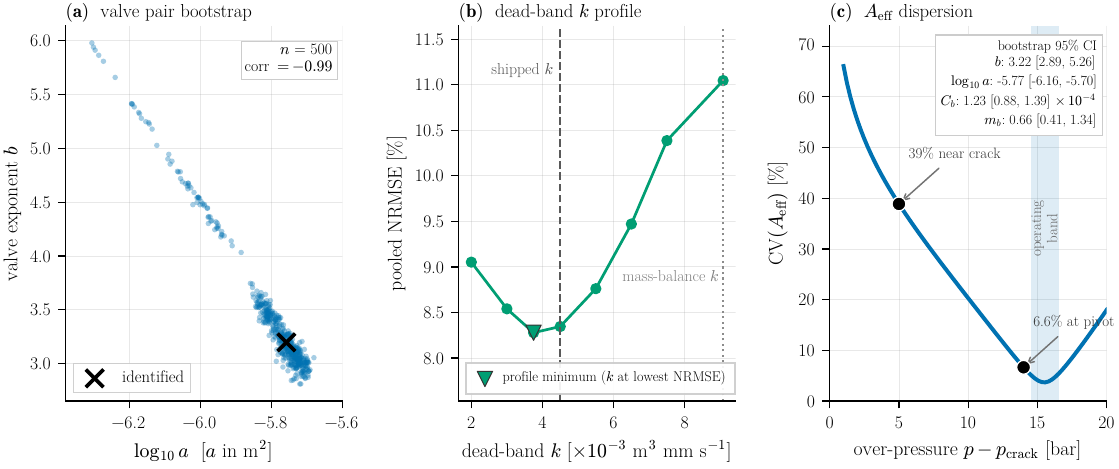}
\caption{Uncertainty and practical identifiability. (a) Bootstrap distribution of the valve pair (500 draws). (b) Trajectory profile of the dead-band constant $k$. (c) Coefficient of variation of the valve effective area against over-pressure, with the operating band marked.}
\label{Figure_17}\end{figure}

\begin{figure}[!ht]\centering
\includegraphics[width=\linewidth]{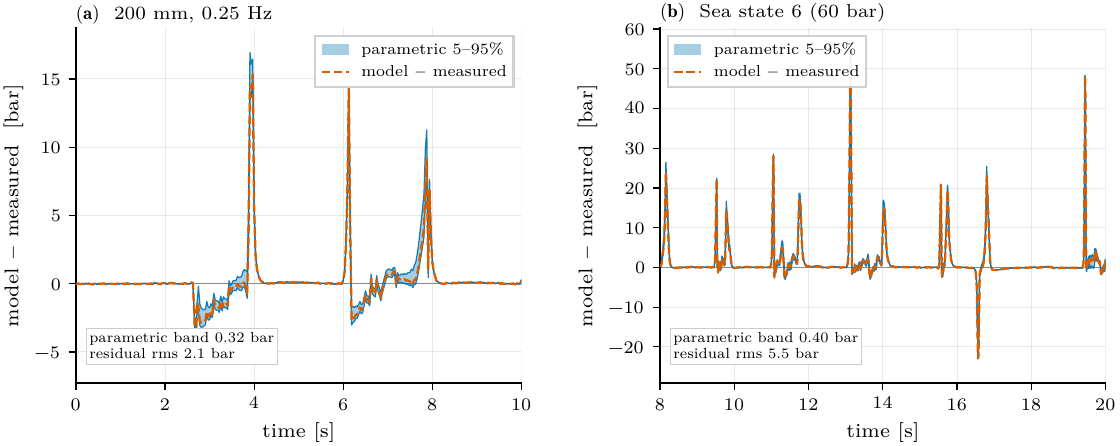}
\caption{Parametric predictive bands on a residual scale, quoted over the windows shown. Model-minus-measured pressure (dashed) with the 5--95\% band propagated from the bootstrap, on (a) a 200\,mm, 0.25\,Hz sinusoid window (an identification record) and (b) a Sea state 6 segment (validation only).}
\label{Figure_18}\end{figure}

Parameter uncertainty was quantified at three levels: Gauss--Newton (Laplace) standard errors at each block solution, a nonparametric bootstrap that resamples the 26 sinusoidal records and re-fits the valve and blow-by pairs 500 times, and forward propagation of the bootstrap draws to predictive bands on validation pressure traces (Fig.~\ref{Figure_17}). At the solution the Gauss--Newton curvature gives local standard errors of $b=3.20\pm0.20$, $\log_{10}a$ to $\pm0.036$, $C_b=(1.24\pm0.065)\times10^{-4}$\,m$^3$\,s$^{-1}$ and $m_b=0.64\pm0.12$, but these understate the intervals several-fold. The honest measure is the bootstrap, whose asymmetric 95\% intervals are $b\in[2.89,\,5.26]$, $\log_{10}a\in[-6.16,\,-5.70]$, $C_b\in[0.88,\,1.39]\times10^{-4}$\,m$^3$\,s$^{-1}$ and $m_b\in[0.41,\,1.34]$. The bootstrap widens these intervals and exposes their structure: the valve pair is strongly anti-correlated (correlation $-0.99$ between $\log_{10}a$ and $b$, Fig.~\ref{Figure_17}a), so the area scale and exponent are individually loose while their combination, the discharge law, is tight over the operating over-pressures. The coefficient of variation of $A_{\mathrm{eff}}$ falls from 39\% just above the crack to 6.6\% at the mid-plateau over-pressure (Fig.~\ref{Figure_17}c): the law is well constrained where the pump spends its cycle and least constrained on the near-crack partial strokes that carry the largest residuals. This is the narrow bootstrap band on $A_{\mathrm{eff}}(p)$ in Fig.~\ref{Figure_2}b, the expected structure when a scale and an exponent jointly set one measured quantity. A trajectory profile examines the dead-band constant pooled over six-cycle windows of six sinusoids spanning the amplitude range (Fig.~\ref{Figure_17}b). The pooled NRMSE is convex in $k$, and its minimum is flat between $3.75\times10^{-3}$ and $4.5\times10^{-3}$. Moving out to the mass-balance estimate of $9.08\times10^{-3}$ raises the error from 8.3\% to 11.0\%. On the pressure trajectories that estimate is therefore ruled out. The shipped $k=4.5\times10^{-3}$ sits on that flat minimum, within 0.1 percentage points of its floor; because the seal target subtracts the tip re-seat mass, moving $k$ from the mass-balance value to the refined one also shifts about two percentage points of inflow from the tip re-seat channel into the seal. Propagated through the forward model, the parameter uncertainty produces 5--95\% predictive bands of 0.32\,bar mean width over the sinusoid window of Fig.~\ref{Figure_18} and 0.40\,bar over its sea-state segment, the latter never used in any fit. Drawn as residuals against the measurement (Fig.~\ref{Figure_18}), these bands are a fraction of the 2--6\,bar free-run residual. Parametric uncertainty in the valve and blow-by pairs is therefore a small fraction of the residual free-run error, which is dominated by structural and measurement effects. The bands are conditional on the frozen ramp-identified block and on the fixed crack pressure, whose uncertainties are not propagated; within that scope the identified set is as resolved as the terminal data allow, and the three identifiability findings of Section~\ref{sec:ident} are what make it so.

\subsection{Verification}
\label{sec:res-verify}

The reported numbers rest on four checks. Two independent implementations of the stiff-record integrator, the reference stepper with callable inputs and the accelerated stepper with pre-sampled inputs, agree to better than $10^{-3}$\,bar on the validation records, and the adaptive LSODA and linearly-implicit schemes agree at plateau level on the smooth regimes, differing only in the timing resolution of the compression fronts at the output grid. The closed-cycle mass budget, assembled from measured signals, closes to $+0.8\%$ of the swept inflow in aggregate and to between $-6\%$ and $+15\%$ per record. One caution applies to reading this number. The blow-by channel is fitted to the non-valve residual, so the closure checks the accounting rather than validating the model independently, and it is reported in that spirit. The gauge/absolute frame convention is enforced in one place in the data layer: the model state is absolute, and all comparisons are made in the sensor frame. Finally, every metric quoted in this paper regenerates from the raw records by the archived identification and validation scripts, in which the ramp-stage parameters enter as stored constants of the pipeline, and the figures are produced from the same cached results the metrics are read from, so text, tables and figures cannot drift apart. The Supplementary Material shows every record the model was run on and collects the diagnostics behind the quoted statistics: complete free-run galleries of the ramp, sinusoidal, sea-state, repeatability, re-sprung-valve and endurance campaigns, the identified parameters with their uncertainties, the imposed-motion spectra, the front-timing distribution, the per-record closure residuals, the blow-by bootstrap geometry and the per-record force-channel metrics.

\subsection{Discussion and limitations}
\label{sec:res-discuss}

The results support the central claim that a parsimonious, physically interpretable grey-box model, identified block by block from terminal measurements, both decomposes the pump loss budget and generalises across excitation classes. The single-state formulation is deliberately spare: one differential state, an algebraic tip-valve switch and a handful of physically labelled parameters per block. It nonetheless reproduces the pumping cycle across ramps, sinusoids, real irregular sea states and thousand-cycle endurance records at both valve settings. For the intended uses, design, energy assessment and model-based control of a seawater-pump PTO, this parsimony is an advantage. The generalisation follows from keying every parameter to a physical state rather than to the label of a test run (Section~\ref{sec:approach}). This is what distinguishes the present model from the black-box identification strand \cite{giorgi2019,zhang2023} and from whole-system models whose loss terms are constructed forward, among them a component-level hydraulic PTO driven by measured wave data \cite{cargo2016}, a computational wave tank coupled to a PTO model \cite{penalba2018} and a stochastic optimisation of a piston-pump PTO across a scatter diagram \cite{zeinali2024}.

These results can be set against the assumptions that comparable studies make, which is where identifying the losses rather than assuming them shows its value. Cargo et al.~\cite{cargo2016} write the cylinder friction as a Coulomb term plus a viscous term, and note that such losses are system specific and are approximated from experience. The rod-force validation here shows what that approximation costs. A constant Coulomb term of 56.4\,N leaves a systematic deficit near 1\,kN along the discharge plateau, and that deficit behaves as seal friction growing with the sealed pressure. The relief valve tells a similar story. Simmons and Van de Ven~\cite{simmons2025} model the pressure relief valve with a flow coefficient varying linearly with poppet position, taken from a force balance between the static pressure and a linear spring, which is the near-linear lift law set aside in Section~\ref{sec:losses}. The area identified here from the weighed discharge follows a power law with exponent 3.20, and the bootstrap places that exponent between 2.89 and 5.26, so a linear opening law sits outside the identified range. With the closest bench-validated study the difference is one of evidence rather than form. Giorgi et al.~\cite{giorgi2024} take the hydraulic motor losses from manufacturer efficiency maps and report agreement with their bench data qualitatively. Every loss channel here is identified from terminal measurements instead, and the agreement is reported as whole-record error, 10.4\% and 11.7\% of range on the two sea-state sets.

Several limitations qualify the scope of the results. The bench fluid is water rather than seawater: seawater is about 2.5\% denser (orifice mass flows scale with $\sqrt{\rho}$) and several percent more viscous, and the temperature dependence of the polymer seal, corrosion and biofouling are outside the present identification, acting on precisely the component found to dominate the losses. The rig imposes a prescribed piston displacement and therefore reproduces the kinematic boundary condition of the energy collector without the two-way fluid--structure coupling of a deployed device, and no field behaviour is claimed. On the suction stroke, the model holds the chamber at atmospheric pressure. In contrast, the measured suction phase sits marginally below atmospheric for roughly a third of the samples on the fastest sea states (median $-0.1$\,bar, excursions to $-0.9$\,bar gauge, below the calibrated range of the transducers), so the refill is treated as complete and cavitation-free. The sub-atmospheric behaviour lies outside the identified envelope. The complete-refill assumption can then over-count the swept inflow; the identification sinusoids reach only $-0.5$\,bar on the fastest strokes, so the bias on the loss budget is bounded, and what remains is booked to the seal residual. The identification records also reach piston speeds of only 0.66\,m\,s$^{-1}$, while Sea state 4 reaches 1.1\,m\,s$^{-1}$, so the pressure-keyed seal law is extrapolated in speed on the most energetic states. The slowest strokes remain the hardest regime. The 0.05\,Hz sinusoid carries the largest in-regime error, its partial stroke lingering near the crack pressure where the area law is least constrained. The static-area valve description omits the nonlinear dynamics and chatter of the valve itself, which are treated in dedicated studies \cite{kadar2023}; the valve rides at low lift over the whole operating band, the regime where direct spring-operated valves are least steady, and the description leaves a frequency-dependent scatter of 21.5\% in the weighed discharge mass, partly attributable to the weighing of the small near-crack masses. The nominal 60\,bar crack carries a set-pressure tolerance that is entangled with the pinned 58\,bar blow-by onset, and the measured 0.15\,Hz plateau at the 32\,bar setting droops in a way the pressure-only seal law does not reproduce, so sub-onset losses cannot be excluded there. A pressure-keyed tip seating would replace the decompression surrogate of Section~\ref{sec:losses} with explicit expansion dynamics, at the cost of a second state. A dwell-dependent seal-creep term was evaluated and rejected as over-fitted, and a small plateau droop over extended holds is not chased. None of these qualifications affects the identified structure or the reported budget. What they mark out is the edge of the validated envelope, and each one points to a concrete next test. A seawater campaign would bring in the fluid effects. Ramp-and-hold tests between 55 and 60\,bar would locate the blow-by onset without the pinning. Collecting the discharge mass at the 32\,bar setting would test the prediction that the blow-by loss disappears there. A post-campaign inspection of the seal would support or overturn the attribution, and a re-identification of the valve area law across spring settings would sharpen the out-of-distribution prediction.

\section{Conclusion}
\label{sec:conclusion}

A block-modular grey-box model of a short-stroke reciprocating seawater pump, the power take-off of a wave energy converter, has been identified and validated at the system level against bench measurements. The pump was decomposed into physical blocks---chamber compliance, relief valve, pressure-energised piston seal and passive tip check valve---each with a structure fixed from component physics and only a few parameters identified from terminal pressure, displacement, force and discharged-mass data. A closed-cycle mass balance was the central identification device. Because the chamber storage term cancels over a periodic cycle, the otherwise degenerate seal and valve losses were separated by construction rather than by regularisation, and every parameter was keyed to a physical state rather than to a test condition.

The identified model recovers the loss budget by mechanism and places 84\% of the lost mass and 97\% of the lost energy on the pressure-energised piston seal, fixing the mechanical-to-hydraulic conversion of the pump at about four-fifths. With every parameter fixed on ramps and sinusoids, it then predicts whole 360\,s records of irregular sea-state-derived motion in free run at 10--12\% of range, capturing about 80\% of several thousand wave peaks within 5\,bar. It transfers to a re-sprung valve on the strokes that crack it, and tracks the thousand-cycle endurance records, whose measured drift stays below 0.1\,bar. Resolved into mechanical terms, it reproduces the actuator force and the absorbed power from about 0.1 to 5.7\,kW with a median error of 4\%. The uncertainty analysis leaves the residual free-run error as structural and experimental rather than parametric, at about four times the point-wise repeatability of the bench. Read together, these results show that the model is compact, physically interpretable and transferable, which makes it suitable for design, energy assessment and model-based control of seawater-pump power take-offs, and its mechanism-resolved loss budget offers a starting point for condition monitoring. The main avenues for further work are a seawater test campaign to capture salinity, temperature and biofouling effects, an independent location of the blow-by onset together with discharge-mass collection at the low valve setting, a re-identification of the valve area characteristic across spring settings to sharpen the out-of-distribution prediction, and the integration of the identified block into a full wave-to-wire model of the device.

\section*{Acknowledgement}
The authors thank Wavepiston A/S for providing the pump and supporting the bench test campaign, the SHY project partners and in particular Dr.~Matt Folley for the energy-collector motion records underlying the irregular sea states, Mr.~Kalle Stengaard J{\o}rgensen for his support of the test campaign, and the DTU Structural Lab (Section of Structures and Safety, Department of Civil and Mechanical Engineering, Technical University of Denmark) for laboratory support. This work was carried out within the EU SHY project (Seawater HYdraulic PTO using dynamic passive controller for wave energy converters), funded by the European Union under grant agreement No.~101147456.

\bibliographystyle{elsarticle-num}
\bibliography{references}

@article{cargo2016,
  author  = {Cargo, C. J. and Hillis, A. J. and Plummer, A. R.},
  title   = {Strategies for active tuning of Wave Energy Converter hydraulic power take-off mechanisms},
  journal = {Renewable Energy},
  year    = {2016},
  volume  = {94},
  pages   = {32--47},
  doi     = {10.1016/j.renene.2016.03.007}
}

@article{penalba2018,
  author  = {Penalba, Markel and Davidson, Josh and Windt, Christian and Ringwood, John V.},
  title   = {A high-fidelity wave-to-wire simulation platform for wave energy converters: Coupled numerical wave tank and power take-off models},
  journal = {Applied Energy},
  year    = {2018},
  volume  = {226},
  pages   = {655--669},
  doi     = {10.1016/j.apenergy.2018.06.008}
}

@article{penalba2019,
  author  = {Penalba, Markel and Ringwood, John V.},
  title   = {Linearisation-based nonlinearity measures for wave-to-wire models in wave energy},
  journal = {Ocean Engineering},
  year    = {2019},
  volume  = {171},
  pages   = {496--504},
  doi     = {10.1016/j.oceaneng.2018.11.033}
}

@article{penalba2020,
  author  = {Penalba, Markel and Ringwood, John V.},
  title   = {Systematic complexity reduction of wave-to-wire models for wave energy system design},
  journal = {Ocean Engineering},
  year    = {2020},
  volume  = {217},
  pages   = {107651},
  doi     = {10.1016/j.oceaneng.2020.107651}
}

@article{zeinali2024,
  author  = {Zeinali, Shokoufa and Wiktorsson, Magnus and Forsberg, Jan and Lindgren, Georg and Lindstr{\"o}m, Johan},
  title   = {Optimizing the hydraulic power take-off system in a wave energy converter},
  journal = {Ocean Engineering},
  year    = {2024},
  volume  = {310},
  pages   = {118636},
  doi     = {10.1016/j.oceaneng.2024.118636}
}

@article{asiikkis2024,
  author  = {Asiikkis, Andreas T. and Grigoriadis, Dimokratis G. E. and Vakis, Antonis I.},
  title   = {Wave-to-wire modelling and hydraulic PTO optimization of a dense point absorber WEC array},
  journal = {Renewable Energy},
  year    = {2024},
  volume  = {237},
  pages   = {121620},
  doi     = {10.1016/j.renene.2024.121620}
}

@article{dang2020,
  author  = {Dang, Tri Dung and Do, Tri Cuong and Ahn, Kyoung Kwan},
  title   = {Experimental assessment of the power conversion of a wave energy converter using hydraulic power take-off mechanism},
  journal = {International Journal of Precision Engineering and Manufacturing-Green Technology},
  year    = {2021},
  volume  = {8},
  number  = {5},
  pages   = {1515--1527},
  doi     = {10.1007/s40684-020-00261-z}
}

@article{giorgi2024,
  author  = {Giorgi, Simone and Coe, Ryan G. and Reasoner, Meagan M. and Bacelli, Giorgio and Forbush, Dominic D. and Jensen, Scott and Cazenave, Fran{\c c}ois and Hamilton, Andrew},
  title   = {Wave energy converter power take-off modeling and validation from experimental bench tests},
  journal = {IEEE Journal of Oceanic Engineering},
  year    = {2024},
  volume  = {49},
  number  = {2},
  pages   = {446--457},
  doi     = {10.1109/joe.2023.3345903}
}

@article{simmons2023,
  author  = {Simmons, II, Jeremy W. and Van de Ven, James D.},
  title   = {A comparison of power take-off architectures for wave-powered reverse osmosis desalination of seawater with co-production of electricity},
  journal = {Energies},
  year    = {2023},
  volume  = {16},
  number  = {21},
  pages   = {7381},
  doi     = {10.3390/en16217381}
}

@incollection{simmons2025,
  author    = {Simmons, Jeremy W. and Van de Ven, James D.},
  title     = {Design of hydraulic power take-offs for wave-powered reverse osmosis desalination: meeting constraints on pressure variation},
  booktitle = {Advancements in Fluid Power Technology: Sustainability, Electrification, and Digitalization: Proceedings of the Global Fluid Power Society PhD Symposium 2024},
  series    = {Lecture Notes in Mechanical Engineering},
  editor    = {Ericson, Liselott and Krus, Petter},
  publisher = {Springer},
  address   = {Cham},
  year      = {2025},
  pages     = {239--258},
  isbn      = {978-3-031-84505-5},
  doi       = {10.1007/978-3-031-84505-5_16}
}

@article{giorgi2019,
  author  = {Giorgi, Simone and Davidson, Josh and Jakobsen, Morten and Kramer, Morten and Ringwood, John V.},
  title   = {Identification of dynamic models for a wave energy converter from experimental data},
  journal = {Ocean Engineering},
  year    = {2019},
  volume  = {183},
  pages   = {426--436},
  doi     = {10.1016/j.oceaneng.2019.05.008}
}

@article{jaramillo2020,
  author  = {Jaramillo-Lopez, Fernando and Flannery, Brian and Murphy, Jimmy and Ringwood, John V.},
  title   = {Modelling of a three-body hinge-barge wave energy device using system identification techniques},
  journal = {Energies},
  year    = {2020},
  volume  = {13},
  number  = {19},
  pages   = {5129},
  doi     = {10.3390/en13195129}
}

@article{bacelli2019,
  author  = {Bacelli, Giorgio and Spencer, Steven J. and Patterson, David C. and Coe, Ryan G.},
  title   = {Wave tank and bench-top control testing of a wave energy converter},
  journal = {Applied Ocean Research},
  year    = {2019},
  volume  = {86},
  pages   = {351--366},
  doi     = {10.1016/j.apor.2018.09.009}
}

@article{zhang2023,
  author  = {Zhang, Jincheng and Zhao, Xiaowei and Greaves, Deborah and Jin, Siya},
  title   = {Modeling of a hinged-raft wave energy converter via deep operator learning and wave tank experiments},
  journal = {Applied Energy},
  year    = {2023},
  volume  = {341},
  pages   = {121072},
  doi     = {10.1016/j.apenergy.2023.121072}
}

@article{davis2020,
  author  = {Davis, Andrew F. and Fabien, Brian C.},
  title   = {Wave excitation force estimation of wave energy floats using extended Kalman filters},
  journal = {Ocean Engineering},
  year    = {2020},
  volume  = {198},
  pages   = {106970},
  doi     = {10.1016/j.oceaneng.2020.106970}
}

@article{hillis2020,
  author  = {Hillis, A. J. and Yardley, J. and Plummer, A. R. and Brask, A.},
  title   = {Model predictive control of a multi-degree-of-freedom wave energy converter with model mismatch and prediction errors},
  journal = {Ocean Engineering},
  year    = {2020},
  volume  = {212},
  pages   = {107724},
  doi     = {10.1016/j.oceaneng.2020.107724}
}

@article{orphin2021,
  author  = {Orphin, Jarrah and Nader, Jean-Roch and Penesis, Irene},
  title   = {Uncertainty analysis of a WEC model test experiment},
  journal = {Renewable Energy},
  year    = {2021},
  volume  = {168},
  pages   = {216--233},
  doi     = {10.1016/j.renene.2020.12.037}
}

@article{zong2020,
  author  = {Zong, Chaoyong and Zheng, Fengjie and Chen, Dianjing and Dempster, William and Song, Xueguan},
  title   = {Computational fluid dynamics analysis of the flow force exerted on the disk of a direct-operated pressure safety valve in energy system},
  journal = {Journal of Pressure Vessel Technology},
  year    = {2020},
  volume  = {142},
  number  = {1},
  pages   = {011702},
  doi     = {10.1115/1.4045131}
}

@article{wuc2021,
  author  = {Wu, Chengshuo and Li, Shiyang and Li, Qianqian and Wu, Peng and Huang, Bin and Wu, Dazhuan},
  title   = {Optimization of nonlinear pressure-flow characteristics of a spring-loaded pressure relief valve based on computational fluid dynamics simulation},
  journal = {Journal of Pressure Vessel Technology},
  year    = {2021},
  volume  = {143},
  number  = {6},
  pages   = {061401},
  doi     = {10.1115/1.4050933}
}

@article{xu2018,
  author  = {Xu, He and Wang, Haihang and Hu, Mingyu and Jiao, Liye and Li, Chang},
  title   = {Optimal design and experimental research of the anti-cavitation structure in the water hydraulic relief valve},
  journal = {Journal of Pressure Vessel Technology},
  year    = {2018},
  volume  = {140},
  number  = {5},
  pages   = {051601},
  doi     = {10.1115/1.4040893}
}

@article{zhangj2018,
  author  = {Zhang, Jian and Yang, Liu and Dempster, William and Yu, Xinhai and Jia, Jiuhong and Tu, Shan-Tung},
  title   = {Prediction of blowdown of a pressure relief valve using response surface methodology and CFD techniques},
  journal = {Applied Thermal Engineering},
  year    = {2018},
  volume  = {133},
  pages   = {713--726},
  doi     = {10.1016/j.applthermaleng.2018.01.079}
}

@article{kadar2023,
  author  = {K{\'a}d{\'a}r, Fanni and St{\'e}p{\'a}n, G{\'a}bor},
  title   = {Nonlinear dynamics and safety aspects of pressure relief valves},
  journal = {Nonlinear Dynamics},
  year    = {2023},
  volume  = {111},
  number  = {13},
  pages   = {12017--12032},
  doi     = {10.1007/s11071-023-08484-w}
}

@article{schmidt2010,
  author  = {Schmidt, T. and Andr{\'e}, M. and Poll, G.},
  title   = {A transient 2D-finite-element approach for the simulation of mixed lubrication effects of reciprocating hydraulic rod seals},
  journal = {Tribology International},
  year    = {2010},
  volume  = {43},
  number  = {10},
  pages   = {1775--1785},
  doi     = {10.1016/j.triboint.2009.11.012}
}

@article{lichen2025,
  author  = {Li, Jiewei and Chen, Guoqiang},
  title   = {Modeling and experimental study of reciprocating seal soft elastohydrodynamic lubrication considering structural thermal coupling},
  journal = {International Journal of Heat and Mass Transfer},
  year    = {2025},
  volume  = {239},
  pages   = {126564},
  doi     = {10.1016/j.ijheatmasstransfer.2024.126564}
}

@article{li2024seal,
  author  = {Li, Yongkang and Yu, Lang and Jiang, Hao and Zhao, Erhui},
  title   = {Transient sealing characteristics of Glyd-ring in the high water-based piston pair under reciprocating pump conditions},
  journal = {Tribology International},
  year    = {2024},
  volume  = {192},
  pages   = {109293},
  doi     = {10.1016/j.triboint.2024.109293}
}

@article{nikas2023,
  author  = {Nikas, George K.},
  title   = {Performance mapping of rectangular-rounded hydraulic reciprocating seals to minimize leakage, frictional work and abrasive wear with the aid of a duty parameter},
  journal = {Tribology International},
  year    = {2023},
  volume  = {179},
  pages   = {108191},
  doi     = {10.1016/j.triboint.2022.108191}
}

@article{wangw2023,
  author  = {Wang, Wei and Liu, Hao and Guo, Chenhao and Jiang, Haoyi and Ouyang, Xiaoping},
  title   = {Effect of vibration on reciprocating sealing performance},
  journal = {Tribology International},
  year    = {2023},
  volume  = {178},
  pages   = {108031},
  doi     = {10.1016/j.triboint.2022.108031}
}

@article{zhaox2022,
  author  = {Zhao, Xiuxu and Appiah, Emmanuel and Xia, Yage and Wang, Jizheng},
  title   = {Degradation process analysis and reliability prediction modeling of hydraulic reciprocating seal based on monitoring data},
  journal = {Engineering Failure Analysis},
  year    = {2022},
  volume  = {140},
  pages   = {106565},
  doi     = {10.1016/j.engfailanal.2022.106565}
}

@article{peng2020,
  author  = {Peng, Chao and Ouyang, Xiaoping and Guo, Shengrong and Zhou, Qinghe and Yang, Huayong},
  title   = {Numerical analysis of the traction effect on reciprocating seals in the hydraulic actuator},
  journal = {Tribology International},
  year    = {2020},
  volume  = {143},
  pages   = {105966},
  doi     = {10.1016/j.triboint.2019.105966}
}

@article{tran2012,
  author  = {Tran, Xuan Bo and Hafizah, Nur and Yanada, Hideki},
  title   = {Modeling of dynamic friction behaviors of hydraulic cylinders},
  journal = {Mechatronics},
  year    = {2012},
  volume  = {22},
  number  = {1},
  pages   = {65--75},
  doi     = {10.1016/j.mechatronics.2011.11.009}
}

@article{sarbu2024,
  author  = {S{\^a}rbu, Flavius Aurelian and Arn{\u a}u{\c t}, Felix and Deaconescu, Andrea and Deaconescu, Tudor},
  title   = {Theoretical and experimental research concerning the friction forces developed in hydraulic cylinder coaxial sealing systems made from polymers},
  journal = {Polymers},
  year    = {2024},
  volume  = {16},
  number  = {1},
  pages   = {157},
  doi     = {10.3390/polym16010157}
}

@article{pan2021,
  author  = {Pan, Qing and Zeng, Yunlong and Li, Yibo and Jiang, Xuepeng and Huang, Minghui},
  title   = {Experimental investigation of friction behaviors for double-acting hydraulic actuators with different reciprocating seals},
  journal = {Tribology International},
  year    = {2021},
  volume  = {153},
  pages   = {106506},
  doi     = {10.1016/j.triboint.2020.106506}
}

@article{schickhofer2022,
  author  = {Schickhofer, Lukas and Wimmer, Johannes},
  title   = {Fluid-structure interaction and dynamic stability of shock absorber check valves},
  journal = {Journal of Fluids and Structures},
  year    = {2022},
  volume  = {110},
  pages   = {103536},
  doi     = {10.1016/j.jfluidstructs.2022.103536}
}

@article{menendezblanco2019,
  author  = {Men{\'e}ndez-Blanco, Alberto and Fern{\'a}ndez Oro, Jes{\'u}s Manuel and Meana-Fern{\'a}ndez, Andr{\'e}s},
  title   = {Numerical methodology for the CFD simulation of diaphragm volumetric pumps},
  journal = {International Journal of Mechanical Sciences},
  year    = {2019},
  volume  = {150},
  pages   = {322--336},
  doi     = {10.1016/j.ijmecsci.2018.10.039}
}

@article{zhaob2018,
  author  = {Zhao, Bin and Jia, Xiaohan and Sun, Shukai and Wen, Jie and Peng, Xueyuan},
  title   = {FSI model of valve motion and pressure pulsation for investigating thermodynamic process and internal flow inside a reciprocating compressor},
  journal = {Applied Thermal Engineering},
  year    = {2018},
  volume  = {131},
  pages   = {998--1007},
  doi     = {10.1016/j.applthermaleng.2017.11.151}
}

@article{bacak2023,
  author  = {Bacak, Aykut and P{\i}narba{\c s}{\i}, Ali and Dalk{\i}l{\i}{\c c}, Ahmet Selim},
  title   = {A 3-D FSI simulation for the performance prediction and valve dynamic analysis of a hermetic reciprocating compressor},
  journal = {International Journal of Refrigeration},
  year    = {2023},
  volume  = {150},
  pages   = {135--148},
  doi     = {10.1016/j.ijrefrig.2023.01.028}
}

@article{qiu2023,
  author  = {Qiu, Guoyi and Zhu, Shaolong and Wang, Kai and Wang, Weibo and Hu, Junhui and Hu, Yun and Zhi, Xiaoqin and Qiu, Limin},
  title   = {Numerical study on the dynamic process of reciprocating liquid hydrogen pumps for hydrogen refueling stations},
  journal = {Energy},
  year    = {2023},
  volume  = {281},
  pages   = {128303},
  doi     = {10.1016/j.energy.2023.128303}
}

@article{zhang2019,
  author  = {Zhang, Zuti and Cao, Shuping and Wang, Huawei and Luo, Xiaohui and Deng, Jia and Zhu, Yuquan},
  title   = {The approach on reducing the pressure pulsation and vibration of seawater piston pump through integrating a group of accumulators},
  journal = {Ocean Engineering},
  year    = {2019},
  volume  = {173},
  pages   = {319--330},
  doi     = {10.1016/j.oceaneng.2018.12.078}
}

@book{merritt1967,
  author    = {Merritt, Herbert E.},
  title     = {Hydraulic Control Systems},
  publisher = {John Wiley \& Sons},
  address   = {New York},
  year      = {1991},
  isbn      = {978-0-471-59617-2},
  note      = {Reissue of the 1967 edition}
}

@book{ljung1999,
  author    = {Ljung, Lennart},
  title     = {System Identification: Theory for the User},
  edition   = {2},
  publisher = {Prentice Hall PTR},
  address   = {Upper Saddle River, NJ},
  year      = {1999},
  isbn      = {978-0-13-656695-3}
}

@article{gholizadeh2014,
  author  = {Gholizadeh, Hossein and Bitner, Doug and Burton, Richard and Schoenau, Greg},
  title   = {Modeling and experimental validation of the effective bulk modulus of a mixture of hydraulic oil and air},
  journal = {Journal of Dynamic Systems, Measurement, and Control},
  year    = {2014},
  volume  = {136},
  number  = {5},
  pages   = {051013},
  doi     = {10.1115/1.4027173}
}

@article{penalba2016,
  author  = {Penalba, Markel and Ringwood, John V.},
  title   = {A review of wave-to-wire models for wave energy converters},
  journal = {Energies},
  year    = {2016},
  volume  = {9},
  number  = {7},
  pages   = {506},
  doi     = {10.3390/en9070506}
}

@article{whittaker2012,
  author  = {Whittaker, Trevor and Folley, Matt},
  title   = {Nearshore oscillating wave surge converters and the development of Oyster},
  journal = {Philosophical Transactions of the Royal Society A: Mathematical, Physical and Engineering Sciences},
  year    = {2012},
  volume  = {370},
  number  = {1959},
  pages   = {345--364},
  doi     = {10.1098/rsta.2011.0152}
}

@techreport{iec62600103,
  author      = {{International Electrotechnical Commission}},
  title       = {Marine energy---Wave, tidal and other water current converters---Part 103: Guidelines for the early stage development of wave energy converters---Best practices and recommended procedures for the testing of pre-prototype devices},
  institution = {International Electrotechnical Commission},
  type        = {Technical Specification},
  number      = {IEC TS 62600-103:2024},
  edition     = {2.0},
  address     = {Geneva, Switzerland},
  year        = {2024},
  note        = {ISBN 978-2-8322-8963-1. Second edition; cancels and replaces IEC TS 62600-103:2018}
}

@article{penalba2019w2w,
  author  = {Penalba, Markel and Ringwood, John V.},
  title   = {A high-fidelity wave-to-wire model for wave energy converters},
  journal = {Renewable Energy},
  year    = {2019},
  volume  = {134},
  pages   = {367--378},
  doi     = {10.1016/j.renene.2018.11.040}
}

@article{gaspar2017,
  author  = {Gaspar, Jos{\'e} F. and Kamarlouei, Mojtaba and Sinha, Ashank and Xu, Haitong and Calv{\'a}rio, Miguel and Fa{\"y}, Fran{\c c}ois-Xavier and Robles, Eider and Guedes Soares, C.},
  title   = {Analysis of electrical drive speed control limitations of a power take-off system for wave energy converters},
  journal = {Renewable Energy},
  year    = {2017},
  volume  = {113},
  pages   = {335--346},
  doi     = {10.1016/j.renene.2017.05.085}
}

@article{windt2020,
  author  = {Windt, Christian and Davidson, Josh and Ransley, Edward J. and Greaves, Deborah and Jakobsen, Morten and Kramer, Morten and Ringwood, John V.},
  title   = {Validation of a {CFD}-based numerical wave tank model for the power production assessment of the {Wavestar} ocean wave energy converter},
  journal = {Renewable Energy},
  year    = {2020},
  volume  = {146},
  pages   = {2499--2516},
  doi     = {10.1016/j.renene.2019.08.059}
}

\clearpage
\setcounter{section}{0}
\setcounter{subsection}{0}
\setcounter{figure}{0}
\setcounter{table}{0}
\setcounter{equation}{0}
\renewcommand{\thesection}{S\arabic{section}}
\renewcommand{\thesubsection}{S\arabic{section}.\arabic{subsection}}
\renewcommand{\thefigure}{S\arabic{figure}}
\renewcommand{\thetable}{S\arabic{table}}
\renewcommand{\theequation}{S\arabic{equation}}
\renewcommand{\theHsection}{S\arabic{section}}
\renewcommand{\theHsubsection}{S\arabic{section}.\arabic{subsection}}
\renewcommand{\theHfigure}{S\arabic{figure}}
\renewcommand{\theHtable}{S\arabic{table}}
\renewcommand{\theHequation}{S\arabic{equation}}
\captionsetup{font=small,labelfont=bf}

\begin{center}
{\LARGE\bfseries Supplementary Material}\\[1ex]
{\large for ``Block-modular grey-box identification of a short-stroke seawater-pump power take-off for wave energy: loss decomposition and multi-regime validation''}\\[1ex]
Mahdi Tayyebati, Amir Hamza Siddiqui, Mian Masoud, Kristian Glejb{\o}l, Christian Berggreen
\end{center}
\vspace{1em}

\noindent This supplement shows every record the model was run on and collects
the per-record metrics behind every summary statistic quoted in the paper. All
57 ramp segments, all 26 sinusoidal records, all fourteen primary sea states,
the nine repeated groups, the four re-sprung-valve runs and both thousand-cycle
endurance runs appear here in full, each with its free-run model trace. No run
is omitted. Every figure and table regenerates from the archived
identification and validation caches by the script
\texttt{fig\_supplementary.py}, which ships with the analysis package, so
nothing in this document is produced by a separate pipeline.
Section~\ref{si:conventions} states the conventions and numerical settings.
Section~\ref{si:params} tabulates the identified parameters and the fixed
constants. Sections~\ref{si:ramps}--\ref{si:endurance} then walk through the
campaigns in the order the paper uses them, and Section~\ref{si:force} covers
the force channel. Section~\ref{si:diag} closes with the diagnostics that
support specific claims in the text.

\section{Conventions, data processing and numerical settings}
\label{si:conventions}

The conventions follow the paper exactly. The bench transducers read gauge
pressure, the model state is absolute pressure, and every comparison in this
document is made in the sensor gauge frame after converting the simulated
state. Free run means the model is integrated forward on the measured piston
motion alone, with no pressure feedback of any kind, from the measured initial
pressure of the record. Errors on whole records are reported as the
range-normalised root-mean-square error,
$\mathrm{NRMSE}=100\,\mathrm{RMSE}/(\max p_{\mathrm{meas}}-\min
p_{\mathrm{meas}})$, together with the RMSE in bar. Whole record means whole
record, including run-in and run-out.

Two forward schemes integrate the same governing equation. The smooth regimes,
ramps and sinusoids, use adaptive LSODA with relative tolerance $10^{-5}$,
absolute tolerance $10^{3}$\,Pa and a 2\,ms step ceiling on interpolated
inputs. The irregular sea states and the endurance records use a
linearly-implicit fixed-step Euler scheme,
$p_{n+1}=p_n+hf/(1-hJ)$ with $h\le1$\,ms and $J$ the numerically evaluated
Jacobian, which is unconditionally stable for the stiff decompression events.
The numerical-verification subsection of the paper shows the two schemes agree at
plateau level on common records. Long-duration figures decimate the plotted
traces for file size; the metrics are always computed on the full grids.

Where a record is long, the galleries below show the whole record beside a
magnified window a few cycles wide, marked by a grey band in the whole-record
panel. The whole-record NRMSE is printed on every panel, so the window never
replaces the record-level assessment.

\section{Identified parameters and fixed constants}
\label{si:params}

Table~\ref{Table_S1} lists the complete parameter set of the identified
model with the provenance of each value, the Laplace standard errors of the
2026 re-fit, and the record-level bootstrap intervals of the valve and blow-by
pairs (500 draws). Table~\ref{Table_S2} lists the fixed constants and
the solver settings. Together with the equations of Section~3 of the paper,
these two tables are sufficient to reproduce every simulation in this
document.

\begin{table}[H]
\centering\footnotesize
\caption{The identified parameter set. Standard errors are Laplace estimates
from the final re-fit; the bracketed intervals are record-level bootstrap 95\%
ranges (500 draws). The asymmetry of the $b$ and $m_b$ intervals is discussed
with Fig.~17 of the paper.}
\label{Table_S1}
\scriptsize\setlength{\tabcolsep}{3.5pt}
\begin{tabular}{lllll}
\toprule
Parameter & Value & Unit & SE / 95\% CI & Identified from \\
\midrule
dead + line volume $V_0$ & $7.09\times10^{-3}$ & m$^3$ & -- & ramp ODE fit (57 segments) \\
entrained-air fraction $\alpha$ & $2.83\times10^{-3}$ & -- & -- & ramp ODE fit (57 segments) \\
film-leak coefficient $C$ & $3.25\times10^{-6}$ & m$^2$ & -- & ramp ODE fit (57 segments) \\
film-leak exponent $\gamma$ & 6.43 & -- & -- & ramp ODE fit (57 segments) \\
tip back-leak coefficient $C_t$ & $2.93\times10^{-6}$ & m$^2$ & -- & ramp ODE fit (57 segments) \\
valve area scale $a$ & $1.75\times10^{-6}$ & m$^2$ & $\pm$$1.5\times10^{-7}$; [$6.9\times10^{-7}$, $2.0\times10^{-6}$] & log-mass fit, 26 records \\
valve area exponent $b$ & 3.20 & -- & $\pm$0.20; [2.9, 5.3] & log-mass fit, 26 records \\
blow-by coefficient $C_b$ & $1.24\times10^{-4}$ & m$^3$\,s$^{-1}$ & $\pm$$6.5\times10^{-6}$; [$8.8\times10^{-5}$, $1.4\times10^{-4}$] & closed-cycle mass balance \\
blow-by exponent $m_b$ & 0.642 & -- & $\pm$0.12; [0.41, 1.3] & closed-cycle mass balance \\
blow-by onset $p_{\mathrm{on}}$ & 58.0 & bar (g) & -- & fixed at observed onset \\
dead-band constant $k$ & $4.50\times10^{-3}$ & m$^3$\,mm\,s$^{-1}$ & -- & trajectory-refined \\
effective piston area $A_P$ & $3.56\times10^{-3}$ & m$^2$ & -- & rod-force fit, effective (held fixed) \\
Coulomb rod friction $F_{\mathrm{fric}}$ & 56.4 & N & -- & force-trace fit \\
\bottomrule
\end{tabular}

\end{table}

\begin{table}[H]
\centering\footnotesize
\caption{Fixed constants and numerical settings. The nominal
``40\,bar'' label of the re-sprung valve is not its true crack; the value used everywhere is the
32\,bar identified from the regulated plateaux (Section~4.5 of the paper).}
\label{Table_S2}
\setlength{\tabcolsep}{4.5pt}
\begin{tabular}{llll}
\toprule
Quantity & Value & Unit & Note \\
\midrule
atmospheric pressure $P_{\mathrm{atm}}$ & $1.01\times10^{5}$ & Pa & sensor gauge frame reference \\
water density $\rho$ & 1000 & kg\,m$^{-3}$ & fresh water at bench temperature \\
pure-water bulk modulus $\beta_L$ & $2.20\times10^{9}$ & Pa & literature value \\
adiabatic gas exponent $\kappa$ & 1.4 & -- & entrained air \\
seal-exponent normaliser $P_{\mathrm{ref}}$ & $6.00\times10^{6}$ & Pa & fixed scaling of Eq. film law \\
valve-area normaliser $P_{\mathrm{ref},v}$ & $1.00\times10^{6}$ & Pa & fixed scaling of the valve area law \\
tip velocity switch $v_\epsilon$ & $2.00\times10^{-3}$ & m\,s$^{-1}$ & tip gate threshold \\
crack pressure, nominal spring & 60 & bar (g) & bench setting \\
crack pressure, re-sprung valve & 32 & bar (g) & identified from regulated plateaux \\
LSODA tolerances (smooth regimes) & rtol $10^{-5}$, atol $10^3$ Pa & -- & max step 2 ms, input sub-sampling 4 \\
implicit Euler step (irregular/long) & $h \le 1$ ms & -- & linearly implicit, $p_{n+1}=p_n+hf/(1-hJ)$ \\
\bottomrule
\end{tabular}

\end{table}

\section{Ramp identification set: all 57 segments}
\label{si:ramps}

\begin{figure}[H]\centering
\includegraphics[width=\linewidth]{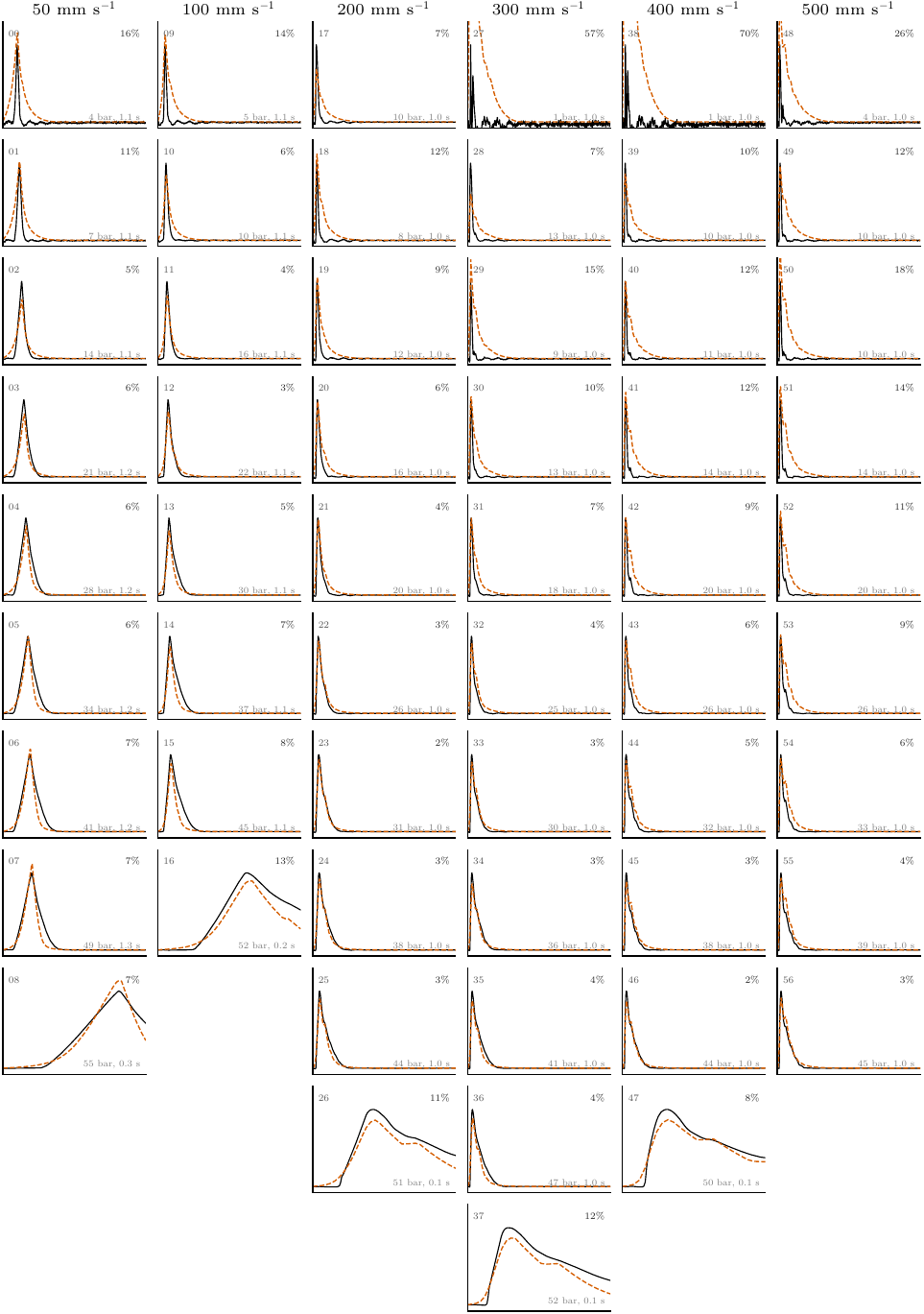}
\caption{All 57 ramp segments, measured (solid) and free-run model (dashed),
one panel per segment and one column per ramp speed. Each panel is scaled to
its own stroke and labelled with the segment index, the whole-segment NRMSE,
the peak pressure and the segment duration. The corresponding numbers are in
Table~\ref{Table_S3}.}
\label{Figure_S1}\end{figure}

Figure~\ref{Figure_S1} shows every ramp segment used for the chamber-block
identification, grouped by ramp speed, with the free-run model on the measured
motion. Table~\ref{Table_S3} gives the matching per-segment numbers. The
two segments with NRMSE above 50\% (300 and 400\,mm\,s$^{-1}$, first row of
the gallery) are the smallest strokes of the set, with peaks near 1\,bar; the
denominator of the normalised error is tiny and the sensor noise band covers
much of the signal. Their absolute errors are near one bar, consistent with
the pooled 1.7\,bar RMSE quoted in the paper.

\begin{table}[H]
\centering\scriptsize
\caption{Per-segment ramp metrics for all 57 segments (free-run on measured
motion). The table is split into two column blocks reading top to bottom,
left block first.}
\label{Table_S3}
\setlength{\tabcolsep}{4pt}
\begin{tabular}{rrrrr@{\hspace{2.2em}}rrrrr}
\toprule
rate [mm\,s$^{-1}$] & id & peak [bar] & NRMSE [\%] & RMSE [bar] & rate [mm\,s$^{-1}$] & id & peak [bar] & NRMSE [\%] & RMSE [bar] \\
\midrule
50 & 00 & 3.6 & 16.5 & 0.64 & 300 & 29 & 8.6 & 15.1 & 1.36 \\
50 & 01 & 6.7 & 10.7 & 0.74 & 300 & 30 & 13.4 & 10.0 & 1.38 \\
50 & 02 & 14.2 & 4.9 & 0.71 & 300 & 31 & 18.4 & 7.3 & 1.37 \\
50 & 03 & 20.7 & 5.6 & 1.17 & 300 & 32 & 24.9 & 4.3 & 1.09 \\
50 & 04 & 27.5 & 6.3 & 1.74 & 300 & 33 & 30.1 & 2.8 & 0.87 \\
50 & 05 & 34.5 & 6.3 & 2.17 & 300 & 34 & 36.4 & 2.7 & 0.99 \\
50 & 06 & 41.1 & 6.7 & 2.74 & 300 & 35 & 41.5 & 3.5 & 1.48 \\
50 & 07 & 48.7 & 7.2 & 3.53 & 300 & 36 & 46.5 & 4.3 & 2.04 \\
50 & 08 & 54.8 & 7.4 & 4.05 & 300 & 37 & 52.1 & 11.8 & 6.18 \\
100 & 09 & 5.1 & 13.9 & 0.74 & 400 & 38 & 1.2 & 70.3 & 1.17 \\
100 & 10 & 10.2 & 6.2 & 0.66 & 400 & 39 & 10.1 & 10.1 & 1.05 \\
100 & 11 & 15.8 & 4.1 & 0.66 & 400 & 40 & 11.5 & 11.9 & 1.41 \\
100 & 12 & 22.0 & 3.4 & 0.76 & 400 & 41 & 14.2 & 12.0 & 1.74 \\
100 & 13 & 30.4 & 5.1 & 1.57 & 400 & 42 & 19.9 & 8.8 & 1.78 \\
100 & 14 & 37.4 & 6.6 & 2.48 & 400 & 43 & 25.7 & 6.4 & 1.68 \\
100 & 15 & 44.9 & 7.8 & 3.53 & 400 & 44 & 31.9 & 4.6 & 1.47 \\
100 & 16 & 52.4 & 12.6 & 6.65 & 400 & 45 & 38.1 & 3.2 & 1.25 \\
200 & 17 & 9.9 & 7.3 & 0.75 & 400 & 46 & 44.4 & 2.5 & 1.10 \\
200 & 18 & 8.3 & 12.0 & 1.04 & 400 & 47 & 50.4 & 8.2 & 4.17 \\
200 & 19 & 11.7 & 9.2 & 1.10 & 500 & 48 & 4.4 & 25.8 & 1.21 \\
200 & 20 & 16.2 & 6.3 & 1.05 & 500 & 49 & 10.1 & 12.1 & 1.26 \\
200 & 21 & 20.4 & 4.4 & 0.91 & 500 & 50 & 9.8 & 18.2 & 1.83 \\
200 & 22 & 26.1 & 2.7 & 0.72 & 500 & 51 & 14.3 & 14.4 & 2.10 \\
200 & 23 & 31.4 & 2.2 & 0.69 & 500 & 52 & 19.9 & 10.8 & 2.18 \\
200 & 24 & 37.7 & 2.6 & 1.00 & 500 & 53 & 25.9 & 9.1 & 2.39 \\
200 & 25 & 44.1 & 3.5 & 1.54 & 500 & 54 & 32.6 & 6.4 & 2.11 \\
200 & 26 & 50.8 & 10.8 & 5.51 & 500 & 55 & 39.2 & 4.4 & 1.76 \\
300 & 27 & 1.3 & 56.9 & 1.00 & 500 & 56 & 45.3 & 2.9 & 1.34 \\
300 & 28 & 12.7 & 7.0 & 0.92 & & & & & \\
\bottomrule
\end{tabular}

\end{table}

\section{Sinusoidal campaign: all 26 records in full}
\label{si:sinus}

\begin{figure}[H]\centering
\includegraphics[width=0.98\linewidth]{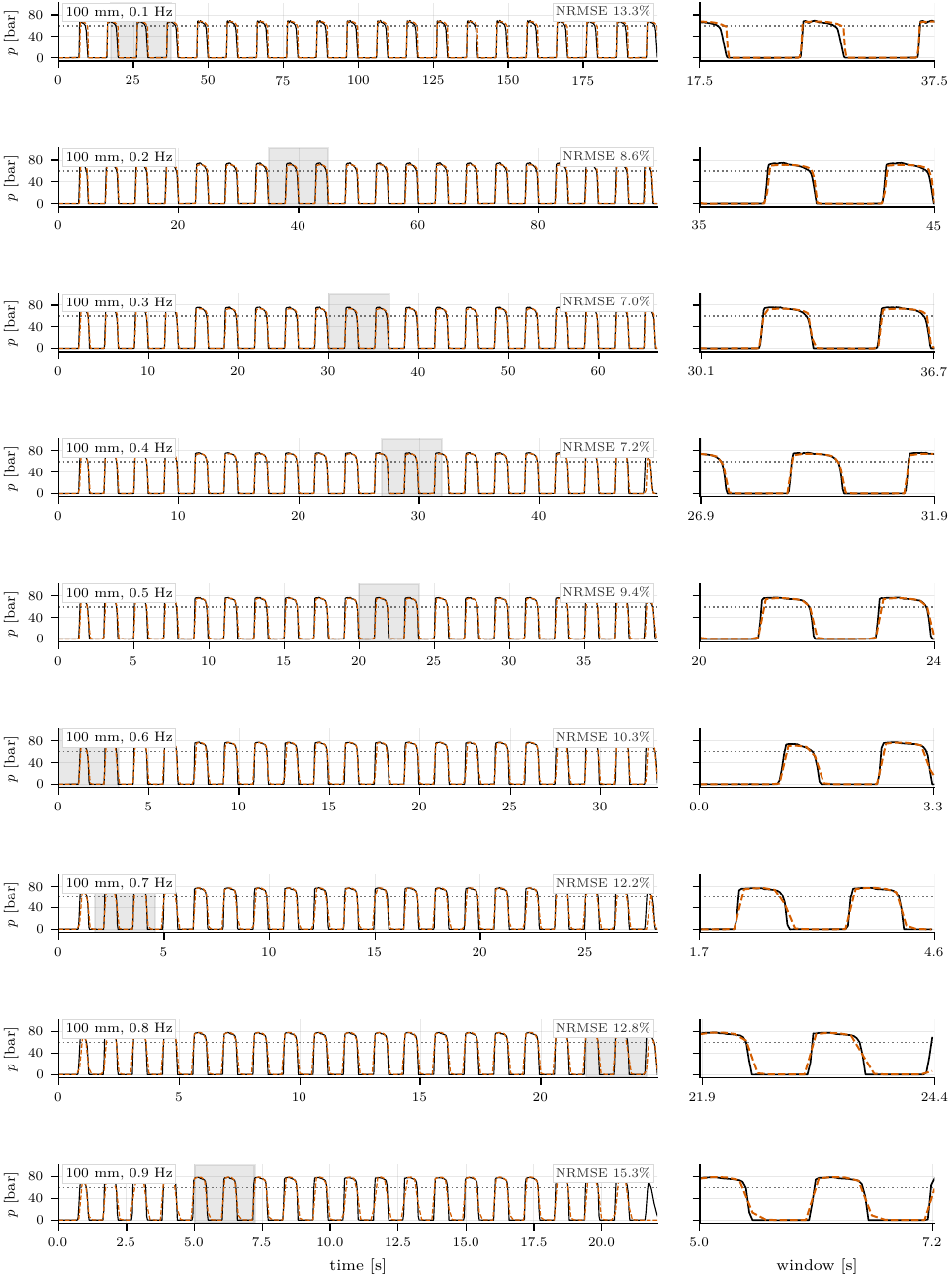}
\caption{The nine 100\,mm records in full (left, twenty cycles) with a
two-cycle window (right, grey band marks its position). Measured solid,
free-run model dashed, 60\,bar crack dotted. The whole-record NRMSE is printed
in each panel.}
\label{Figure_S2}\end{figure}

\begin{figure}[H]\centering
\includegraphics[width=0.98\linewidth]{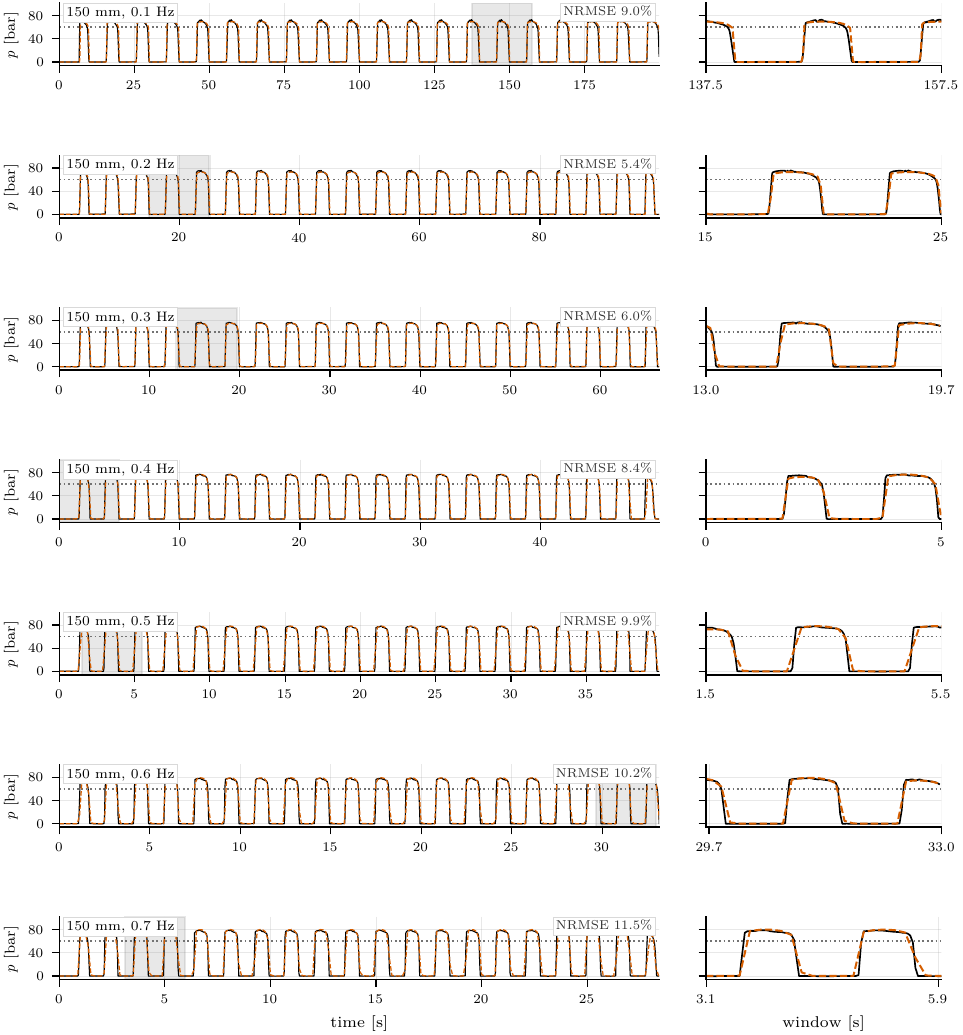}
\caption{As Fig.~\ref{Figure_S2}, for the seven 150\,mm records.}
\label{Figure_S3}\end{figure}

\begin{figure}[H]\centering
\includegraphics[width=0.94\linewidth]{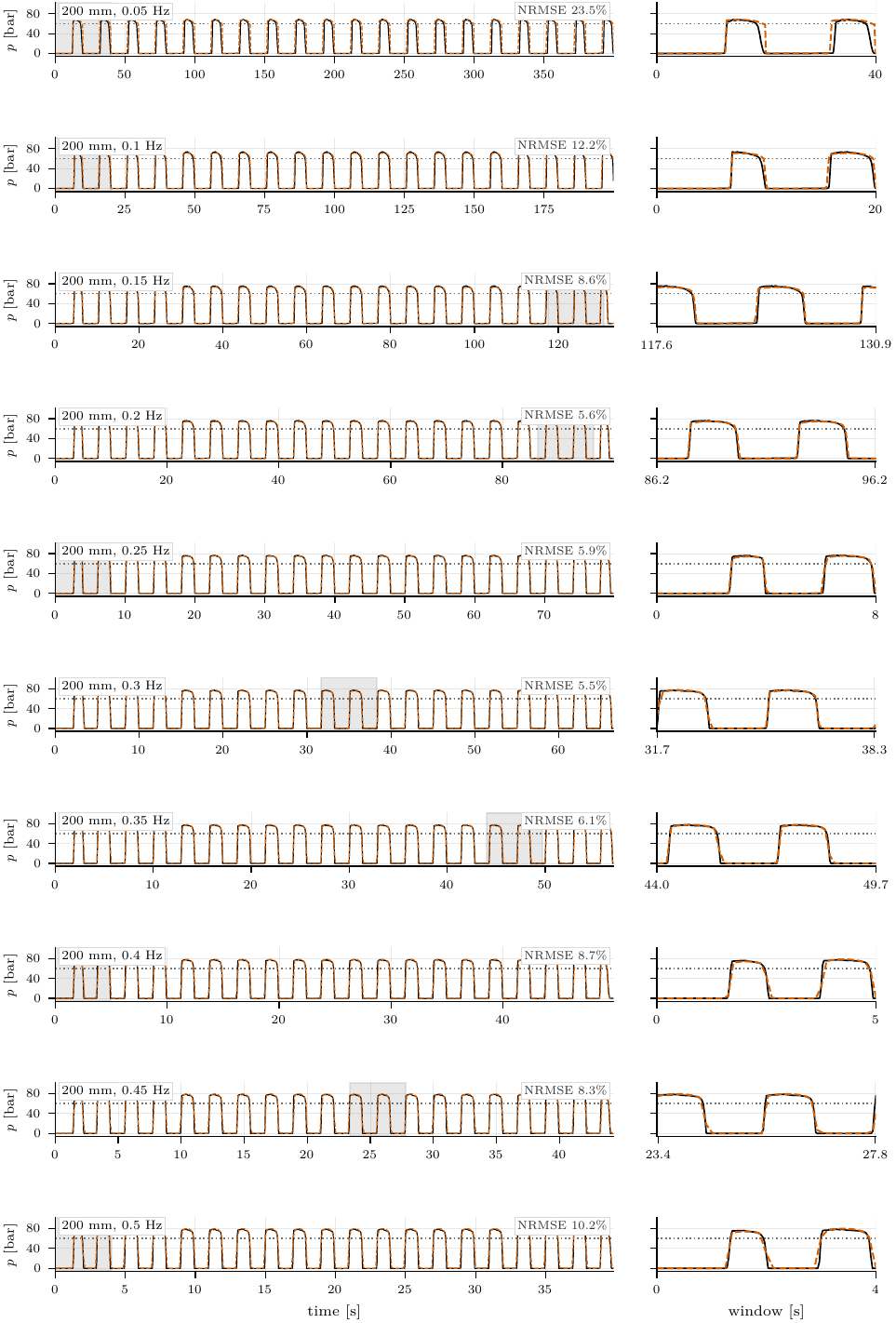}
\caption{As Fig.~\ref{Figure_S2}, for the ten 200\,mm records.}
\label{Figure_S4}\end{figure}

Figures~\ref{Figure_S2}--\ref{Figure_S4} show the complete
sinusoidal campaign by amplitude class. Every record appears whole, all twenty
cycles including run-in, with a two-cycle magnified window beside it.
Table~\ref{Table_S4} gives the per-record numbers, including the
steady-window error and the measured against modelled 20-cycle discharge
masses, and Table~\ref{Table_S5} gives the per-record energy budget. The
0.05\,Hz partial stroke carries the largest error of the campaign, consistent
with the near-crack regime identified as the weakest throughout the paper.
These records identify the valve and blow-by pairs (through the masses) and
the dead-band constant (through six of the trajectories), so the trace errors
here are partly in-sample, as stated with Table~4 of the paper.

\begin{table}[H]
\centering\footnotesize
\caption{All 26 sinusoidal records: whole-record and steady-window free-run
pressure error, stroke peaks missed at the 5\,bar criterion, and the measured
versus modelled 20-cycle valve discharge mass.}
\label{Table_S4}
\scriptsize\setlength{\tabcolsep}{3.5pt}
\begin{tabular}{rrrrrrrrr}
\toprule
$A$ [mm] & $f$ [Hz] & NRMSE [\%] & steady [\%] & RMSE [bar] & peaks missed ($>$5 bar) & $m_{20}$ meas [kg] & $m_{20}$ model [kg] & mass err [\%] \\
\midrule
100 & 0.1 & 13.3 & 12.8 & 9.4 & 0/30 & 4.82 & 3.49 & -27.6 \\
100 & 0.2 & 8.6 & 8.1 & 6.5 & 0/25 & 7.97 & 11.82 & +48.3 \\
100 & 0.3 & 7.0 & 7.1 & 5.3 & 0/22 & 9.30 & 12.73 & +36.8 \\
100 & 0.4 & 7.2 & 5.0 & 5.5 & 0/24 & 9.65 & 11.56 & +19.8 \\
100 & 0.5 & 9.4 & 9.6 & 7.3 & 0/21 & 10.13 & 10.95 & +8.0 \\
100 & 0.6 & 10.3 & 10.1 & 8.1 & 0/21 & 10.41 & 10.23 & -1.8 \\
100 & 0.7 & 12.2 & 10.2 & 9.6 & 0/20 & 10.35 & 9.47 & -8.5 \\
100 & 0.8 & 12.8 & 12.2 & 10.1 & 0/20 & 10.38 & 9.01 & -13.2 \\
100 & 0.9 & 15.3 & 12.0 & 12.1 & 1/21 & 10.25 & 8.55 & -16.6 \\
150 & 0.1 & 9.0 & 9.0 & 6.6 & 0/30 & 10.30 & 12.19 & +18.3 \\
150 & 0.2 & 5.4 & 5.3 & 4.1 & 0/25 & 14.15 & 18.82 & +33.0 \\
150 & 0.3 & 6.0 & 5.2 & 4.6 & 0/22 & 15.66 & 17.60 & +12.4 \\
150 & 0.4 & 8.4 & 7.6 & 6.5 & 0/22 & 15.96 & 15.44 & -3.3 \\
150 & 0.5 & 9.9 & 9.8 & 7.8 & 0/21 & 16.40 & 14.63 & -10.8 \\
150 & 0.6 & 10.2 & 10.8 & 8.1 & 0/20 & 16.67 & 13.88 & -16.7 \\
150 & 0.7 & 11.5 & 10.9 & 9.2 & 0/20 & 16.18 & 12.71 & -21.5 \\
200 & 0.05 & 23.5 & 23.0 & 16.3 & 0/43 & 8.33 & 6.17 & -26.0 \\
200 & 0.1 & 12.2 & 12.4 & 9.1 & 0/29 & 15.96 & 20.84 & +30.6 \\
200 & 0.15 & 8.6 & 10.0 & 6.5 & 0/28 & 18.36 & 24.51 & +33.5 \\
200 & 0.2 & 5.6 & 6.5 & 4.3 & 0/25 & 20.11 & 24.09 & +19.8 \\
200 & 0.25 & 5.9 & 6.4 & 4.6 & 0/24 & 21.21 & 22.50 & +6.1 \\
200 & 0.3 & 5.5 & 5.7 & 4.2 & 0/23 & 21.86 & 21.06 & -3.7 \\
200 & 0.35 & 6.1 & 6.3 & 4.8 & 0/22 & 22.33 & 20.37 & -8.8 \\
200 & 0.4 & 8.7 & 8.7 & 6.8 & 0/23 & 22.07 & 18.88 & -14.5 \\
200 & 0.45 & 8.3 & 8.3 & 6.5 & 0/22 & 22.37 & 18.62 & -16.8 \\
200 & 0.5 & 10.2 & 10.4 & 8.1 & 0/21 & 22.50 & 17.84 & -20.7 \\
\bottomrule
\end{tabular}

\end{table}

\begin{table}[H]
\centering\footnotesize
\caption{Per-record energy budget over the sinusoidal campaign: mechanical
indicator work $E_{\mathrm{in}}$, the delivered (relief-valve) fraction, and
the loss channels as shares of $E_{\mathrm{in}}$. The campaign-aggregate
delivered fraction is 78.6\% (bootstrap 95\% interval 72--84\%); the
operating-point dependence quoted in Section~4.3 of the paper is the spread of
the delivered column.}
\label{Table_S5}
\setlength{\tabcolsep}{5pt}
\begin{tabular}{rrrrrrr}
\toprule
$A$ [mm] & $f$ [Hz] & $E_{\mathrm{in}}$ [kJ] & delivered [\%] & blow-by [\%] & film [\%] & tip (both) [\%] \\
\midrule
100 & 0.1 & 76.5 & 31.1 & 50.4 & 0.76 & 0.13 \\
100 & 0.2 & 88.9 & 97.0 & 38.8 & 0.29 & 0.64 \\
100 & 0.3 & 92.6 & 102.2 & 29.4 & 0.18 & 0.77 \\
100 & 0.4 & 91.5 & 94.6 & 24.1 & 0.13 & 1.53 \\
100 & 0.5 & 92.8 & 89.0 & 20.4 & 0.10 & 1.56 \\
100 & 0.6 & 93.4 & 83.0 & 17.7 & 0.08 & 1.40 \\
100 & 0.7 & 91.5 & 78.8 & 15.7 & 0.07 & 1.24 \\
100 & 0.8 & 91.8 & 75.1 & 14.1 & 0.05 & 1.63 \\
100 & 0.9 & 89.6 & 73.3 & 13.0 & 0.06 & 1.41 \\
150 & 0.1 & 132.0 & 65.3 & 45.0 & 0.38 & 0.08 \\
150 & 0.2 & 141.7 & 98.3 & 30.0 & 0.15 & 0.60 \\
150 & 0.3 & 145.7 & 90.6 & 22.2 & 0.09 & 0.75 \\
150 & 0.4 & 143.3 & 81.4 & 17.8 & 0.06 & 1.07 \\
150 & 0.5 & 144.9 & 77.0 & 14.8 & 0.05 & 0.95 \\
150 & 0.6 & 145.8 & 73.1 & 12.8 & 0.04 & 0.58 \\
150 & 0.7 & 142.3 & 69.0 & 11.3 & 0.03 & 0.62 \\
200 & 0.05 & 152.2 & 27.3 & 51.4 & 0.75 & 0.01 \\
200 & 0.1 & 181.9 & 82.6 & 38.4 & 0.23 & 0.02 \\
200 & 0.15 & 186.8 & 97.0 & 29.9 & 0.13 & 0.24 \\
200 & 0.2 & 192.9 & 93.2 & 24.2 & 0.10 & 0.47 \\
200 & 0.25 & 196.5 & 86.0 & 20.4 & 0.07 & 0.53 \\
200 & 0.3 & 198.2 & 80.2 & 17.6 & 0.06 & 0.51 \\
200 & 0.35 & 199.6 & 77.5 & 15.6 & 0.05 & 0.47 \\
200 & 0.4 & 195.2 & 73.8 & 14.1 & 0.04 & 0.66 \\
200 & 0.45 & 196.4 & 72.8 & 12.8 & 0.04 & 0.46 \\
200 & 0.5 & 197.3 & 69.7 & 11.7 & 0.04 & 0.70 \\
\bottomrule
\end{tabular}

\end{table}

\section{Sea states: all fourteen primary records in full}
\label{si:seastate}

\begin{figure}[H]\centering
\includegraphics[width=0.72\linewidth]{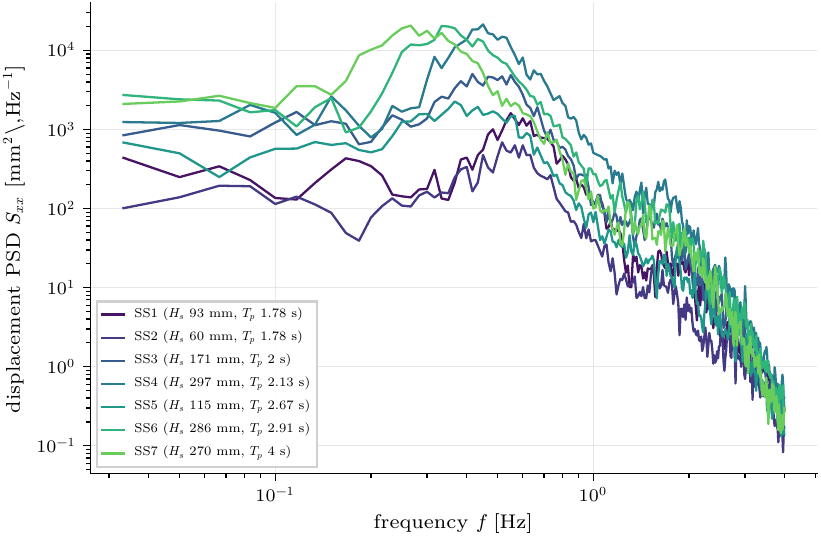}
\caption{Displacement power spectral density of the seven achieved sea-state
motion records (60\,bar primaries; Welch estimate, 60\,s segments). The
spectra correspond to the imposed-displacement statistics of Table~2 in the
paper; they describe the imposed piston motion, not a wave field.}
\label{Figure_S5}\end{figure}

Figures~\ref{Figure_S6} and \ref{Figure_S7} show the fourteen primary
irregular records whole, all 360\,s of each, with a 14\,s window beside each
record. Table~\ref{Table_S6} gives the per-record metrics at both peak
thresholds. Figure~\ref{Figure_S5} characterises the imposed motion itself.
The 32\,bar rows are transfer conditional on the crack pressure identified
from the regulated plateaux of these records, as Section~4.5 of the paper states.

\begin{table}[H]
\centering\footnotesize
\caption{All fourteen primary sea-state runs: free-run error, per-wave peak
capture at both thresholds, and the signed mean peak bias.}
\label{Table_S6}
\setlength{\tabcolsep}{5pt}
\begin{tabular}{llrrrrrr}
\toprule
Setting & Case & NRMSE [\%] & RMSE [bar] & waves $n$ & $\le$5 bar [\%] & $\le$2 bar [\%] & mean bias [bar] \\
\midrule
60 bar & Sea state 1 & 10.7 & 8.5 & 593 & 77 & 67 & +1.85 \\
60 bar & Sea state 2 & 8.7 & 6.7 & 575 & 80 & 72 & -0.00 \\
60 bar & Sea state 3 & 12.9 & 10.3 & 511 & 75 & 72 & +3.34 \\
60 bar & Sea state 4 & 11.4 & 9.4 & 429 & 88 & 76 & +1.71 \\
60 bar & Sea state 5 & 11.0 & 8.6 & 538 & 70 & 67 & +2.13 \\
60 bar & Sea state 6 & 9.5 & 7.7 & 430 & 87 & 72 & +1.28 \\
60 bar & Sea state 7 & 8.7 & 7.0 & 447 & 85 & 72 & +0.46 \\
32 bar & Sea state 1 & 13.8 & 6.8 & 576 & 76 & 61 & +2.36 \\
32 bar & Sea state 2 & 10.3 & 5.1 & 585 & 81 & 72 & +0.60 \\
32 bar & Sea state 3 & 12.0 & 6.1 & 529 & 83 & 60 & +2.14 \\
32 bar & Sea state 4 & 11.4 & 6.1 & 448 & 85 & 42 & +2.90 \\
32 bar & Sea state 5 & 11.5 & 5.7 & 558 & 77 & 65 & +0.93 \\
32 bar & Sea state 6 & 11.3 & 5.9 & 451 & 86 & 54 & +2.09 \\
32 bar & Sea state 7 & 11.4 & 5.9 & 471 & 89 & 62 & +1.37 \\
\bottomrule
\end{tabular}

\end{table}

\begin{landscape}
\begin{figure}[p]\centering
\includegraphics[width=\linewidth]{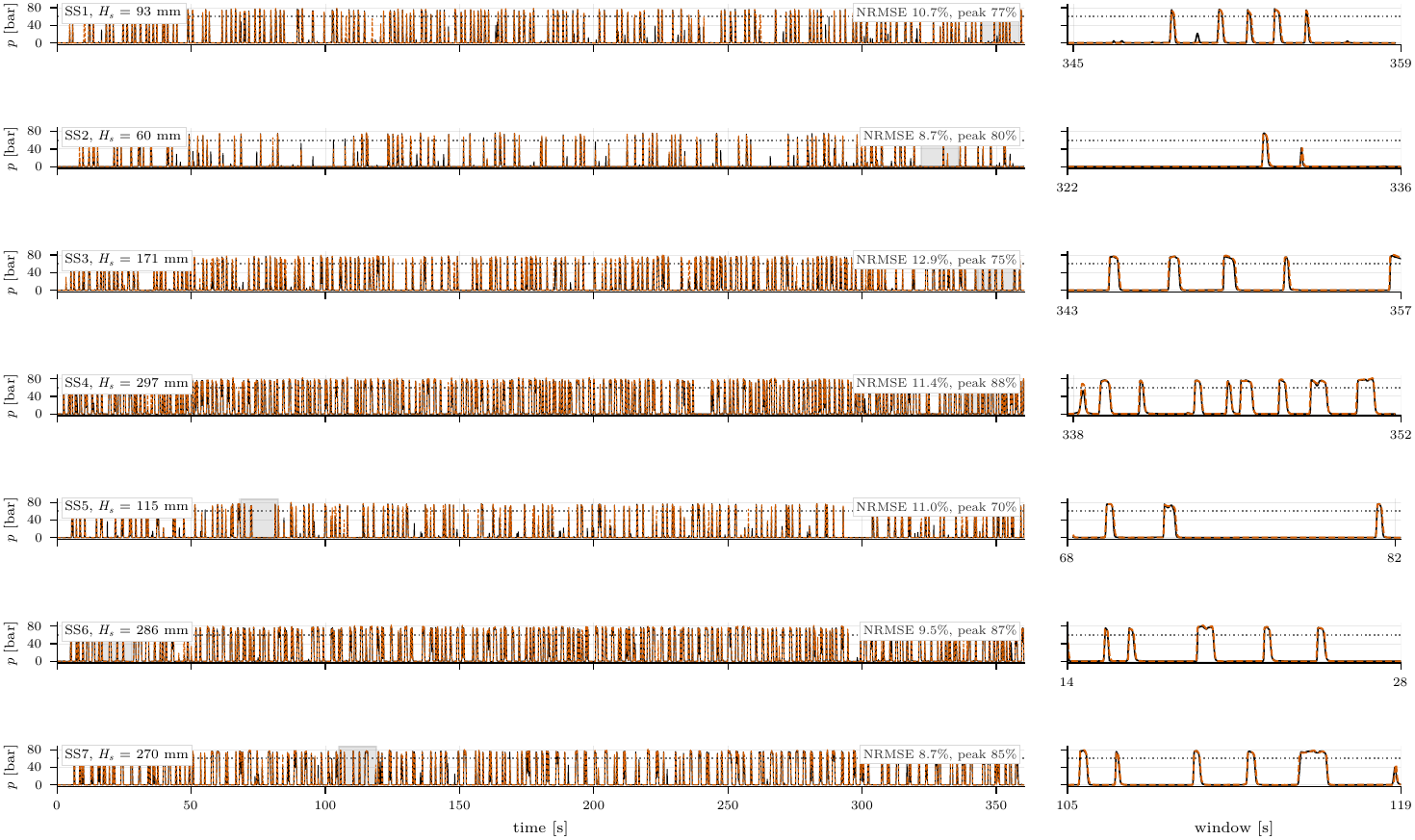}
\caption{All seven 60\,bar sea-state records in full (left, 360\,s) with a
14\,s window (right; the grey band in the whole-record panel marks its
position). Measured solid, free-run model dashed, 60\,bar crack dotted;
whole-record NRMSE and peak capture printed per record.}
\label{Figure_S6}\end{figure}

\begin{figure}[p]\centering
\includegraphics[width=\linewidth]{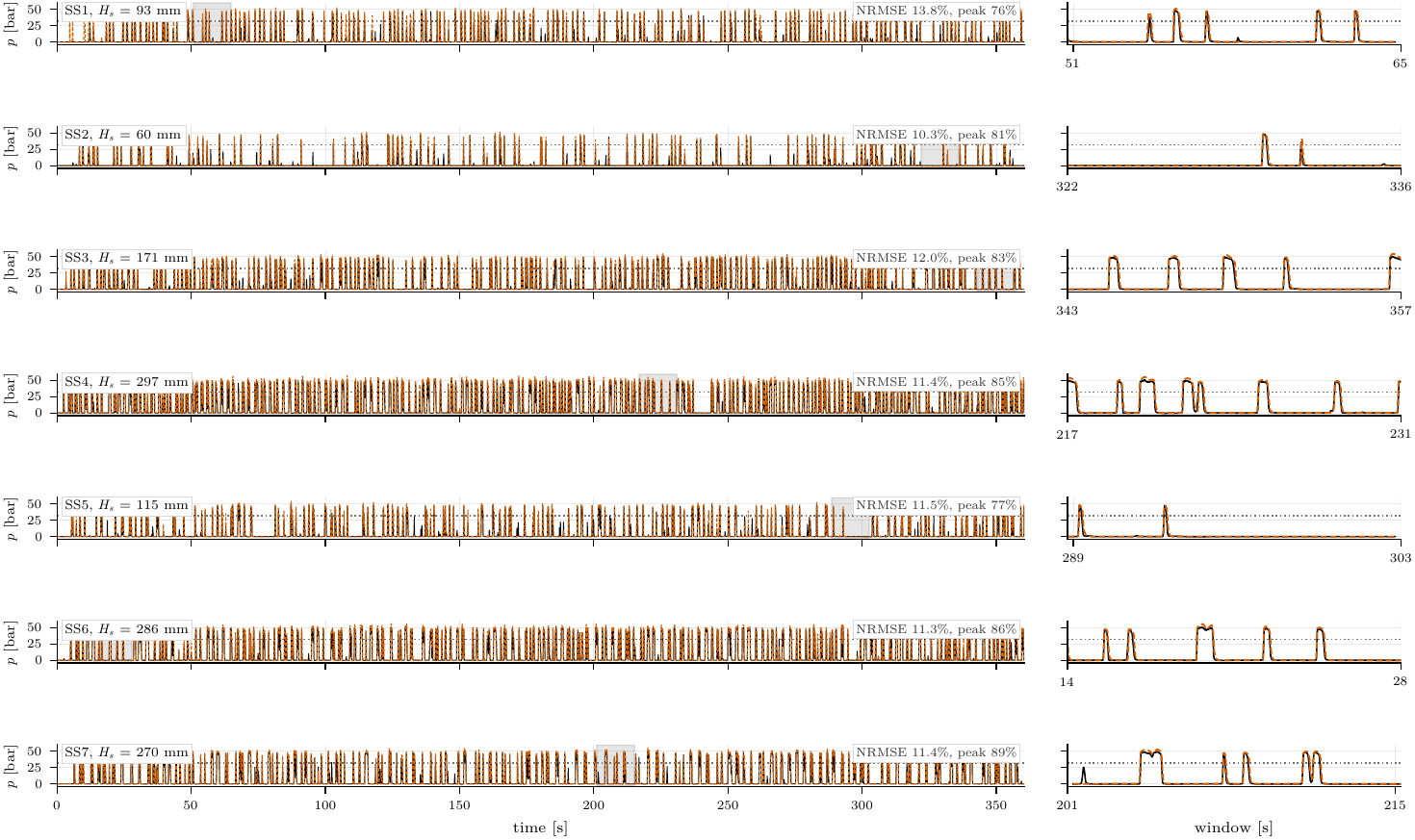}
\caption{As Fig.~\ref{Figure_S6}, for the seven re-sprung-valve records
(identified crack 32\,bar, dotted).}
\label{Figure_S7}\end{figure}
\end{landscape}

\section{Repeatability: the nine repeated groups}
\label{si:repeat}

\begin{figure}[H]\centering
\includegraphics[width=\linewidth]{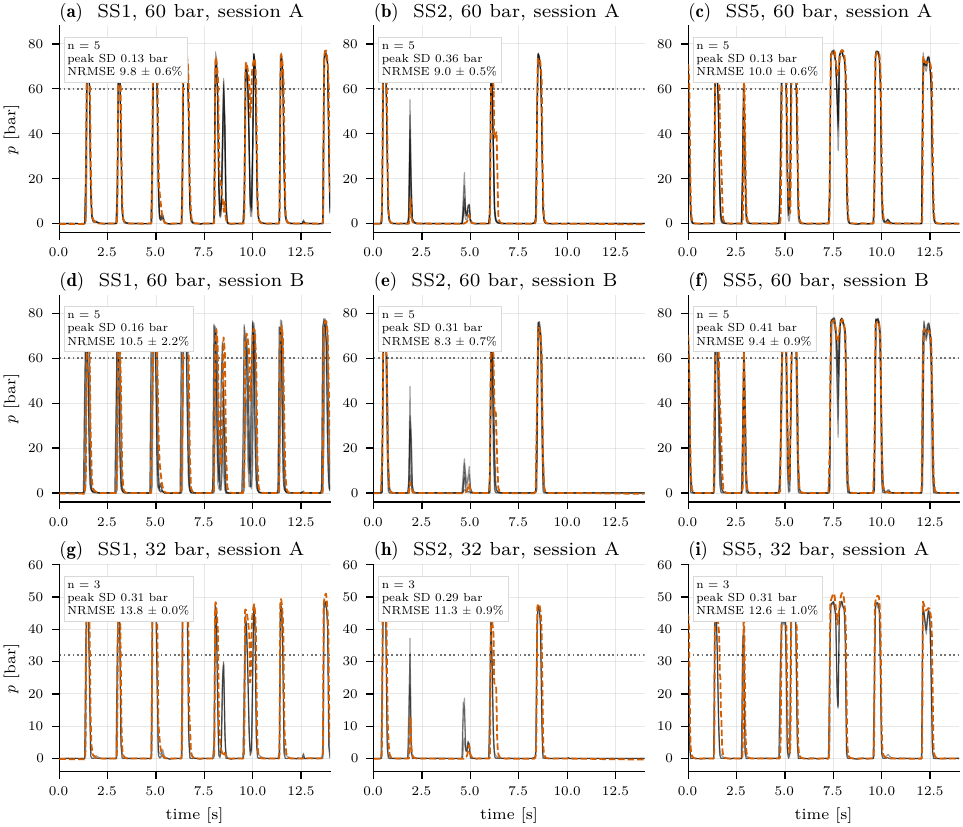}
\caption{The nine repeated sea-state groups. Grey, every measured repeat of
the group overlaid on a common 14\,s window; dashed, the free-run model of the
first record of the group; dotted, the crack pressure. The boxes quote the group
size, the record-peak standard deviation and the NRMSE of the model across the
repeats, from the same cache as Table~\ref{Table_S7}.}
\label{Figure_S8}\end{figure}

Figure~\ref{Figure_S8} overlays every repeat of each repeated group on a
common 14\,s window, with the free-run model of the first record of the group.
Table~\ref{Table_S7} gives the group-level statistics behind the
experimental floor used throughout the paper, and
Table~\ref{Table_S8} lists each of the 37 repeat records beyond
the primaries individually. Sessions A and B are separated by a
pressure-controller reset.

\begin{table}[H]
\centering\footnotesize
\caption{The nine repeated sea-state groups: run-to-run standard deviations of
the discharge-plateau level and of the record peak pressure, and the stability
of the model free-run error across repeats.}
\label{Table_S7}
\setlength{\tabcolsep}{5pt}
\begin{tabular}{lrrrrr}
\toprule
Group & $n$ & plateau SD [bar] & record-peak SD [bar] & model NRMSE [\%] & run-to-run SD [\%] \\
\midrule
SS1, 32 bar, session A & 3 & 0.26 & 0.31 & 13.8 & 0.0 \\
SS2, 32 bar, session A & 3 & 0.16 & 0.29 & 11.3 & 0.9 \\
SS5, 32 bar, session A & 3 & 0.30 & 0.31 & 12.6 & 1.0 \\
SS1, 60 bar, session A & 5 & 0.11 & 0.13 & 9.8 & 0.6 \\
SS2, 60 bar, session A & 5 & 0.20 & 0.36 & 9.0 & 0.5 \\
SS5, 60 bar, session A & 5 & 0.06 & 0.13 & 10.0 & 0.6 \\
SS1, 60 bar, session B & 5 & 0.23 & 0.16 & 10.5 & 2.2 \\
SS2, 60 bar, session B & 5 & 0.06 & 0.31 & 8.3 & 0.7 \\
SS5, 60 bar, session B & 5 & 0.22 & 0.41 & 9.4 & 0.9 \\
\bottomrule
\end{tabular}

\end{table}

\begin{table}[H]
\centering\scriptsize
\caption{Every repeat record beyond the primaries (37 records), reading top to
bottom, left block first. Groups are labelled sea state / valve setting /
session.}
\label{Table_S8}
\setlength{\tabcolsep}{4pt}
\begin{tabular}{llrrr@{\hspace{2.0em}}llrrr}
\toprule
Rec. & group & NRMSE [\%] & $\hat p_{\mathrm{meas}}$ [bar] & $\hat p_{\mathrm{model}}$ [bar] & Rec. & group & NRMSE [\%] & $\hat p_{\mathrm{meas}}$ [bar] & $\hat p_{\mathrm{model}}$ [bar] \\
\midrule
TC08 & SS1/60/A & 9.1 & 78.3 & 78.9 & TC44 & SS2/60/B & 7.8 & 78.4 & 78.4 \\
TC11 & SS1/60/A & 9.3 & 78.2 & 78.9 & TC47 & SS2/60/B & 7.6 & 78.5 & 78.5 \\
TC14 & SS1/60/A & 9.8 & 78.3 & 78.9 & TC50 & SS2/60/B & 8.2 & 78.9 & 78.5 \\
TC17 & SS1/60/A & 10.1 & 78.2 & 78.9 & TC36 & SS3/60/B & 9.8 & 79.8 & 81.0 \\
TC09 & SS2/60/A & 8.9 & 78.0 & 78.4 & TC39 & SS4/60/B & 12.0 & 82.0 & 83.6 \\
TC12 & SS2/60/A & 8.5 & 78.0 & 78.4 & TC35 & SS5/60/B & 10.1 & 79.1 & 80.1 \\
TC15 & SS2/60/A & 9.1 & 77.8 & 78.4 & TC42 & SS5/60/B & 10.3 & 79.4 & 80.1 \\
TC18 & SS2/60/A & 9.8 & 78.1 & 78.5 & TC45 & SS5/60/B & 8.0 & 79.4 & 80.1 \\
TC10 & SS5/60/A & 9.3 & 78.9 & 80.1 & TC48 & SS5/60/B & 8.8 & 79.3 & 80.1 \\
TC13 & SS5/60/A & 9.7 & 78.7 & 80.1 & TC51 & SS5/60/B & 9.8 & 79.5 & 80.1 \\
TC16 & SS5/60/A & 10.1 & 79.0 & 80.1 & TC38 & SS6/60/B & 10.5 & 80.9 & 82.5 \\
TC19 & SS5/60/A & 10.1 & 78.8 & 80.1 & TC37 & SS7/60/B & 9.8 & 80.1 & 81.7 \\
TC33 & SS1/60/B & 11.0 & 78.7 & 78.9 & TC27 & SS1/32/A & 13.8 & 49.5 & 52.9 \\
TC40 & SS1/60/B & 11.7 & 78.8 & 78.9 & TC30 & SS1/32/A & 13.8 & 49.6 & 52.9 \\
TC43 & SS1/60/B & 13.3 & 79.0 & 78.9 & TC28 & SS2/32/A & 11.8 & 49.2 & 52.4 \\
TC46 & SS1/60/B & 7.9 & 79.0 & 78.9 & TC31 & SS2/32/A & 11.8 & 49.5 & 52.4 \\
TC49 & SS1/60/B & 8.6 & 79.0 & 78.9 & TC29 & SS5/32/A & 13.3 & 50.0 & 54.0 \\
TC34 & SS2/60/B & 9.0 & 78.4 & 78.5 & TC32 & SS5/32/A & 13.2 & 50.1 & 54.0 \\
TC41 & SS2/60/B & 9.0 & 78.7 & 78.4 & & & & & \\
\bottomrule
\end{tabular}

\end{table}

\section{The re-sprung valve: all four runs}
\label{si:ood}

\begin{figure}[H]\centering
\includegraphics[width=\linewidth]{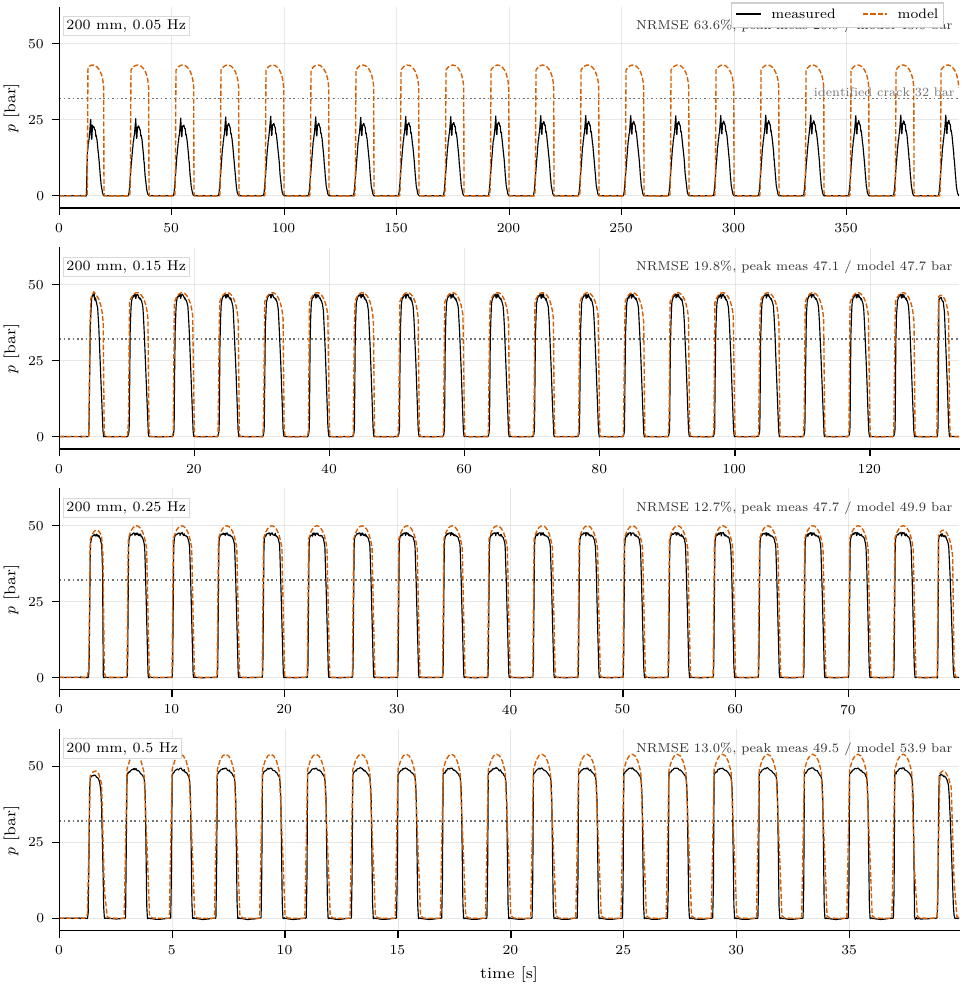}
\caption{All four re-sprung-valve runs (200\,mm) in full, ordered by
frequency: measured (solid) and free-run model (dashed) with the identified
32\,bar crack (dotted). The 0.05\,Hz record in the first row is the disclosed
out-of-distribution failure discussed in Section~4.7 of the paper.}
\label{Figure_S9}\end{figure}

Figure~\ref{Figure_S9} shows the four sinusoidal runs on the hand-re-adjusted
relief valve in full, and Table~\ref{Table_S9} the matching numbers. Three
of the four transfer with errors comparable to the in-distribution campaign.
The fourth is the 0.05\,Hz partial stroke, shown first, whose measured peak
(26.9\,bar) never reaches the identified 32\,bar crack while the model
over-cracks to 43.0\,bar. This is the partial-stroke regime identified as the
weakest regime of the model, exercised where the valve area law is pure extrapolation; it
bounds the transfer claim of Section~4.7 of the paper to strokes that reach
the regulated plateau, and it is disclosed rather than excluded.

\begin{table}[H]
\centering\footnotesize
\caption{The four re-sprung-valve runs. The 0.05\,Hz row is the disclosed
failure; the other three reach the regulated plateau and transfer.}
\label{Table_S9}
\setlength{\tabcolsep}{5pt}
\begin{tabular}{rrrrrr}
\toprule
$f$ [Hz] & duration [s] & NRMSE [\%] & RMSE [bar] & peak meas / model [bar] & $\le$5 bar [\%] \\
\midrule
0.05 & 400 & 63.6 & 17.2 & 26.9 / 43.0 & 51 \\
0.15 & 133 & 19.8 & 9.4 & 47.1 / 47.7 & 100 \\
0.25 & 80 & 12.7 & 6.1 & 47.7 / 49.9 & 100 \\
0.5 & 40 & 13.0 & 6.5 & 49.5 / 53.9 & 100 \\
\bottomrule
\end{tabular}

\end{table}

\section{Endurance: the thousand-cycle runs}
\label{si:endurance}

\begin{figure}[H]\centering
\includegraphics[width=\linewidth]{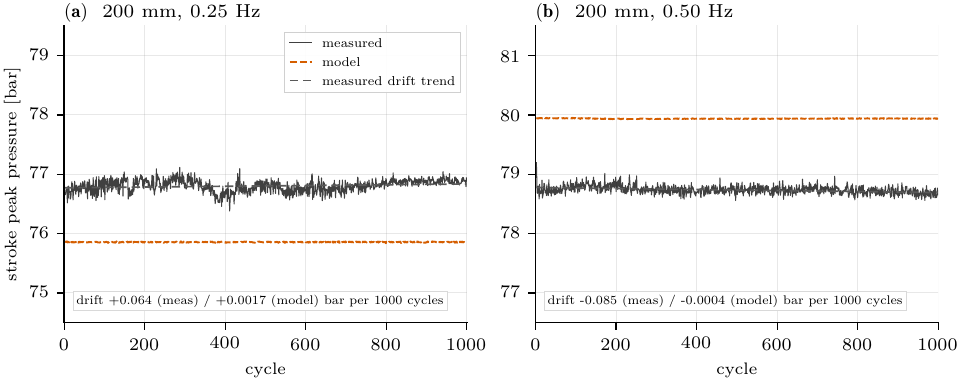}
\caption{Per-cycle stroke peak pressure over the two thousand-cycle endurance
runs, measured (solid) and free-run model (dashed), with the fitted measured
drift trend (grey). The first and last cycles of each record are partial and
are not drawn.}
\label{Figure_S10}\end{figure}

Figure~\ref{Figure_S10} shows the per-cycle stroke peaks of the two
thousand-cycle records with the model run over the same horizon, and
Table~\ref{Table_S10} the drift numbers. The measured drift stays below
0.1\,bar per thousand cycles at both frequencies, and the model, which
contains no wear or thermal state, correctly predicts an essentially flat
peak line.

\begin{table}[H]
\centering\footnotesize
\caption{The thousand-cycle endurance runs, drift per $10^3$ cycles.}
\label{Table_S10}
\setlength{\tabcolsep}{5pt}
\begin{tabular}{lrrrrr}
\toprule
Run & cycles & mean peak meas / model [bar] & drift meas [bar/10$^3$ cyc] & drift model & NRMSE [\%] \\
\midrule
200 mm, 0.25 Hz & 1000 & 76.8 / 75.9 & +0.064 & +0.0017 & 3.9 \\
200 mm, 0.50 Hz & 1000 & 78.7 / 79.9 & -0.085 & -0.0004 & 5.1 \\
\bottomrule
\end{tabular}

\end{table}

\section{Force channel: every record}
\label{si:force}

\begin{figure}[H]\centering
\includegraphics[width=\linewidth]{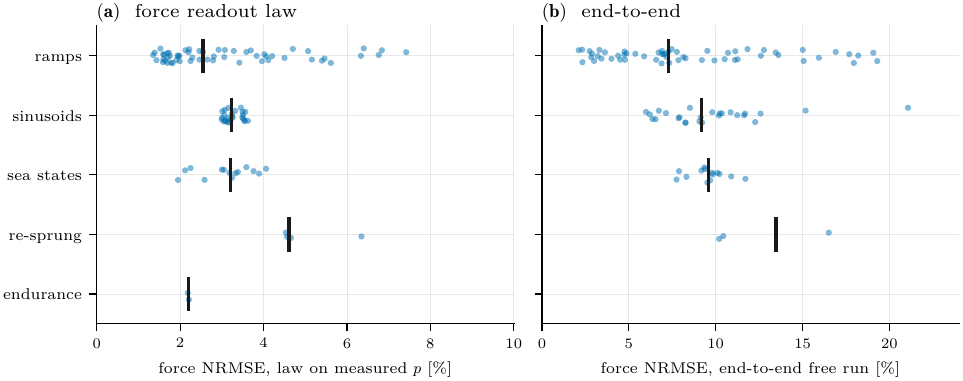}
\caption{Force-channel NRMSE of every record by regime (dots; black bar,
regime median). (a) Readout law on the measured pressure. (b) End-to-end on
the free-run pressure.}
\label{Figure_S11}\end{figure}

The paper validates the force readout law in Section~4.9; this section gives
the per-record picture across all 103 records. The law error applies the
force readout law of Section~3 of the paper to the measured pressure,
isolating the readout itself;
the end-to-end error applies it to the free-run pressure, so it inherits the
phase penalty of the pressure model. Figure~\ref{Figure_S11} summarises both by
regime, and Tables~\ref{Table_S11} and \ref{Table_S12} list
every record. The endurance records archive no end-to-end force trace, so
that column is empty for them.

\begin{table}[H]
\centering\scriptsize
\caption{Force-channel metrics for all 57 ramp segments: readout law on
measured pressure, end-to-end on the free run, peak capture at the force
criterion and signed peak bias. Two column blocks, left block first.}
\label{Table_S11}
\setlength{\tabcolsep}{4pt}
\begin{tabular}{lrrrr@{\hspace{1.6em}}lrrrr}
\toprule
Segment & law [\%] & e2e [\%] & cap.\ [\%] & bias [kN] & Segment & law [\%] & e2e [\%] & cap.\ [\%] & bias [kN] \\
\midrule
R50\_00 & 6.3 & 15.9 & 100 & +0.47 & R300\_29 & 5.6 & 17.9 & 100 & +0.93 \\
R50\_01 & 3.4 & 9.9 & 100 & +0.82 & R300\_30 & 3.8 & 11.1 & 50 & +1.09 \\
R50\_02 & 2.1 & 4.9 & 100 & +0.01 & R300\_31 & 2.9 & 7.9 & 50 & +1.08 \\
R50\_03 & 1.8 & 5.9 & 100 & -1.13 & R300\_32 & 2.3 & 4.5 & 100 & +0.89 \\
R50\_04 & 1.7 & 6.9 & 50 & -1.15 & R300\_33 & 1.9 & 2.8 & 100 & +0.60 \\
R50\_05 & 1.6 & 6.9 & 0 & -2.02 & R300\_34 & 1.7 & 2.3 & 100 & -0.24 \\
R50\_06 & 1.6 & 7.3 & 50 & -1.22 & R300\_35 & 1.6 & 3.4 & 100 & -0.41 \\
R50\_07 & 1.6 & 7.8 & 50 & -1.33 & R300\_36 & 1.5 & 4.4 & 0 & -2.16 \\
R50\_08 & 2.8 & 7.5 & -- & -- & R300\_37 & 4.2 & 11.8 & -- & -- \\
R100\_09 & 6.4 & 16.9 & 100 & +0.57 & R400\_38 & 6.8 & 18.2 & 100 & +0.69 \\
R100\_10 & 2.5 & 6.6 & 100 & +0.32 & R400\_39 & 4.7 & 11.1 & 100 & +0.72 \\
R100\_11 & 2.0 & 4.0 & 100 & +0.26 & R400\_40 & 4.5 & 12.8 & 100 & +1.75 \\
R100\_12 & 1.7 & 3.3 & 100 & -0.14 & R400\_41 & 4.0 & 12.6 & 50 & +1.28 \\
R100\_13 & 1.4 & 5.4 & 100 & -0.94 & R400\_42 & 3.1 & 9.2 & 50 & +1.43 \\
R100\_14 & 1.4 & 7.0 & 100 & -0.89 & R400\_43 & 2.6 & 6.7 & 50 & +1.18 \\
R100\_15 & 1.4 & 8.3 & 50 & -1.65 & R400\_44 & 2.2 & 4.8 & 100 & +0.82 \\
R100\_16 & 3.1 & 13.6 & -- & -- & R400\_45 & 2.0 & 3.2 & 100 & +1.26 \\
R200\_17 & 4.0 & 8.1 & 100 & +0.58 & R400\_46 & 1.8 & 2.1 & 100 & +0.29 \\
R200\_18 & 5.1 & 15.1 & 100 & +0.84 & R400\_47 & 5.2 & 7.3 & -- & -- \\
R200\_19 & 3.6 & 10.5 & 100 & +0.83 & R500\_48 & 6.8 & 17.7 & 100 & +0.76 \\
R200\_20 & 2.7 & 7.1 & 100 & +0.86 & R500\_49 & 5.4 & 13.5 & 100 & +1.51 \\
R200\_21 & 2.2 & 4.7 & 100 & +0.41 & R500\_50 & 5.5 & 19.3 & 50 & +1.14 \\
R200\_22 & 1.9 & 2.9 & 100 & +0.28 & R500\_51 & 4.1 & 15.0 & 50 & +1.44 \\
R200\_23 & 1.8 & 2.3 & 100 & +0.24 & R500\_52 & 3.3 & 11.3 & 0 & +3.19 \\
R200\_24 & 1.7 & 2.7 & 100 & +0.13 & R500\_53 & 2.8 & 9.5 & 50 & +1.52 \\
R200\_25 & 1.6 & 3.6 & 100 & -0.18 & R500\_54 & 2.5 & 6.9 & 50 & +1.36 \\
R200\_26 & 4.1 & 10.7 & -- & -- & R500\_55 & 2.2 & 4.7 & 50 & +1.07 \\
R300\_27 & 7.4 & 19.1 & 100 & +0.63 & R500\_56 & 2.0 & 3.0 & 100 & +1.46 \\
R300\_28 & 3.7 & 7.2 & 100 & +0.52 & & & & & \\
\bottomrule
\end{tabular}

\end{table}

\begin{table}[H]
\centering\scriptsize
\caption{As Table~\ref{Table_S11}, for the sinusoidal, sea-state,
re-sprung and endurance records.}
\label{Table_S12}
\setlength{\tabcolsep}{4pt}
\begin{tabular}{lrrrr@{\hspace{1.6em}}lrrrr}
\toprule
Record & law [\%] & e2e [\%] & cap.\ [\%] & bias [kN] & Record & law [\%] & e2e [\%] & cap.\ [\%] & bias [kN] \\
\midrule
0.25 Hz & 2.2 & -- & 100 & -1.10 & D100\_F0,40 & 3.5 & 7.9 & 12 & -2.38 \\
0.50 Hz & 2.2 & -- & 100 & -1.05 & D100\_F0,50 & 3.5 & 10.2 & 10 & -2.30 \\
D200\_F0,05\_Adj\_Leser & 6.3 & 50.7 & 51 & +2.10 & D100\_F0,60 & 3.5 & 10.9 & 10 & -2.13 \\
D200\_F0,15\_Adj\_Leser & 4.5 & 16.5 & 100 & -1.15 & D100\_F0,70 & 3.5 & 12.6 & 15 & -2.03 \\
D200\_F0,25\_Adj\_Leser & 4.7 & 10.2 & 100 & -0.80 & D100\_F0,80 & 3.5 & 12.3 & 15 & -2.04 \\
D200\_F0,50\_Adj\_Leser & 4.6 & 10.4 & 100 & -0.21 & D100\_F0,90 & 3.6 & 15.2 & 67 & -3.01 \\
TC01 & 2.1 & 9.5 & 66 & -0.16 & D150\_F0,10 & 3.3 & 8.3 & 37 & -2.01 \\
TC02 & 1.9 & 7.9 & 76 & -0.72 & D150\_F0,20 & 3.3 & 6.5 & 20 & -2.19 \\
TC03 & 2.2 & 9.7 & 59 & -0.09 & D150\_F0,30 & 3.2 & 7.1 & 23 & -1.90 \\
TC04 & 2.6 & 10.9 & 57 & +0.11 & D150\_F0,40 & 3.2 & 9.1 & 91 & -1.71 \\
TC05 & 3.0 & 7.7 & 57 & -1.04 & D150\_F0,50 & 3.1 & 10.2 & 95 & -1.58 \\
TC06 & 3.2 & 8.3 & 53 & -0.92 & D150\_F0,60 & 3.1 & 10.4 & 95 & -1.42 \\
TC07 & 3.4 & 9.8 & 48 & -0.96 & D150\_F0,70 & 3.1 & 11.6 & 95 & -1.16 \\
TC20 & 3.2 & 11.7 & 78 & +0.02 & D200\_F0,05 & 3.2 & 21.1 & 86 & -1.09 \\
TC21 & 3.0 & 9.3 & 84 & -0.49 & D200\_F0,10 & 3.2 & 11.2 & 34 & -1.86 \\
TC22 & 3.3 & 10.2 & 80 & -0.49 & D200\_F0,15 & 3.2 & 8.3 & 29 & -1.92 \\
TC23 & 3.6 & 10.1 & 84 & -0.17 & D200\_F0,20 & 3.1 & 6.0 & 24 & -1.82 \\
TC24 & 3.8 & 9.8 & 91 & -0.46 & D200\_F0,25 & 3.1 & 6.4 & 83 & -1.63 \\
TC25 & 3.9 & 9.4 & 90 & -0.30 & D200\_F0,30 & 3.1 & 6.2 & 96 & -1.41 \\
TC26 & 4.1 & 9.2 & 90 & -0.13 & D200\_F0,35 & 3.0 & 6.7 & 100 & -1.37 \\
D100\_F0,10 & 3.6 & 11.7 & 37 & -1.65 & D200\_F0,40 & 3.0 & 9.1 & 91 & -1.33 \\
D100\_F0,20 & 3.5 & 9.2 & 20 & -2.65 & D200\_F0,45 & 3.0 & 8.5 & 91 & -1.29 \\
D100\_F0,30 & 3.5 & 7.9 & 14 & -2.69 & D200\_F0,50 & 3.0 & 9.8 & 95 & -1.03 \\
\bottomrule
\end{tabular}

\end{table}

\section{Diagnostics behind specific claims}
\label{si:diag}

\begin{figure}[H]\centering
\includegraphics[width=0.72\linewidth]{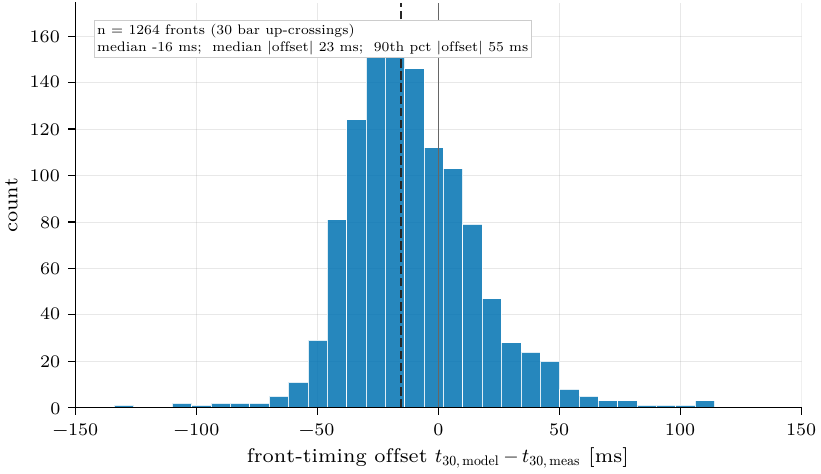}
\caption{Distribution of the compression-front timing offset between model and
measurement across the seven 60\,bar sea-state records, measured at the
30\,bar up-crossings (n = 1264 matched fronts). The median offset is
$-16$\,ms (model slightly early), the median absolute offset 23\,ms and the
90th percentile 55\,ms; at the measured front slopes of 0.4--0.8\,bar\,ms$^{-1}$
this timing alone produces instantaneous errors of several bar, which is the
phase penalty carried by the whole-record NRMSE (Sections~4.5--4.6 of the
paper).}
\label{Figure_S12}\end{figure}

\begin{figure}[H]\centering
\includegraphics[width=\linewidth]{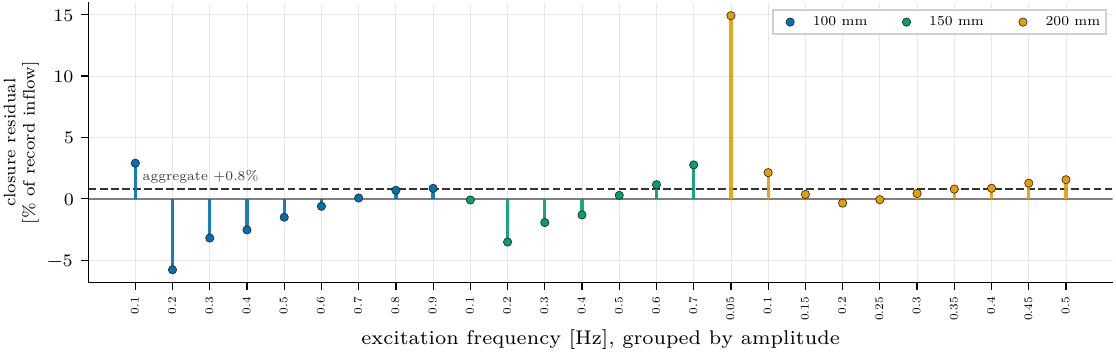}
\caption{Closure residual of the closed-cycle mass balance for each sinusoidal
record (positive = inflow not accounted for by the identified channels plus
the weighed discharge). The aggregate closure is $+0.8\%$ of the swept inflow;
per record the residual spans $-6\%$ to $+15\%$, the extreme being the
0.05\,Hz partial stroke. Because the blow-by pair is fitted to the campaign
residual, the closure checks the accounting rather than validating the model
independently (Section~4.11 of the paper).}
\label{Figure_S13}\end{figure}

\begin{figure}[H]\centering
\includegraphics[width=0.62\linewidth]{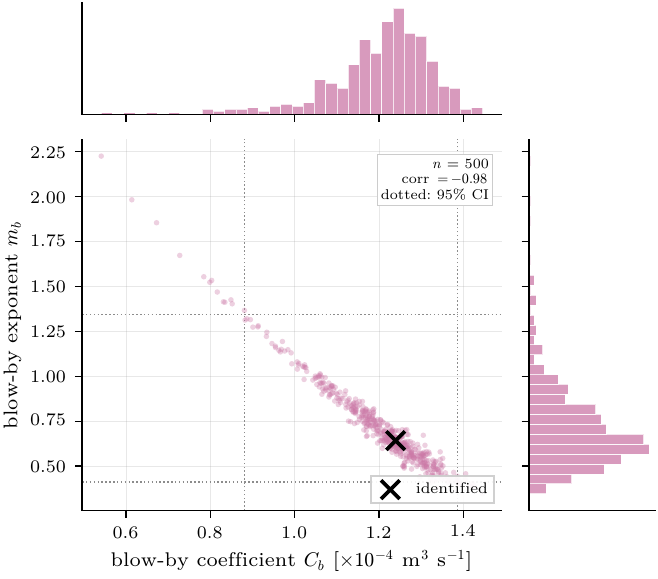}
\caption{Record-level bootstrap (500 draws) of the blow-by pair
$(C_b, m_b)$ with marginal histograms and the asymmetric 95\% intervals
(dotted). The mild negative correlation and the long upper tail of $m_b$
mirror the valve-pair geometry of Fig.~17a in the paper.}
\label{Figure_S14}\end{figure}

Three diagnostics stand behind numbers quoted in the paper. The compression-front
timing offset that the whole-record NRMSE carries as a phase penalty is resolved in
Fig.~\ref{Figure_S12}. Figure~\ref{Figure_S13} distributes the mass-balance
closure across the sinusoidal records rather than reporting it only in aggregate, and
the record-level bootstrap behind the blow-by intervals is given in
Fig.~\ref{Figure_S14}.

\end{document}